\documentclass{aa}
\usepackage{graphics}

\begin{document}

\thesaurus{06
          (02.01.3; 
           02.01.4; 
           02.12.1; 
           03.20.8; 
           11.01.2;
           11.17.2; 
           13.25.2)}

\title{X-ray photoionized plasma diagnostics with Helium-like ions. Application to Warm Absorber-Emitter in Active Galactic Nuclei.}
\author{Delphine Porquet\inst{1,2} \& Jacques Dubau\inst{3}}

\offprints{D. Porquet}
\mail{Delphine.Porquet@obspm.fr}
\institute{DAEC, Observatoire de Paris, Section Meudon, F-92195 Meudon Cedex, France 
\and CEA/DSM/DAPNIA, Service d'Astrophysique, CEA Saclay, F-91191 Gif sur Yvette Cedex, France
\and DARC, Observatoire de Paris, Section Meudon, F-92195 Meudon Cedex, France}
\date{Received ...; accepted ...}

\titlerunning{Photoionized plasma diagnostics with He-like ions}

\authorrunning{Porquet \& Dubau}
\maketitle

%%%%%%%%%%%%%%%%%%%%%%%%%%%%%%%%%%%%%%%%%%%%%%%%%%%%%%%%%%%%%%%%%%%%%%%%%%%%%%%%%%%%%%%%%%%%
%\large
\begin{abstract}

We present He-like line ratios (resonance, intercombination and forbidden lines) for totally and partially photoionized media. For solar plasmas, these line ratios are already widely used for density and temperature diagnostics of coronal (collisional) plasmas. In the case of  totally and partially photoionized plasmas, He-like line ratios allow for the determination of the ionization processes involved in the plasma (photoionization with or without an additional collisional ionization process), as well as the density and the electronic temperature. \\ 
With the new generation of X-ray satellites, Chandra / AXAF, XMM and Astro-E, it will be feasible to obtain both high spectral resolution and high sensitivity observations. Thus in the coming years, the ratios of these three components will be measurable for a large number of non-solar objects.\\ 
In particular, these ratios could be applied to the Warm Absorber-Emitter, commonly present in Active Galactic Nuclei (AGN). A better understanding of the Warm Absorber connection to other regions (Broad Line Region, Narrow Line Region) in AGN (Seyferts type-1 and type-2, low- and high-redshift quasars...) will be an important key to obtaining strong constraints on unified schemes.\\
 We have calculated He-like line ratios, for $Z$=6, 7, 8, 10, 12 and 14, taking into account the upper level radiative cascades which we have computed for radiative and dielectronic recombinations and collisional excitation. The atomic data are tabulated over a wide range of temperatures in order to be used for interpreting a large variety of astrophysical plasmas.\\
\end{abstract}
%%%%%%%%%%%%%%%%%%%%%%%%%%%%%%%%%%%%%%%%%%%%%%%%%%%%%%%%%%%%%%%%%%%%%%%%%%%%%%%%%%%%%%%%%%%%
\section{Introduction} \label{sec:Introduction}
The new X-ray satellites (Chandra, XMM and Astro-E) will offer unprecedented high spectral resolution and high sensitivity spectra. Indeed, it will be possible to observe and to separate, in the X-ray range, the three most intense lines of He-like ions: the {\bf resonance} line ({\bf w}: 1s$^{2}$\,\element[][1]{S}$_{\mathrm{0}}$ -- 1s2p\,\element[][1]{P}$_{\mathrm{1}}$), the {\bf intercombination} lines ({\bf x,y}: 1s$^{2}$\,\element[][1]{S}$_{0}$ -- 1s2p \element[][3]{P}$_{2,1}$ respectively) and the {\bf forbidden} line ({\bf z}: 1s$^{2}$\,\element[][1]{S}$_{0}$ -- 1s2s\,\element[][3]{S}$_{1}$). They correspond to transitions between the $n$=2 shell and the $n$=1 ground state shell (see Figure~\ref{gotrian}).\\
%%%%%%%%%%%%%%%%%%%%%%%%%%%%%%
%PLEASE INSERT FIG 1
%%%%%%%%%%%%%%%%%%%%%%%%%%%%%%
The ratios of these lines are already widely used for collisional (coronal) plasma diagnostics of various types of objects: solar flares, supernovae remnants, the interstellar medium and tokamak plasmas, i.e. for very hot collisional plasmas (Mewe \& Schrijver \cite{Mewe78a} \cite{Mewe78b}, Winkler et al. \cite{Winkler81}, Doyle \& Schwob \cite{Doyle82}, and Pradhan \& Shull \cite{Pradhan81}). 
As shown by Gabriel \& Jordan (\cite{Gabriel69}, \cite{GabrielJordan72}, \cite{Gabriel73}), these ratios are sensitive to electron density (R(n$_{\mathrm{e}}$), equation~\ref{eq:R}) and to electronic temperature (G(T$_{\mathrm{e}}$), equation~\ref{eq:G}):
\begin{equation}
\mathrm{ R~(n_e)~=~\frac{z}{(x+y)}} \label{eq:R}
\end{equation}
%and
\begin{equation}
\mathrm{ G~(T_e)=\frac{z+(x+y)}{w}} \label{eq:G}
\end{equation}
As emphasized by Pradhan (\cite{Pradhan85}), Liedahl (\cite{Liedahl99}) and Mewe (\cite{Mewe99}) (see also Paerels et al. \cite{Paerelsetal98}), these plasma diagnostics could be also extended to study photoionized plasmas. Indeed, Pradhan has calculated the {\bf R} and {\bf G} ratios for highly charged ions (\ion{Ar}{xvii} and \ion{Fe}{xxv}) in ``recombination dominated non-coronal plasmas''. We present numerical calculations of these ratios, for six lighter ions, which could be applied directly for the first time to Chandra and XMM observations of the Warm Absorber present in Active Galactic Nuclei (AGN), and especially in Seyfert\,1.\\

\indent The Warm Absorber (WA) is a totally or a partially photoionized medium (with or without an additional ionization process), first proposed by Halpern (\cite{Halpern84}) in order to explain the shape of the X-ray spectrum of the \object{QSO MR2251-178}, observed with the Einstein Observatory. Its main signatures are the two high-ionization oxygen absorption edges, \ion{O}{vii} and \ion{O}{viii} at 0.74 keV and 0.87 keV respectively, seen in fifty percent of Seyfert 1 galaxies at least (Nandra $\&$ Pounds \cite{Nandra94}, Reynolds \cite{Reynolds97}, George et al. \cite{George98}). According to Netzer (\cite{Netzer93}), an emission line spectrum from the WA should also be observed. 
Indeed, He-like ion lines have been observed in different types of Seyfert galaxies (\object{NGC 3783}: George et al. \cite{George95}, \object{MCG-6-30-15}: Otani et al. \cite{Otani96}, \object{E 1615+061}: Piro et al. \cite{Piro97}, \object{NGC 4151}: Leighly et al. {\cite{Leighly97}, \object{NGC 1068}: Ueno et al. {\cite{Ueno94}, Netzer \& Turner \cite{Netzer97}, and Iwasawa et al. \cite{Iwasawa97}). 
The WA is supposed to be at least a two-zone medium with an inner part (called the ``inner WA'') associated with \ion{O}{viii} and an outer part (called the ``outer WA''), less ionized, associated with \ion{O}{vii} (Reynolds \cite{Reynolds97}, Porquet et al. \cite{Porquet99}). Furthermore, the \ion{O}{vii} line is predicted to be the strongest line associated with the outer WA; the \ion{Ne}{ix} line is predicted to be one of the strongest lines formed in the inner WA (Porquet et al. \cite{Porquet98}).\\
The ionization processes, that occur in the Warm Absorber, are still not very well known. Indeed, even though the WA is commonly thought to be a photoionized gas, an additional ionization process cannot be ruled out (Porquet \& Dumont \cite{PorquetDumont98}, Porquet et al. \cite{Porquet99}, Nicastro et al. \cite{Nicastro99}). Thus, in the present paper, we do not restrict ourselves to only a single type of plasma, but rather study the following cases.\\ %
We consider a ``pure photoionized plasma'' to be a plasma ionized by high energy photons (external ionizing source). For such a plasma, H-like radiative recombination (and dielectronic recombination at high temperature) are dominant compared to electronic excitation from the ground level (1s$^{2}$) of He-like ions. The lines are formed by recombination.\\
A ``hybrid plasma'' is a partially photoionized plasma, but with an additional ionization process, e.g. collisional (internal ionizing source). For this case, He-like electronic excitation processes from the ground level are usually as important as H-like recombinations, and may even dominate. The lines are formed by collisional excitation from the ground level with or without recombination.\\
 
In the next section, we introduce the atomic data calculations needed for such plasmas and we emphasize the role of upper-level radiative cascade contributions calculated in this paper for the populations of the $n$=2 shell levels. In section~\ref{sec:diagnostics}, we develop line diagnostics of the ionization process (temperature) and the density for pure photoionized and hybrid plasmas. We give the corresponding numerical calculations of the line ratios for \ion{C}{v}, \ion{N}{vi}, \ion{O}{vii}, \ion{Ne}{ix}, \ion{Mg}{xi}, and \ion{Si}{xiii}. In section~\ref{sec:practical}, we give a practical method for using these results to determine the physical parameters of the WA, in the context of the expected data from the new X-ray satellites (section~\ref{sec:satellite}).

%%%%%%%%%%%%%%%%%%%%%%%%%%%%%%%%%%%%%%%%%%%%%%%%%%%%%%%%%%%%%%%%%%%%%%%%%%%%%%%%%%%%%%%%%%%%
%%%%%%%%%%%%%%%%%%%%%%%%%%%%%%%%%%%%%%%%%%%%%%%%%%%%%%%%%%%%%%%%%%%%%%%%%%%%%%%%%%%%%%%%%%%%
\section{Atomic data}\label{sec:atomicdata}

Liedahl (\cite{Liedahl99}) described the basic mechanisms of density diagnostics for X-ray photoionized plasmas from He-like ions. As he noted, a proper calculation of the population of the $n$=2 shell levels depends upon a number of additional levels. We propose in this article to use extensive calculations of atomic data taking into account upper level (n$>$2) radiative cascade contribution on $n$=2 shell levels for \ion{C}{v}, \ion{N}{vi}, \ion{O}{vii}, \ion{Ne}{ix}, \ion{Mg}{xi}, and \ion{Si}{xiii}, to give a much more precise treatment of this plasma diagnostic. \\
We consider in this paper, the main atomic processes involved in pure photoionized and hybrid plasmas: radiative recombination and dielectronic recombination (only important for high temperature plasmas), collisional excitation inside the $n$=2 shell, and collisional excitation from the ground level (important for high temperature plasmas). 

\subsection{Energy levels, radiative transition probabilities}

Using the SUPERSTRUCTURE code (Eissner et al. \cite{Eissner74}), we have calculated the energy levels for the first 49 fine-structure levels ($\mathrm{^{2S+1}{L}_{J}}$) for the six ions. This corresponds to the levels of the first 15 configurations (from 1s$^{2}$ to 1s5g). Nevertheless, for the first seven levels, we have preferred to use the Vainshtein \& Safronova (\cite{Vainshtein85}) data which have a slightly better accuracy ($\sim$10$^{-3}$).\\
%%%%%%%%%%%%%%%%%%%%%%%%%%%%%%
%PLEASE INSERT TAB 1
%%%%%%%%%%%%%%%%%%%%%%%%%%%%%%
In Table~\ref{energielevel}}, in order to reduce the amount of data, we only report the energy levels for the first 17 levels ($n$=1 to $n$=3 shell). The values for the others levels are available on request.
The transition probabilities (A$_{\mathrm{ki}}$ in s$^{-1}$) for the ``allowed'' transition (E1), are also calculated by the SUPERSTRUCTURE code; for the other transitions (M1, M2 \& 2E1) the A$_{\mathrm{ki}}$ values are from Lin et al. (\cite{Lin77}). In a same way, only direct radiative contributions of the first 17 levels onto the first 7 levels are given in Table~\ref{Aki1}.\\
%%%%%%%%%%%%%%%%%%%%%%%%%%%%%%
%PLEASE INSERT TAB 2
%%%%%%%%%%%%%%%%%%%%%%%%%%%%%%

\subsection{Recombination coefficient rates}

Blumenthal et al. (\cite{Blumenthal72}) have noted that radiative and dielectronic recombination can have a significant effect on the populations of the $n$=2 states in He-like ions through radiative cascades from higher levels as well as through direct recombination. 

\subsubsection{Radiative recombination (RR) coefficients rates}\label{sec:RR}

\noindent For radiative recombination rate coefficients, we have used the method of Bely-Dubau et al. (\cite{Bely-Dubau82a}). This method is based on $(Z-0.5)$ screened hydrogenic approximation of the Burgess (\cite{Burgess58}) formulae, as we explain below.\\

\noindent For recombination of a bare nucleus of charge $Z$ to form H-like ions, Burgess (\cite{Burgess58}) fitted simple power law expressions to the ``exact'' theoretical hydrogenic photoionization cross-sections $\sigma_{\mathrm{nl}}$(E) (in cm$^{2}$) for the {$n l$} levels ($1\le n\le 12$ and $0\le l \le n-1$). According to Burgess ``for moderately small $n$, the errors should be not more than about 5\%. Such accuracy should be sufficient for most astrophysical applications''. 
\begin{eqnarray}\label{eq:sectBurgess}
&\sigma_{nl}(E)&=0.55597\, \frac {Z^2}{n^2}\frac 1{2(2l+1)}\nonumber\\
&&\times \left[l~|\sigma(nl,o~l-1)|^{2}\left(\frac {I_H Z^2}{n^2 E}\right)^{\gamma(nl,l-1)}
\right . \nonumber \\ 
&& \left . +~(l+1)~|\sigma(nl,o~l+1)|^{2} 
\left(\frac{I_H Z^2}{n^2 E}\right)^{\gamma(nl,l+1)}\right] 
\end{eqnarray} 
\noindent where E is the photon energy $ E\ge I_H Z^2/n^2$.\\
Bely-Dubau et al. (\cite{Bely-Dubau82a}) used this equation for He-like 1s$nl$ levels by replacing $Z$ with $(Z-0.5)$. The quantity $(Z-0.5)$ was chosen to take into account the screening of the $1s$ orbital. To check the validity of this assumption we compared the photoionization cross sections obtained from equation (3) to the recent calculations of the Opacity Project by Fernley et al. (\cite{Fernley87}). In Figure~\ref{photoionisation} are plotted photoionisation cross sections for 1s2s $^1$S $^3$S, 1s2p $^1$P $^3$P and 1s10d $^1$D $^3$D for $Z$ =6, 10, 14 (continuous curves), scaled as $(Z-0.5)$. With the exception of 1s2p $^1$P, the three continuous curves can hardly be distinguished. Furthermore, the curves do not differ when passing from singlet to triplet cases. This is strong evidence that for 1s$nl$, it is possible to use screened hydrogenic calculations. For comparison, we give the present calculation corresponding to formulae (3) modified (empty circles). \\
%%%%%%%%%%%%%%%%%%%%%%%%%%%%%%
%PLEASE INSERT FIG 2
%%%%%%%%%%%%%%%%%%%%%%%%%%%%%%

\noindent  The Opacity Project data were taken from the Topbase Bank (Cunto et al. \cite{Cunto93}). This bank includes the 1s$nl$ photoionization cross sections for $1 \le n \le 10$ and $l=0,1,2$. The Burgess data, $\sigma(nl,o~l\pm 1)$ and $\gamma(nl,l\pm 1)$, are more complete since they also include $3\le l\le n-1$. Formula (3) is also more convenient since being analytic one can derive directly the radiative recombination rates (cm$^{3}$\,s$^{-1}$) from it.  

\begin{equation}\label{eq:alphaBurgess}
\mathrm{\alpha_{nl}(Z,T_{e})~=~8.9671\times10^{-23}~T_e^{3/2}~Z~f_{nl}(T_e)}
\end{equation}
where
T$_{\mathrm{e}}$ is the electronic temperature, $Z$ is the atomic number and
\begin{eqnarray}
&\mathrm{f_{nl}(T_{e})}& = \frac{x_\mathrm{n}^3}{\mathrm{n}^2}~[~\mathrm{l}|\sigma(\mathrm{nl,o~l-1})|^{2}~\Gamma_{\mathrm{c}}(\mathrm{x_n},3-\gamma(\mathrm{nl,l-1)})~ \nonumber\\
    &+ (\mathrm{l+1})& |\sigma(\mathrm{nl,o~l+1})|^{2}~\Gamma_{\mathrm{c}}(\mathrm{x_{n}},3-\gamma(\mathrm{nl,l+1}))]
\end{eqnarray}
\begin{equation}\label{eq:xn}
{\rm with} \qquad \mathrm{\mathrm{x_{n}}~=~\frac{\mathrm{Z^{2}~I_{H}}}{\mathrm{k~T_{e}~n}^{2}}~~~~\left(\frac{\mathrm{I_H}}{\mathrm{k}}=157\,890 \right)}
\end{equation}
\noindent The quantities $|\sigma(\mathrm{nl,o~l}\pm1)|/{\mathrm{n}}^{2}$ and $\gamma(\mathrm{nl,l}\pm1)$ are given in Table\,I of Burgess and 
\begin{equation}
\mathrm{\Gamma_{c}(x,p)=\frac{e^{x}}{x^p}~\int_{x}^{\infty}t^{(p-1)}~e^{-t}~dt}
\end{equation}

\noindent Finally, to transform H-like data to He-like data, we used the two following expressions for 1s$^{2}$ and 1s\,nl:
\begin{equation}
\mathrm{\noindent \alpha_{1s^{2}}~=~\frac{1}{2}~\alpha_{1s}\mathrm{(Z,T_{e})}~~~(n=1, ground~level)}
\end{equation}
\begin{equation}
\mathrm{\noindent \alpha_{1s~nl~(LSJ)}=\frac{(2J+1)}{(2L+1)~(2S+1)}~\alpha_{nl}(Z,T_{e})~~(n\geq2)}
\end{equation}
\noindent And we replace $Z$ by ($Z$-0.5) in formula (\ref{eq:alphaBurgess}) and (\ref{eq:xn}).\\

For 10$<$n$< \infty$, we have used the Seaton (\cite{Seaton59}) formula (see below) which gives RR rates for each quantum number $n$ (shell) of H-like ions. We have assumed that the $l$ recombination for such high $n$ is the same as for $n$=10. \\
Seaton derived his formula by expanding the Gaunt factor, usually taken to be one, to third order (Menzel \& Pekeris \cite{MenzelPekeris35}, Burgess \cite{Burgess58}). 
According to Seaton, the radiative recombination rates (in cm$^{3}$\,s$^{-1}$) for the $n$ shell of H-like ions can be written as:
\begin{equation}\label{eq:alphanl}
\mathrm{\alpha_{n(Z,T)}~=~5.197\times10^{-14}~Z~x_{n}^{3/2}~S_{n}(x_{n})}
\end{equation}
\begin{equation}
\mathrm{S(x_n)~=~X_0(x_n)+\frac{0.1728}{n^{\frac{2}{3}}}~{X_1(x_n})-\frac{0.0496}{n^{\frac{4}{3}}}~X_2(x_n)}
\end{equation}
\begin{equation}
\mathrm{X_0(x_n)=\Gamma_{c}(x,0)}\\
\end{equation}
\begin{equation}
\mathrm{X_{1}(x_{n})=\Gamma_{c}\!\left(x,\frac{1}{3}\right)-2~\Gamma_{c}\!\left(x,-\frac{2}{3}\right)}\\
\end{equation}
\begin{equation}
\mathrm{X_2(x_n)=\Gamma_{c}\!\left(x,\frac{2}{3}\right)-\frac{2}{3}~\Gamma_{c}\!\left(x,-\frac{1}{3}\right)+\frac{2}{3}~\Gamma_{c}\!\left(x,-\frac{4}{3}\right)}
\end{equation}

Next, we have computed the effects of cascades from $n>$2 levels on each 1s2l level (1s\,2s~$^{3}\mathrm{S}_{1}$,\,$^{1}\mathrm{S}_{0}$; 1s\,2p $^{3}\mathrm{P}_{0}$, $^{3}\mathrm{P}_{1}$, $^{3}\mathrm{P}_{2}$, $^{1}\mathrm{P}_{1}$; $n$=2 shell levels). The present study has shown that the radiative recombination (RR) is slowly convergent with $n$, thus the first 49 levels (n$\leq$5) are considered as fine-structure levels (LSJ), the levels from $n$=6 to $n$=10 (l=9) shells are separated in LS term (Bely-Dubau et al. \cite{Bely-Dubau82a}, \cite{Bely-Dubau82b}), and finally levels from $n$=11 to $n$=$\infty$ are taken into account inside $n$=10. Figure~\ref{delphine2.ps} shows the scaled direct plus upper (n$>$2) level radiative cascade RR rates $\alpha^{\mathrm{s}}$=T$^{1/2}$\,$\alpha$/($Z$-0.5)$^{2}$ versus T$^{\mathrm{s}}$=T/($Z$-0.5)$^{2}$ for 1s2l levels ($Z$= 8, 10, 12 and 14), and for comparison the direct RR contribution. T is in Kelvin. This points out the importance of the cascade contribution at low temperature. The $\alpha^{\mathrm{s}}$ curves are very well superposed and thus allows us to deduce the RR rate coefficients for other Z, as for example $Z=$ 9,11,13.\\
\noindent Tables~\ref{cascadeC6},\,~\ref{cascadeO8},\,~\ref{cascadeNe9},\,~\ref{cascadeMg11} and\,~\ref{cascadeSi13} report separately the direct and the cascade contribution to the RR rate coefficients for each 1s2l level.\\

%%%%%%%%%%%%%%%%%%%%%%%%%%%%%%
%PLEASE INSERT FIG 3
%%%%%%%%%%%%%%%%%%%%%%%%%%%%%%

We checked that the calculated rates summed over $n\ge 2$ and $l$, added to the rate of the 1s$^2$ (ground level) level, are similar to the total RR rates calculated by Arnaud \& Rothenflug (\cite{Arnaud85}), Pequignot et al. (\cite{Pequignot91}), Mazzotta et al. (\cite{Mazzotta98}), Jacobs et al. (\cite{Jacobs77}) (for He-like Fe ion) and Nahar (\cite{Nahar99}) (for \ion{O}{vii}). Since these authors used hydrogenic formulae, the RR rate coefficient depends on which screening value was used. As already noted, we have taken for our calculations a screening of 0.5 which is a realistic screening of the atomic nuclei by the 1s inner electron. Most probably, some of these authors have used a ($Z$-1) scaling. For example for \ion{C}{v}, a screening of unity implies a lower value by some 20\% with respect to the value obtained with a screening of 0.5. \\

\subsubsection{Dielectronic recombination (DR) coefficient rates}

For the low temperature range (photoionized plasma) considered in this paper the dielectronic recombination can be neglected. However at high temperatures, the contribution of DR is no longer negligible. Therefore, we have calculated DR coefficients rates (direct plus upper ($n>$2) level radiative cascade contribution).\\

\noindent We used the same method as Bely-Dubau et al. (\cite{Bely-Dubau82a}). The AUTOLSJ code (including the SUPERSTRUCTURE code) was run with 42 configurations belonging to 1s$nl$, 2s$nl$ and 2p$nl$, with $n\le 5$. All the fine-structure radiative and autoionization probabilities were calculated. For low $Z$ ions, it was necessary to do an extrapolation to higher $n$ autoionizing levels. Specifically, we extrapolate autoionization probabilities, as 1/$n^3$, while keeping the radiative probabilities constant. This extrapolation is not perfectly accurate, and we can estimate that the RD for C, N and O might be slightly over or under estimated.\\

In Table~\ref{cascadeC6}~,\ref{cascadeO8},~\ref{cascadeNe9},~\ref{cascadeMg11}, and \ref{cascadeSi13}, the DR rates are reported for $Z$=6, 8, 10, 12, 14 over a wide range of temperature.\\

%%%%%%%%%%%%%%%%%%%%%%%%%%%%%%
%PLEASE INSERT TAB 3
%%%%%%%%%%%%%%%%%%%%%%%%%%%%%%
%%%%%%%%%%%%%%%%%%%%%%%%%%%%%%
%PLEASE INSERT TAB 4
%%%%%%%%%%%%%%%%%%%%%%%%%%%%%%
%%%%%%%%%%%%%%%%%%%%%%%%%%%%%%
%PLEASE INSERT TAB 5
%%%%%%%%%%%%%%%%%%%%%%%%%%%%%%
%%%%%%%%%%%%%%%%%%%%%%%%%%%%%%
%PLEASE INSERT TAB 6
%%%%%%%%%%%%%%%%%%%%%%%%%%%%%%
%%%%%%%%%%%%%%%%%%%%%%%%%%%%%%
%PLEASE INSERT TAB 7
%%%%%%%%%%%%%%%%%%%%%%%%%%%%%%
%%%%%%%%%%%%%%%%%%%%%%%%%%%%%%
%PLEASE INSERT TAB 8
%%%%%%%%%%%%%%%%%%%%%%%%%%%%%%

\subsection{Electron excitation rate coefficients}\label{sec:exc}

The collisional excitation (CE) rate coefficient (in cm$^3$\,s$^{-1}$) for each 
transition is given by:
\begin{equation}
\mathrm{C_{ij}(T_{e})=\frac{8.60 \times 10^{-6}}{g_i~T^{1/2}}~exp\left(-\frac{\Delta E_{ij}}{k~T}\right)~\Upsilon_{ij}(T_{e})}
\end{equation}
Where $\Delta$E$_{\mathrm{ij}}$ is the energy of the transition, g$_{\mathrm{i}}$ is the statistical weight of the lower level of the transition, and $\Upsilon_{\mathrm{ij}}$ is the so-called effective collision strength of the transition i$\to$j.\\

The 1s\,2l--1s\,2l${'}$ transitions (i.e. inside the $n$=2 shell) are very important for density diagnostic purpose. The data are from Zhang \& Sampson (\cite{Zhang87}). \\
\noindent Below, we report scaled effective collision strength $\Upsilon^{\mathrm{s}}_{\mathrm{ij}}$=($Z$-0.5)$^{2}$\,$\Upsilon_{\mathrm{ij}}$. We also use a scaled electronic temperature T$^{\mathrm{s}}$ = T(K)/(1000 $Z^{2}$). The ($Z$-0.5)$^{2}$ coefficient has been chosen to obtain scaled $\Upsilon^{\mathrm{s}}$ almost independent of $Z$ (for 6\,$\leq Z \leq$\,14). In Figure~\ref{figure1s21s2l}, $\Upsilon^{\mathrm{s}}$(T$^{\mathrm{s}}$) is displayed for the four most important transitions (between 2$^{3}$S$_{1}$ and 2$^{3}$P$_{0,1,2}$ levels, and between 2$^{1}$S$_{0}$ and 2$^{1}$P$_{1}$ levels) including both direct and resonant contribution, and for comparison the direct contribution alone is shown for $Z$=8. We remark that the curves $\Upsilon^{\mathrm{s}}$(T$^{\mathrm{s}}$) are nearly identical for different $Z$, and for these transitions the resonant contribution is quite negligible since the two curves for $Z$=8 are superposed. The rates for the transitions between 2$^{3}$S$_{1}$ and 2$^{3}$P$_{0,1,2}$ levels are proportional to their statistical weight. The curves for transitions 2$^{3}$S$_{1}$--2$^{3}$P$_{1}$ 2$^{1}$S$_{0}$--2$^{1}$P$_{1}$ are nearly identical.\\ 
%%%%%%%%%%%%%%%%%%%%%%%%%%%%%%
%PLEASE INSERT FIG 4
%%%%%%%%%%%%%%%%%%%%%%%%%%%%%%
These high values of $\Upsilon^{\mathrm{s}}$ inside the $n$=2 shell and the small energy difference between these levels, favour transitions by excitation between the $n$=2 shell levels. Thus the excitation inside the $n$=2 shell should be taken into account even for low temperature plasmas.\\
Excitation from $n$=2 levels to higher shell levels can be neglected due to the weak population of the $n$=2 shell compared to the ground level ($n$=1) in a moderate density plasma and also due to the high $\Upsilon^{\mathrm{s}}$(T$^{\mathrm{s}}$) values inside the $n$=2 shell which favour transitions between the $n$=2 levels, as we see below.\\
\indent CE from the 1s$^{2}$ (ground) level to excited levels are only important for high temperature such as the hybrid case, due to the high energy difference between these levels. For 1s$^{2}$--1s2l transitions, we have used the effective collision strength values from Zhang \& Sampson (\cite{Zhang87}). These values include both non-resonant and resonant contributions.\\ 
\indent CE rates for the 1s$^{2}$--1snl (3$\leq$n$\leq$5) transitions are from Sampson et al. (\cite{Sampson83}). Their calculations do not include resonance effects but these are expected to be relatively small (Dubau \cite{Dubau94}). The rates converge as $n^{-3}$.\\

We have calculated the radiative cascade contribution from n$>$2 levels for each $n$=2 level. We have considered the first 49 levels, as fine-structure levels (LSJ); the contributions from the $n$=6 to $n$=$\infty$ levels are considered to converge as $n^{-3}$. The cascade contributions become more important for high temperatures. The cascade contribution (from $n>$2 levels) increases steadily with temperature and has an effect mostly on the 1s\,2s$^{3}$S$_{1}$ level. The resonant contribution increases then decreases with temperature. For high temperature plasmas, cascade effects should be taken into account. For very low temperature plasmas only the direct non-resonant contribution is important, except for the 1s\,2s\,$^{3}$S$_{1}$ level which also receives cascade from within the $n$=2 level, i.e. from 1s\,2p\,$^{3}$P$_{0,1,2}$ levels, as long as the density does not redistribute the level population, i.e. the density is not above the critical density. \\
Tables~\ref{collcascadeC5},~\ref{collcascadeO8},~\ref{collcascadeNe9},~\ref{collcascadeMg11} and \ref{collcascadeSi13} report data which correspond respectively to the direct (b), the resonance (c), and the $n>$2 cascade (d) contributions.
%%%%%%%%%%%%%%%%%%%%%%%%%%%%%%
%PLEASE INSERT TAB 9
%%%%%%%%%%%%%%%%%%%%%%%%%%%%%%
%%%%%%%%%%%%%%%%%%%%%%%%%%%%%%
%PLEASE INSERT TAB 10
%%%%%%%%%%%%%%%%%%%%%%%%%%%%%%
%%%%%%%%%%%%%%%%%%%%%%%%%%%%%%
%PLEASE INSERT TAB 11
%%%%%%%%%%%%%%%%%%%%%%%%%%%%%%
%%%%%%%%%%%%%%%%%%%%%%%%%%%%%%
%PLEASE INSERT TAB 12
%%%%%%%%%%%%%%%%%%%%%%%%%%%%%%
%%%%%%%%%%%%%%%%%%%%%%%%%%%%%%
%PLEASE INSERT TAB 13
%%%%%%%%%%%%%%%%%%%%%%%%%%%%%%

%%%%%%%%%%%%%%%%%%%%%%%%%%%%%%%%%%%%%%%%%%%%%%%%%%%%%%%%%%%%%%%%%%%%%%%%%%%%%%%%%%%%%
\section{Plasma diagnostics} \label{sec:diagnostics}

\subsection{Computation of the line ratios}
The intensities of the three component lines (resonance, forbidden and intercombination) are calculated from atomic data presented in the former section. The ratios R(n$_{\mathrm{e}}$) and G(T$_{\mathrm{e}}$) are calculated for \ion{C}{v}, \ion{N}{vi}, \ion{O}{vii}, \ion{Ne}{ix}, \ion{Mg}{xi}, and \ion{Si}{xiii}. The wavelengths of these three lines for each He-like ion treated in this paper are reported in Table~\ref{lambda}.\\
 
%%%%%%%%%%%%%%%%%%%%%%%%%%%%%%
%PLEASE INSERT TAB 14
%%%%%%%%%%%%%%%%%%%%%%%%%%%%%%

We note that for all temperatures (low and high), we have included in the line ratio calculations, RR contribution (direct + upper-level radiative cascade), and collisional excitations inside the $n$=2 shell. For high temperature plasmas, the CE contribution (direct + resonance + cascade) from the ground level ($n$=1 shell, 1s$^{2}$) should be included in the calculations as well as DR (direct + cascade). Figure~\ref{exc_recomb} displays these different contributions which populate a given $n$=2 level.\\

%%%%%%%%%%%%%%%%%%%%%%%%%%%%%%
%PLEASE INSERT FIG 5
%%%%%%%%%%%%%%%%%%%%%%%%%%%%%%

As emphasized previously, the cascade contribution from n$>$2 levels, especially for the $^{3}$S$_{1}$ level, should be taken into account in line ratio calculations since this level is responsible for the forbidden component ({\bf z}) line, which appears in both ratios {\bf R} and {\bf G}. 
For a pure photoionized plasma, when no upper level radiative cascade contribution is included in the RR rates, {\bf R} and {\bf G} could be underestimated by 6--10\% (for \ion{O}{vii}). In a hybrid plasma, where collisional processes from the ground level are not negligible, the ratio {\bf R} is lower by $\sim$20$\%$ at T=3.6\,10$^{6}$\,K, when no cascades from upper levels are taken into account. In a similar way, the value of {\bf G} would be underestimated.\\

We also point out the importance of taking into account the branching ratios in the calculations of {\bf x} and {\bf y} lines. B$_{\mathrm{x}}$ = A$_{5\to1}$ / (A$_{5\to1}$ + A$_{5\to2}$), and B$_{\mathrm{y}}$ = A$_{4\to1}$ / (A$_{4\to1}$ + A$_{4\to2}$) are respectively the branching ratios of the {\bf x} and {\bf y} lines (A$_{\mathrm{j \to i}}$ being the transition probability from level j to level i, see Fig.~\ref{gotrian}). Branching ratios are very important in the case of light nuclear charge ($Z$), as shown in Figure~\ref{BRCVSiXIII}, for \ion{C}{v}, A$_{5\to1}<<$A$_{5\to2}$ as well A$_{4\to1}<$A$_{4\to2}$. When $Z$ increases, most branching ratios become less important but nevertheless some of them should be included in the calculations. Without these branching ratios the intensities of the intercombination lines {\bf x} and {\bf y} could be overestimated, resulting in an underestimate of the ratio {\bf R}. This could lead to huge discrepancies for the value of {\bf R} as well as for {\bf G}.

%%%%%%%%%%%%%%%%%%%%%%%%%%%%%%
%PLEASE INSERT FIG 6
%%%%%%%%%%%%%%%%%%%%%%%%%%%%%%

\subsection{Ionizing process diagnostics}\label{sec:process}
First of all, the ionization processes that occur should be determined. High resolution spectra enable us to measure the intensities of the forbidden ({\bf z}), intercombination ({\bf x+y}) and resonance ({\bf w}) lines of a He-like ion. They give an indication of the ionization processes which occur in the gas using the relative intensity of the resonance line {\bf w} compared to those of the forbidden {\bf z} and the intercombination {\bf x+y} lines. This corresponds to the {\bf G} ratio (see eq.~\ref{eq:G}).\\
RR to the $^{3}$S and $^{3}$P (triplet) levels is more than a factor 4 greater than the $^{1}$P (singlet) level, due to the higher statistical weights of the triplet levels. When RR dominates compared to CE from the ground level (1s$^{2}$), this results in a very intense forbidden {\bf z} ($^{3}$S$_{1}$ level) or {\bf (x+y)} ($^{3}$P$_{1,2}$ levels) lines, compared to the resonance {\bf w} line ($^{1}$P$_{1}$ level). On the contrary, when CE from the ground level dominates compared to RR, the $^{1}$P$_{1}$ level is preferentially populated (high value of $\Upsilon$(1s$^{2}$\,$^{1}$S$_{0}\to$1s2p\,$^{1}$P$_{1}$)), thus implying an intense resonance {\bf w} line.\\
We also introduce the parameter {\bf X$_{\mathrm{ion}}$} which is the relative ionic abundance of the H-like and He-like ions. As an example for oxygen, it corresponds to the ratio of \ion{O}{viii}/\ion{O}{vii} ground state population. A low value of {\bf X$_{\mathrm{ion}}$} means that the H-like ion relative abundance is small compared to the He-like one and thus CE from the 1s$^{2}$ ground level is dominant compared to RR (H-like$\to$He-like), when the temperature is high enough to permit excitation from the ground level.\\
 Figure~\ref{1surG} displays the ratios {\bf G} as a function of electronic temperature (T$_{\mathrm{e}}$) for different values of {\bf X$_{\mathrm{ion}}$}. The range of temperatures (low values) where the ratio ($>$4 see $\S$\ref{sec:process}) is almost independent of T$_{\mathrm{e}}$ and {\bf X$_{\mathrm{ion}}$} occurs for a plasma dominated by RR (pure photoionized plasmas).\\
%%%%%%%%%%%%%%%%%%%%%%%%%%%%%%
%PLEASE INSERT FIG 7
%%%%%%%%%%%%%%%%%%%%%%%%%%%%%%
At higher temperatures, i.e. large enough to permit excitation from the ground level to upper levels, {\bf G} becomes sensitive to both parameters (T$_{\mathrm{e}}$, {\bf X$_{\mathrm{ion}}$}). High values of {\bf X$_{\mathrm{ion}}$} favour mainly DR (H-like ions towards He-like ions), but for photoionized plasma such high temperatures (where {\bf G}$<$4) are probably extreme cases (i.e not realistic) for WA plasmas.\\
On the contrary, for lower values of {\bf X$_{\mathrm{ion}}$} the lines are produced mainly by collisional excitation (``hybrid'' plasma in our nomenclature). A value of {\bf G$<$4} will be the signature of a plasma where collisional processes are no longer negligible and even be dominant compared to recombination. We should notice that this is no more the case when {\bf G} is sensitive to n$_{\mathrm{e}}$, i.e. when the resonance {\bf w} line becomes sensitive to density due to the depopulation of the 1s2s\,$^{1}$S$_{0}$ level to the 1s2p$^{1}$P$_{1}$ level (see also fig. 4--6--9 in Gabriel \& Jordan \cite{GabrielJordan72}).\\
In conclusion, the relative intensity of the resonance {\bf w} line, compared to the forbidden {\bf z} and the intercombination ({\bf x+y}) lines, contains informations about the ionization processes that occur: a weak {\bf w} line compared to the {\bf z} or the ({\bf x+y}) lines corresponds to a pure photoionized plasma. It leads to a ratio of {\bf G}=(z+x+y)/w$>$4. On the contrary a strong {\bf w} line corresponds to a hybrid plasma (or even a collisional plasma), where collisional processes are not negligible and may even dominate (see $\S$\ref{sec:hybrid}). In this case, {\bf w} is at least as intense as the {\bf z} or {\bf x+y} lines.

\subsection{Density diagnostic}\label{sec:calculations}
\indent In the low density limit, all $n$=2 states are populated by electron impact directly or via upper-level radiative cascade from He-like ground state and by H-like recombination (see Figure~\ref{gotrian} and \ref{exc_recomb}). These states decay radiatively directly or by cascade to the ground level. The relative intensities of the three intense lines are then independent of density. As n$_{\mathrm{e}}$ increases from the low density limit, some of these states (1s2s\,$^{3}$S$_{1}$ and $^{1}$S) are depleted by collision to the nearby states where n$_{\mathrm{crit}}$\,C\,$\sim$A, with C being the collisional coefficient rate, A being the radiative transition probabilities from $n$=2 to $n$=1 (ground state), and n$_{\mathrm{crit}}$ being the critical density. Collisional excitation depopulates first the 1s2s\,\element[][3]{S}$_{1}$ level (metastable) to the 1s2p  \element[][3]{P}$_{0,1,2}$ levels. The intensity of {\bf z} decreases and those of {\bf x} and {\bf y} increase, hence implying a reduction of the ratio {\bf R} (according to eq.\ref{eq:R}). For much higher densities, 1s2s\,$^{1}$S$_{0}$ is also depopulated to 1s2p$^{1}$P$_{1}$.
\subsubsection{Pure photoionized plasmas}
As explained previously, pure photoionized plasmas are characterized by a weak resonance {\bf w} line compared to the forbidden {\bf z} or the intercombination ({\bf x+y}) lines. The ratio {\bf R} as a function of electronic density n$_{\mathrm{e}}$ is reported in Figure~\ref{Rphoto} for \ion{C}{v}, \ion{N}{vi}, \ion{O}{vii}, \ion{Ne}{ix}, \ion{Mg}{xi}, \ion{Si}{xiii} at different values of T$_{\mathrm{e}}$.\\
%%%%%%%%%%%%%%%%%%%%%%%%%%%%%%
%PLEASE INSERT FIG 8
%%%%%%%%%%%%%%%%%%%%%%%%%%%%%%
 For low values of T$_{\mathrm{e}}$ corresponding to the density range where {\bf G} is independent of {\bf X$_{\mathrm{ion}}$} (fig.~\ref{Rphoto}), {\bf R} is almost insensitive to temperature.\\
 But in the case of high temperature (with a high {\bf X$_{\mathrm{ion}}$} value so that the medium is dominated by recombination), the value of {\bf R} is larger. Thus in the density range where {\bf R} takes a constant value (i.e. low density values), a high value of {\bf R} corresponds to a high temperature. This also imply a very intense H-like line (K$_{\alpha}$) since the ratio X$_{\mathrm{ion}}$=H-like/He-like need to be large enough so that the gas is dominated by recombinations (see caption of the Figure~\ref{1surG}).

\subsubsection{Hybrid plasmas}\label{sec:hybrid}
Hybrid plasmas, where both recombination and collisional processes occur, are characterized by {\bf G}$<$4, i.e. an intense resonance {\bf w} line.\\
For high temperature, the ratio {\bf R} as a function of electronic density n$_{\mathrm{e}}$ is reported in Figure~\ref{Rhybrid} for \ion{C}{v}, \ion{N}{vi}, \ion{O}{vii}, \ion{Ne}{ix}, \ion{Mg}{xi}, \ion{Si}{xiii} for different values of {\bf X$_{\mathrm{ion}}$}. {\bf R} is calculated at the temperature corresponding to the maximum  abundance of the He-like ion for a collisional plasma (see Arnaud \& Rothenflug \cite{Arnaud85}). In the low density limit, in the range where {\bf R} is independent of density, its value is correlated with {\bf X$_{\mathrm{ion}}$}. However for intermediate values of {\bf X$_{\mathrm{ion}}$}, {\bf R} is similar to the {\bf R} calculated for photoionized plasmas (see also Figure~\ref{Rphoto}), especially for low charge ions ({\ion{C}{v}, \ion{N}{vi} and \ion{O}{vii}). Thus discriminating between ionization processes is difficult using this {\bf R} ratio. As one can also see, at higher densities this ratio is almost insensitive to the {\bf X$_{\mathrm{ion}}$} value.

%%%%%%%%%%%%%%%%%%%%%%%%%%%%%%
%PLEASE INSERT FIG 9
%%%%%%%%%%%%%%%%%%%%%%%%%%%%%%

%%%%%%%%%%%%%%%%%%%%%%%%%%%%%%%%%%%%%%%%%%%%%%%%%%%%%%%%%%%%%%%%%%%%%%%%%%%%%%%%%%%%%%%%%%%%

\section{Practical use of the diagnostics} \label{sec:practical}
\noindent The physical parameters which could be inferred are numerous:\\
\noindent - Firstly, we can determine which ionization processes occur in the medium, i.e. a whether photoionization dominates or if an additional process competes (such as a collisional one). Indeed, in the case of a pure photoionized plasma, the intensity of the resonance line {\bf w}, is weak compared to those of the intercombination {\bf x+y} and forbidden {\bf z} lines. On the contrary, if there is a strong {\bf w} line, this means that collisional processes are not negligible and may even dominate. This combined with the relative intensity of the K$_{\alpha}$ line (H-like) can give an estimate of the ratio of the ionic abundance of H-like/He-like and according to Figure~\ref{1surG}, this can also give an indication of the electronic temperature T$_{\mathrm{e}}$ in the case of a hybrid plasma, since {\bf G} is sensitive to T$_{\mathrm{e}}$. Figure~\ref{nedomaine} gives the temperature range where {\bf G} is insensitive to {\bf X$_{\mathrm{ion}}$} and T$_{\mathrm{e}}$ for pure photoionized plasmas. \\
%%%%%%%%%%%%%%%%%%%%%%%%%%%%%%
%PLEASE INSERT FIG 10
%%%%%%%%%%%%%%%%%%%%%%%%%%%%%%
\noindent - Next, density diagnostics can be used. The ratio {\bf R = z/(x+y)} changes rapidly over approximatively two decades of density, around the critical value, which is different for each He-like ion (see Fig.\ref{nedomaine}). In this narrow density range, when the density increases the 1s2s\,\element[][3]{S}$_{1}$ level (metastable) is depopulated by electron impact excitation to the 1s2p\,\element[][3]{P}$_{0,1,2}$ levels which imply that the intensity of the forbidden {\bf z} line decreases while the intensity of the intercombination {\bf x+y} lines increases (see Figure~\ref{XMM}). Outside this range, at the low density limit (intense {\bf z} and a constant {\bf R} value), {\bf R} gives an upper limit for the value of the gas density producing the He-like ion. At higher densities (the forbidden {\bf z} line disappears since the density value is greater than the critical density and hence {\bf R} tends to zero), {\bf R} gives a lower density limit. Thus if the physical parameters deduced from each He-like ion do not correspond, this could be the signature of stratification of the WA.\\
%%%%%%%%%%%%%%%%%%%%%%%%%%%%%%
%PLEASE INSERT FIG 11
%%%%%%%%%%%%%%%%%%%%%%%%%%%%%%
\noindent - Once the density is determined from the ratio {\bf R}, an estimate of the size of the medium ($\Delta r$) becomes possible, since N$_{\mathrm{H}}$=n$_{\mathrm{H}}~\Delta r$, where N$_{\mathrm{H}}$ is the column density of the WA.\\
\noindent - In addition, the distance $r$ of the medium from the central ionizing source could be deduced, since the density and the distance are related by the ``ionization parameter'' $\xi$=L/n$_{\mathrm{H}}$\,r$^{2}$. Note that the determination of $\xi$ is dependant of the shape of the incident continuum.\\
%\end{itemize}
%newpage
%%%%%%%%%%%%%%%%%%%%%%%%%%%%%%%%%%%%%%%%%%%%%%%%%%%%%%%%%%%%%%%%%%%%%%%%%%%%%%%%%%%%%%%%%%%%
\section{Current and future X-ray satellites opportunities}\label{sec:satellite}

The high spectral resolution of the next generation of X-ray telescopes (Chandra, XMM and Astro-E) will enable us to detect and to separate the three main X-ray lines (resonance, intercombination and forbidden) of He-like ions.\\
Concerning Chandra (AXAF), all the main lines (w, x+y, z) for the He-like ions treated in this paper (\ion{C}{v} to \ion{Si}{xiii}), can be resolved using either the HRC-S combined with the LETG (0.08--6.0 keV; 2--160\,\AA), or the ACIS-S with  HETG (0.4-10\,keV; 1.2--31\AA).\\
The XMM mission, due to its high sensitivity and high spectral resolution (RGS: 0.35--2.5\,keV; 5--35\,\AA), will enable us to detect these He-like ions, except for \ion{C}{v} which is outside the detector's energy range. See Figure~\ref{XMM} for cases illustrating a pure photoionized plasma and a hybrid plasma for \ion{O}{vii} near 0.57\,keV.\\
The Astro-E XRS will be the first X-ray micro-calorimeter in space. 
It will have an energy resolution of 12\,eV (FWHM) over a broad energy range, 0.4 - 10 keV. Although, this is not sufficient for detailed spectroscopy at low energies, it will be very useful for the study of He-like ions (see figure 8 in Paerels \cite{Paerels99}) with E$>$2.5\,keV (i.e. $Z>$16), i.e.  complementary to the Chandra and XMM capabilities. \\
At some future date, XEUS (X-Ray Evolving Universe Spectroscopy mission), which is a potential follow-on to ESA's cornerstone XMM (Turner et al. \cite{Turner97}), will offer to observers a high energy astrophysics facility with high resolving power (E/$\Delta$E$\sim$1\,eV  near 1\,keV with its narrow field imager) combined with a unprecedented collecting area (initial mirror area of 6\,m$^{2}$). This will enable observers to use these types of plasma diagnostics for Carbon to Iron He-like ions. And, in addition, for high $Z$, ($Z$=26 for iron) He-like lines and their corresponding dielectronic satellite lines will be resolved and give accurate temperature diagnostics in the case of hybrid plasmas. Satellite lines to the He-like 1s$^{2}$--1s2l$^{'}$ parent line are due to transitions of the type:
\begin{equation}
1s^{2}nl-1s2l^{'}nl~~~~~~~n\geq2
\end{equation}
The main dielectronic satellite lines of \ion{Ca}{xix} and \ion{Fe}{xxv} He-like ions correspond to $n$=2 and 3 and are most important for temperature diagnostic purposes. For more details see the review by Dubau \& Volonte (\cite{DubauVolonte80}).

%%%%%%%%%%%%%%%%%%%%%%%%%%%%%%%%%%%%%%%%%%%%%%%%%%%%%%%%%%%%%%%%%%%%%%%%%%%%%%%%%%%%%%%%%%%%
\section{Conclusion}\label{sec:conclusion}
We have shown that the ratios of the three main lines (forbidden, intercombination and resonance) of He-like ions provide very powerful diagnostics for totally or partially photoionized media. For the first time, these diagnostics can be applied to non solar plasmas thanks to the high spectral resolution and the high sensitivity of the new X-ray satellites Chandra/AXAF, XMM and Astro-E.\\
These diagnostics have strong advantages. The lines are emitted by the same ionization stage of one element, thus eliminating any uncertainties due to elemental abundances. In addition, since the line energies are relatively close together, this minimizes wavelength dependent instrumental calibration uncertainties, thus ensuring that observed photon count rates can be used almost directly.\\
 For example, the determination of the physical parameters of the Warm Absorber component in AGN, such as the ionization process, the density and in some case the electronic temperature (``hybrid plasma''), will allow observers to deduce the size and the location (from the ionizing source) of the WA. In addition, since He-like ions are sensitive to different range of parameters (density, temperature), it could permit confirmation of the idea that the WA comes from a stratified, or a multi-zone medium (Reynolds \cite{Reynolds97}, Porquet et al. \cite{Porquet99}). As a consequence, a better understanding of the WA will be important for relating the WA to other regions (Broad Line Region, Narrow Line Region) in different AGN classes (Seyferts type-1 and type-2, low- and high-redshift quasars...). This will offer strong constraints on unified schemes.

\begin{acknowledgements}
The authors wish to acknowledge M. Cornille, J. Hughes and the anonymous referee for their careful reading of this paper. The authors greatly thank R. Mewe for his interest in this work and for very fruitful comparisons.
\end{acknowledgements}

%*****************************************************************************************
\begin{figure*}[h]
\vspace{1cm}\resizebox{8cm}{!}{\includegraphics{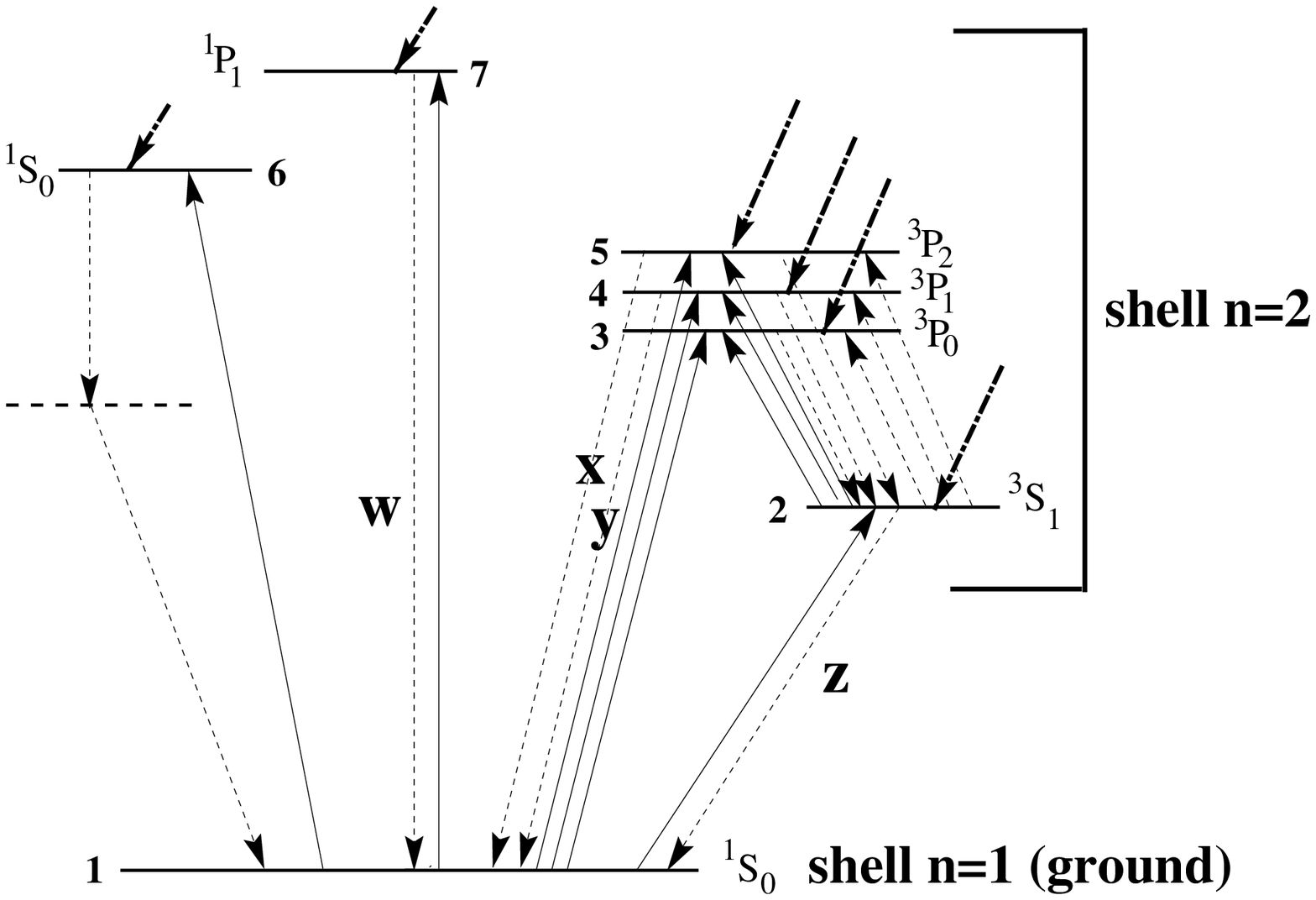}} 
\caption{Simplified Gotrian diagram for He-like ions. {\bf w}, {\bf x,y} and {\bf z} correspond respectively to the resonance, intercombination and forbidden lines. {\it Full curves}: collisional excitation transitions, {\it broken curves}: radiative transitions and {\it thick dot-dashed curves}: recombination (radiative and dielectronic). {\it Note: the broken arrow ($^{1}$S$_{0}$ to the ground level) corresponds to the 2-photon continuum}.}
\label{gotrian}
\end{figure*}

\begin{figure*}[h]
\begin{center}
\rotatebox{90}{\resizebox{8cm}{!}{\includegraphics{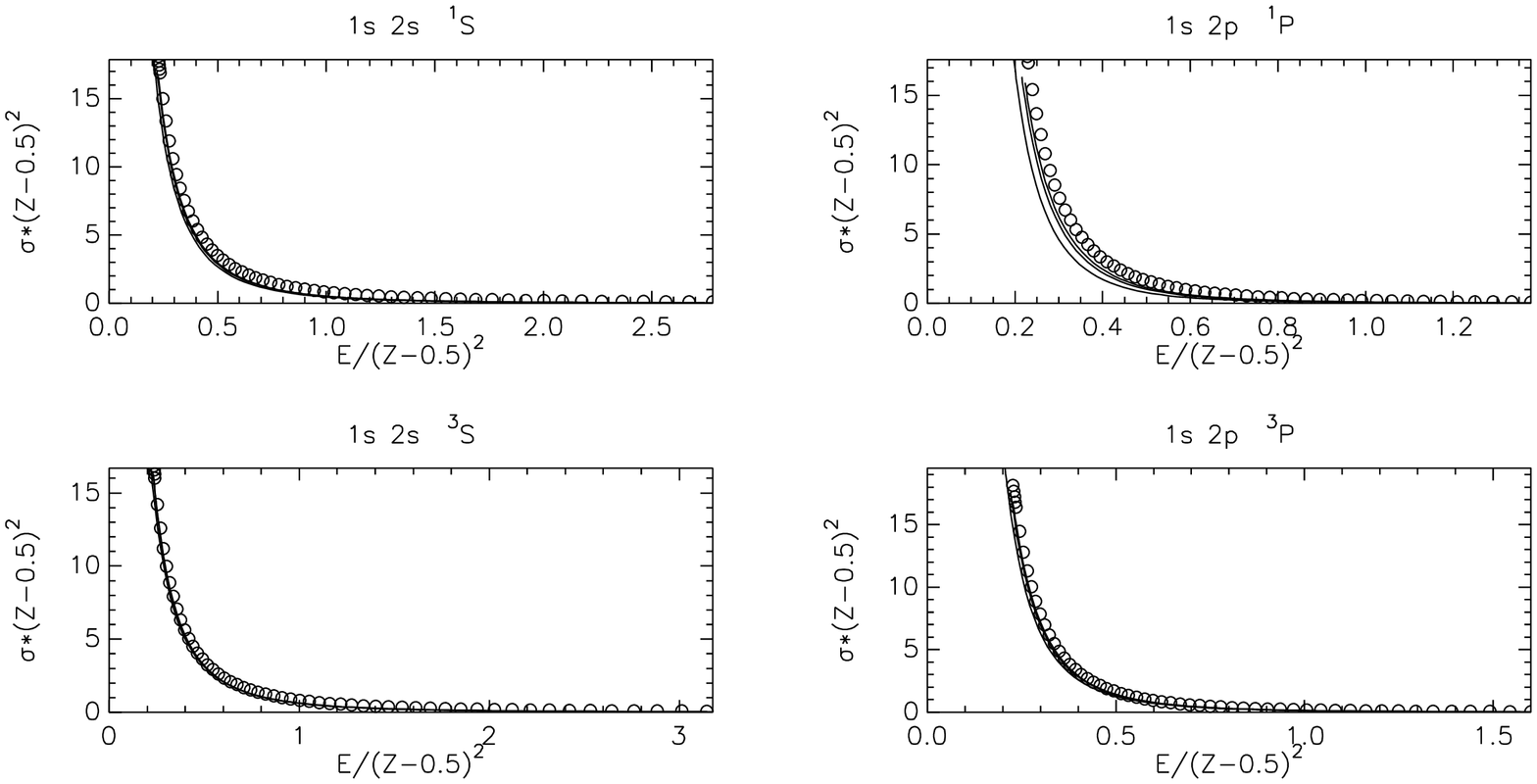}}}\vspace{0.3cm}
\rotatebox{90}{\resizebox{3.80cm}{!}{\includegraphics{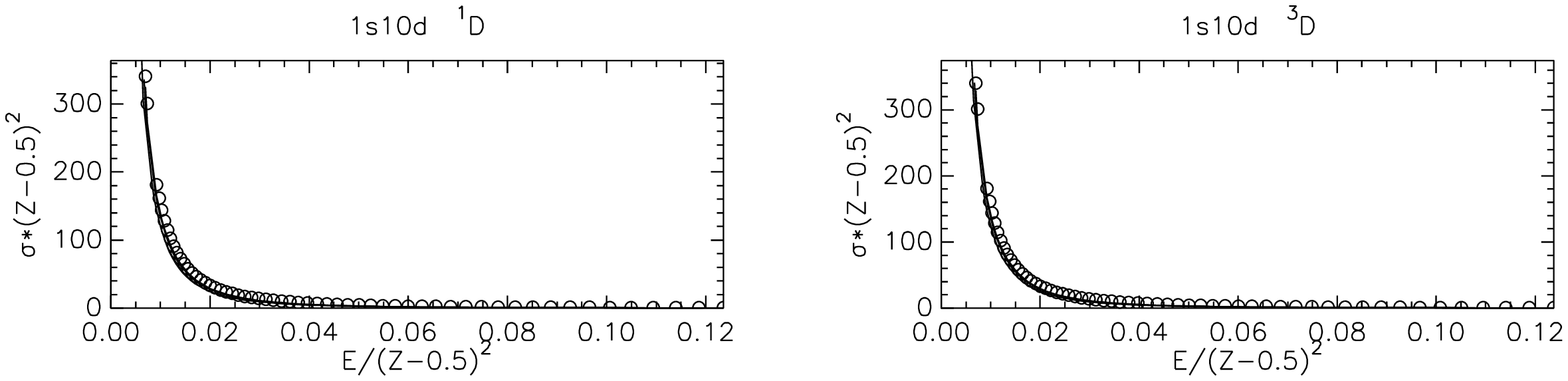}}}
\end{center}
\caption{Scaled photoionization cross sections $\sigma_{\mathrm{s}}$=$\sigma$\,($Z$-0.5)$^{2}$ (in cm$^{-3}$\,s$^{-1}$) as a function of E/($Z$-0.5)$^{2}$ (E is in Rydberg). {\it Empty circles}: photoionisation cross sections calculated in the present work; {\it solid lines}: photoionisation cross sections available in Topbase for different values of $Z$=6, 10, 14.}
\label{photoionisation}
\end{figure*}

\begin{figure*}[h]
\begin{center}
\rotatebox{90}{\resizebox{12cm}{!}{\includegraphics{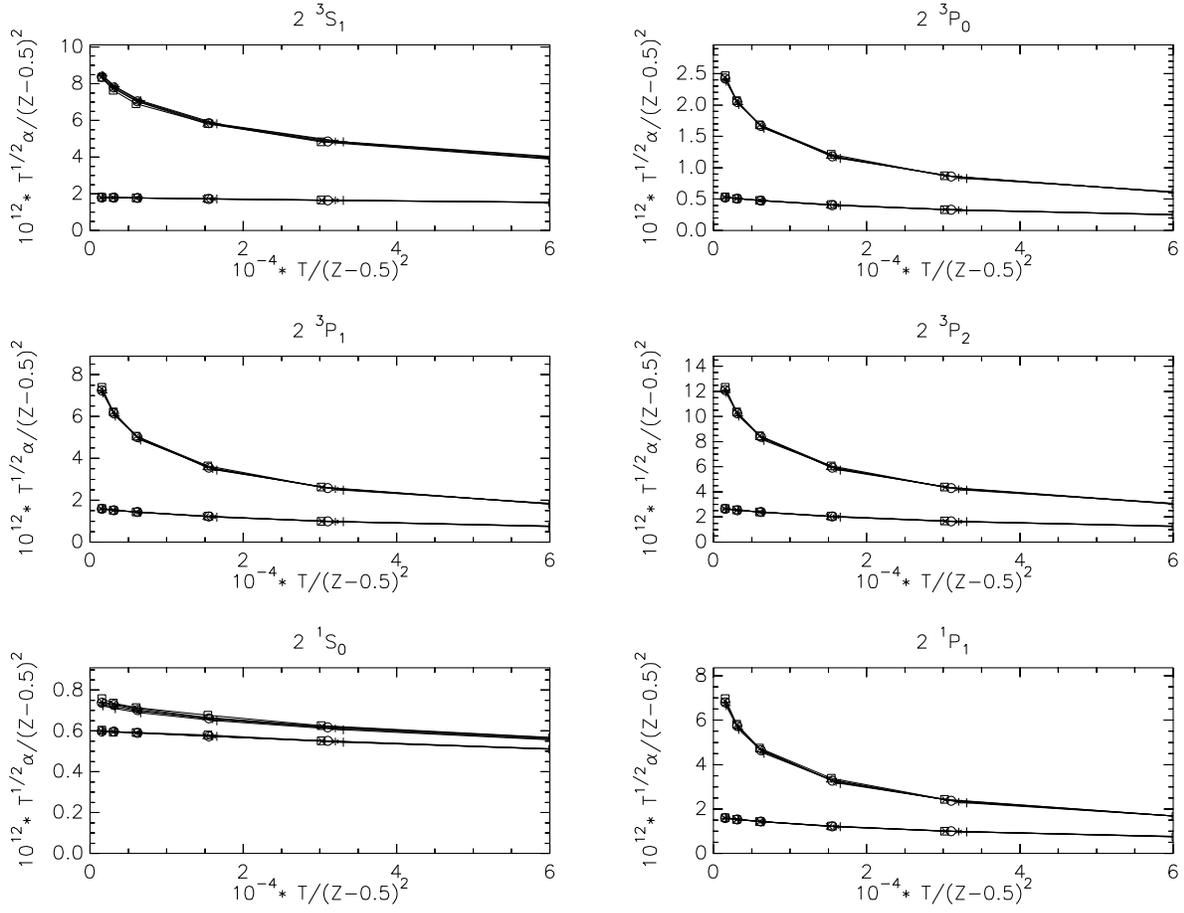}}}
\end{center}
\caption{Scaled total radiative recombination rates (upper curves: direct plus cascade contribution from n$>$2 levels) $\alpha^{\mathrm{s}}$=T$^{1/2}$\,$\alpha$/($Z$-0.5)$^{2}$ ($\times$10$^{12}$ cm$^{3}$\,s$^{-1}$) versus T$^{\mathrm{s}}$=T/($Z$-0.5)$^{2}$ ($\times$10$^{-4}$) towards each $n$=2 level (Plus, star, circle and cross are respectively for $Z$=8,10,12,14), and for comparison the direct contribution (lower curve in each graph). T is in Kelvin.}
\label{delphine2.ps}
\end{figure*}

\begin{figure*}[h]
\rotatebox{180}{\resizebox{16.5cm}{!}{\includegraphics{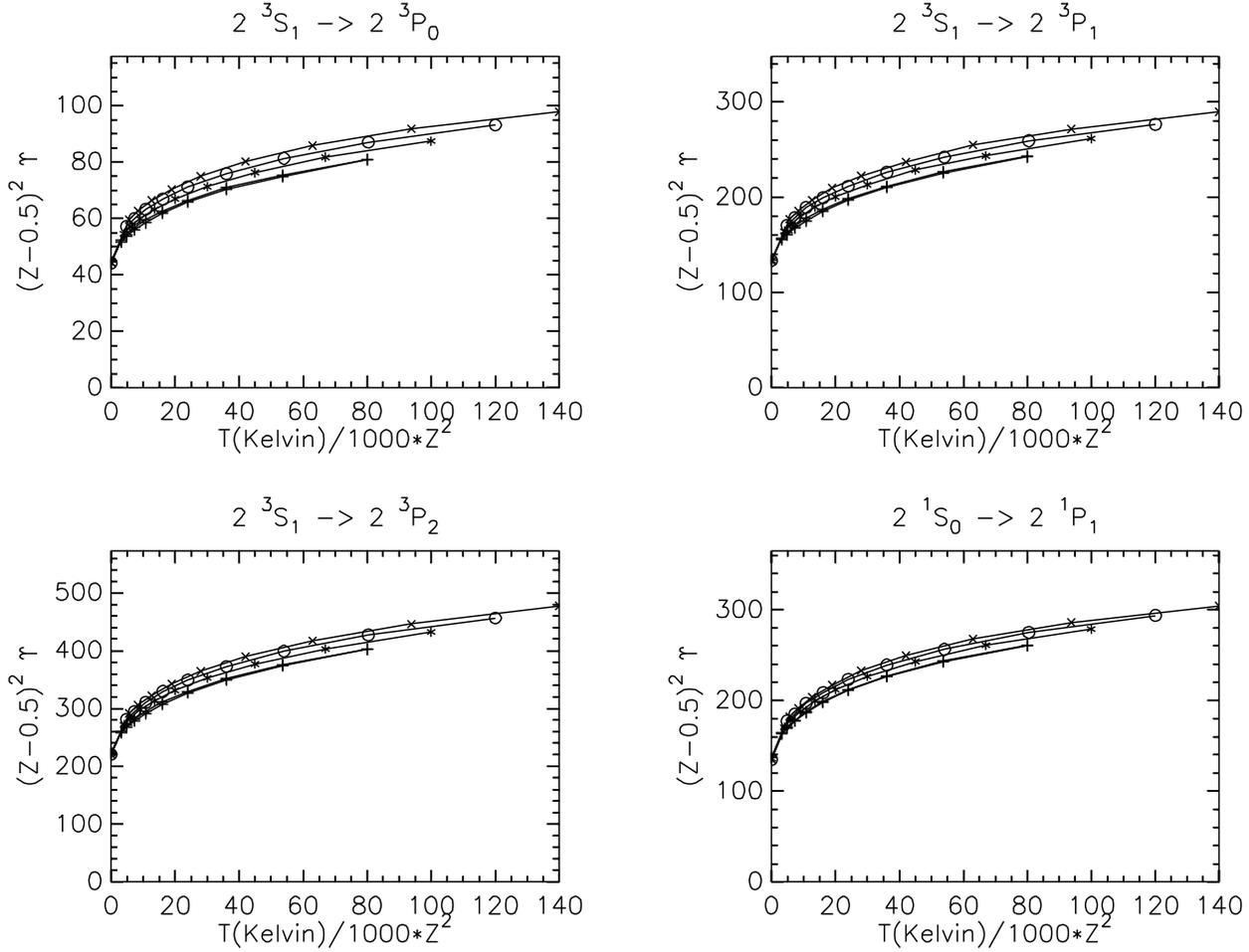}}}
\caption{Scaled effective collision strengths $\Upsilon^{\mathrm{s}}$=($Z$-0.5)$^{2}$\,$\Upsilon$ versus T$^{\mathrm{s}}$=T(K)/(1000\,$Z^{2}$) for He-like ions with $Z$=\,8,\,10,\,12,\,14 inside the $n$=2 level. The upper curves represent $\Upsilon^{\mathrm{s}}$ with the resonance effect taken into account (plus, star, circle and cross respectively for $Z$=\,8,\,10,\,12\,and 14) and for comparison the lower curve (with plus) corresponds to $\Upsilon^{\mathrm{s}}$ without resonance effect for $Z$=8. {\it Note: the two Z=8 curves (with plus) are superposed.}}
\label{figure1s21s2l}
\end{figure*}

\begin{figure*}[h]
\resizebox{8cm}{!}{\includegraphics{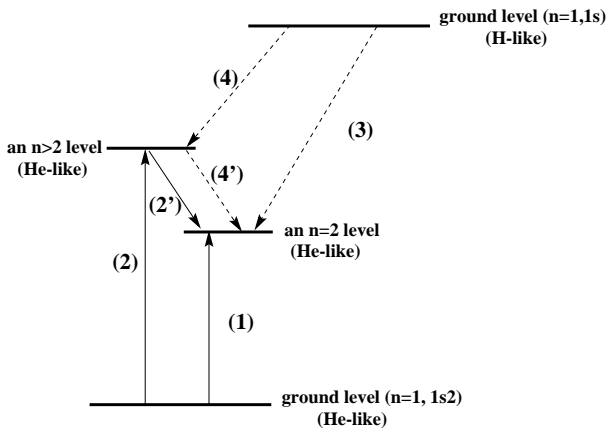}}
\caption{Simplified Gotrian diagram reporting the different contributions for the population of a given $n$=2 shell level. 
(1): direct contribution due to collisional excitation (CE) from the ground level (1s$^{2}$) of He-like ions; (2)+(2'): CE upper level radiative cascade contribution; (3): direct RR or direct DR from H-like ions contribution; and (4)+(4'): RR or DR upper level radiative cascade contribution. {\it Note: CE and DR are only effective at high temperature}.}
\label{exc_recomb}
\end{figure*}

\begin{figure*}[h]
\begin{tabular}{cc}
\resizebox{7.5cm}{!}{\includegraphics{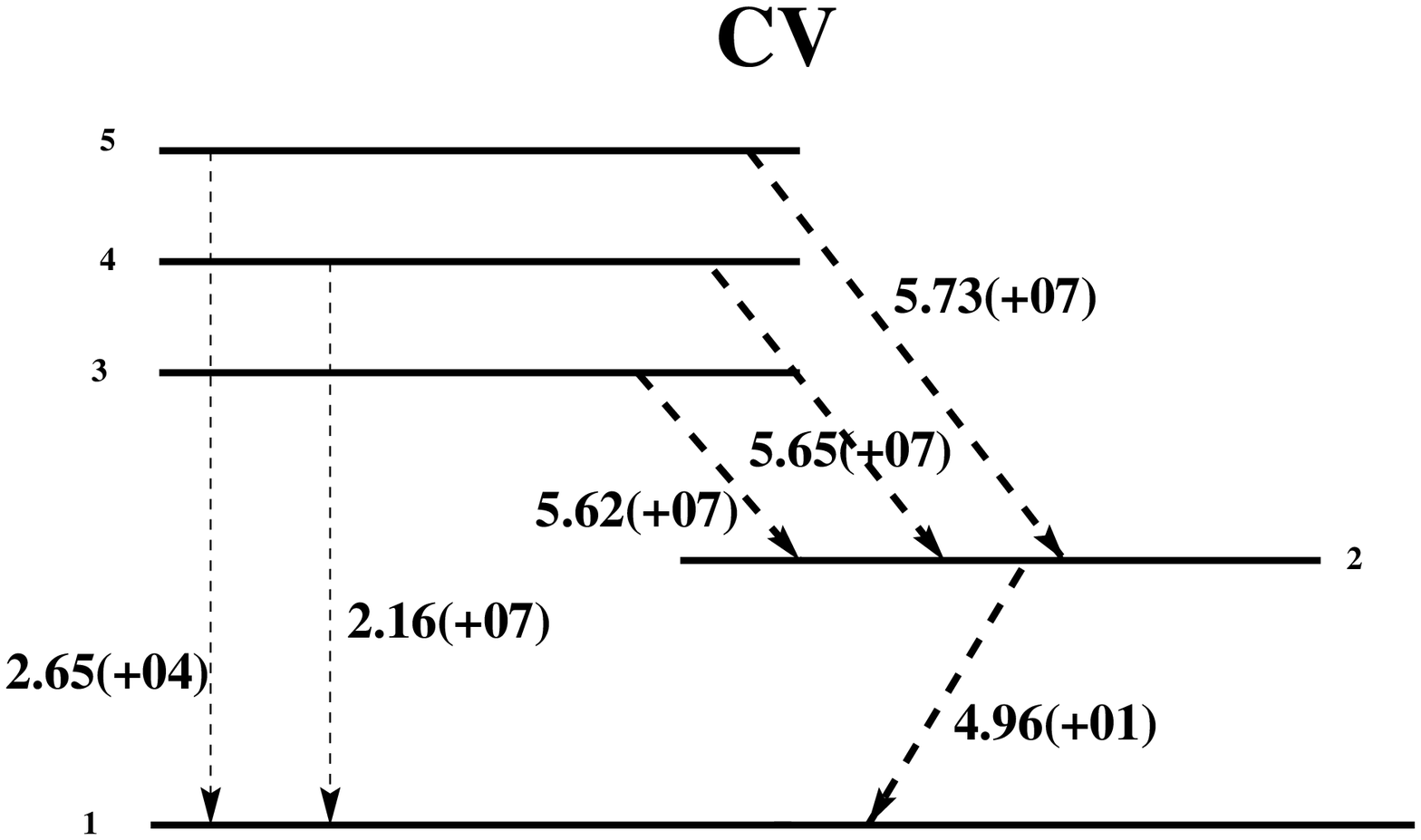}} & \resizebox{7.5cm}{!}{\includegraphics{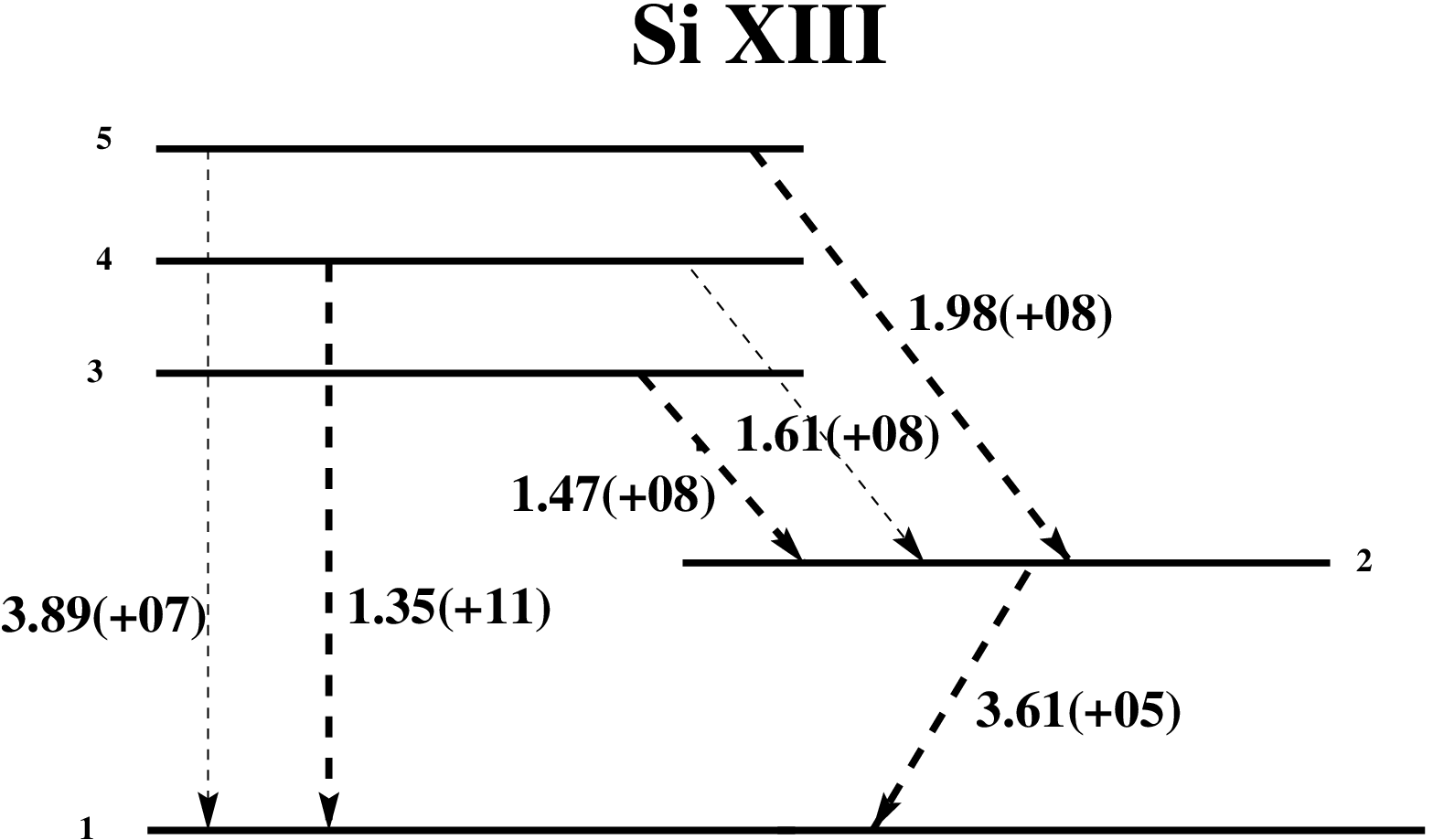}}\\
\end{tabular}
\caption{Simplified Gotrian diagrams for \ion{C}{v} and \ion{Si}{xiii}. Thick curves correspond to the strongest radiative transitions (A$_{\mathrm{i} \to {\mathrm{j}}}$ in s$^{-1}$), and thin curves correspond to lower values.}
\label{BRCVSiXIII}
\end{figure*}
%\clearpage
%----------------------------------------------------------------------------------------
\begin{figure*}[h]
\begin{tabular}{cc}
\resizebox{7.75cm}{!}{\includegraphics{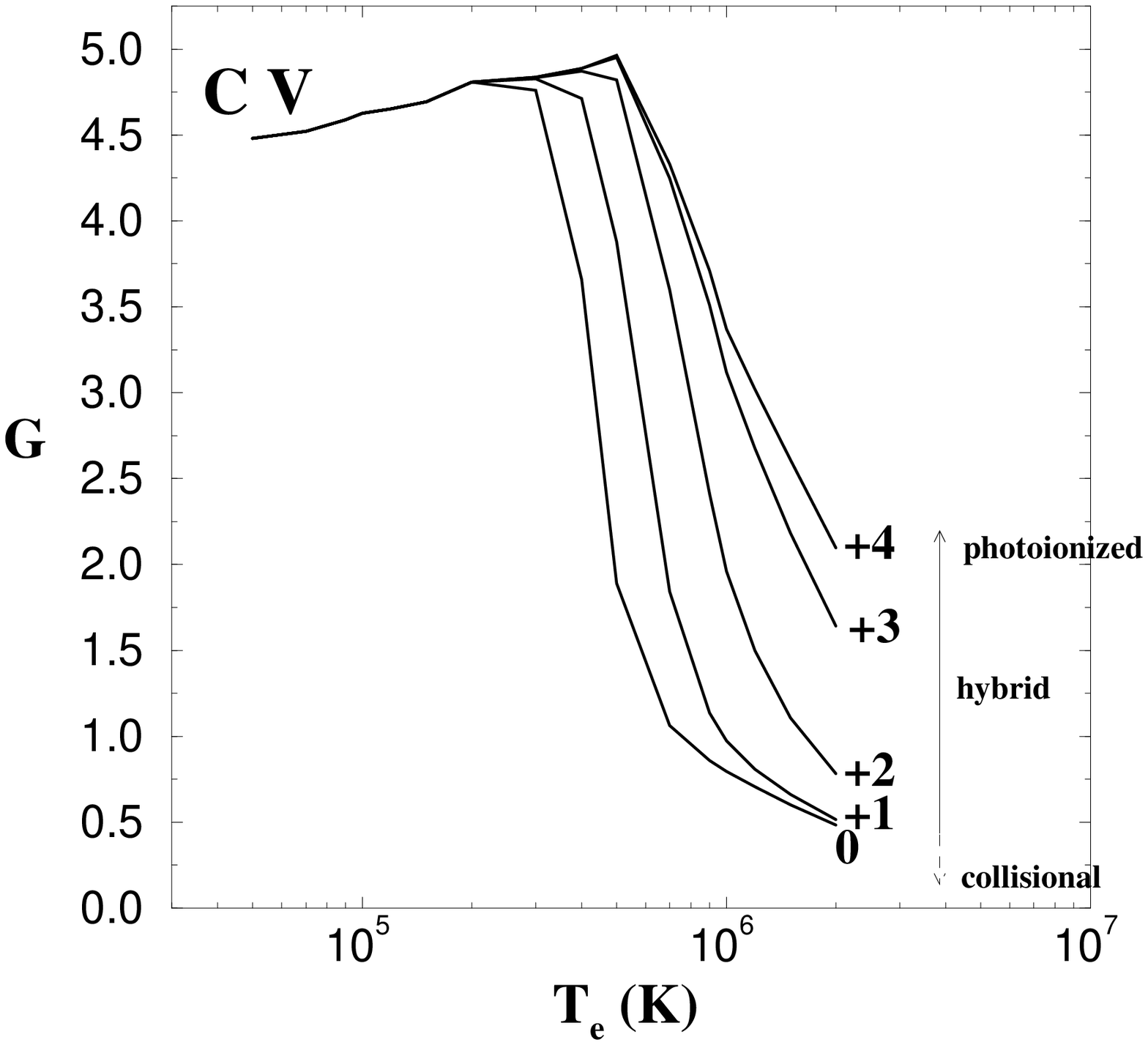}} &\resizebox{7.5cm}{!}{\includegraphics{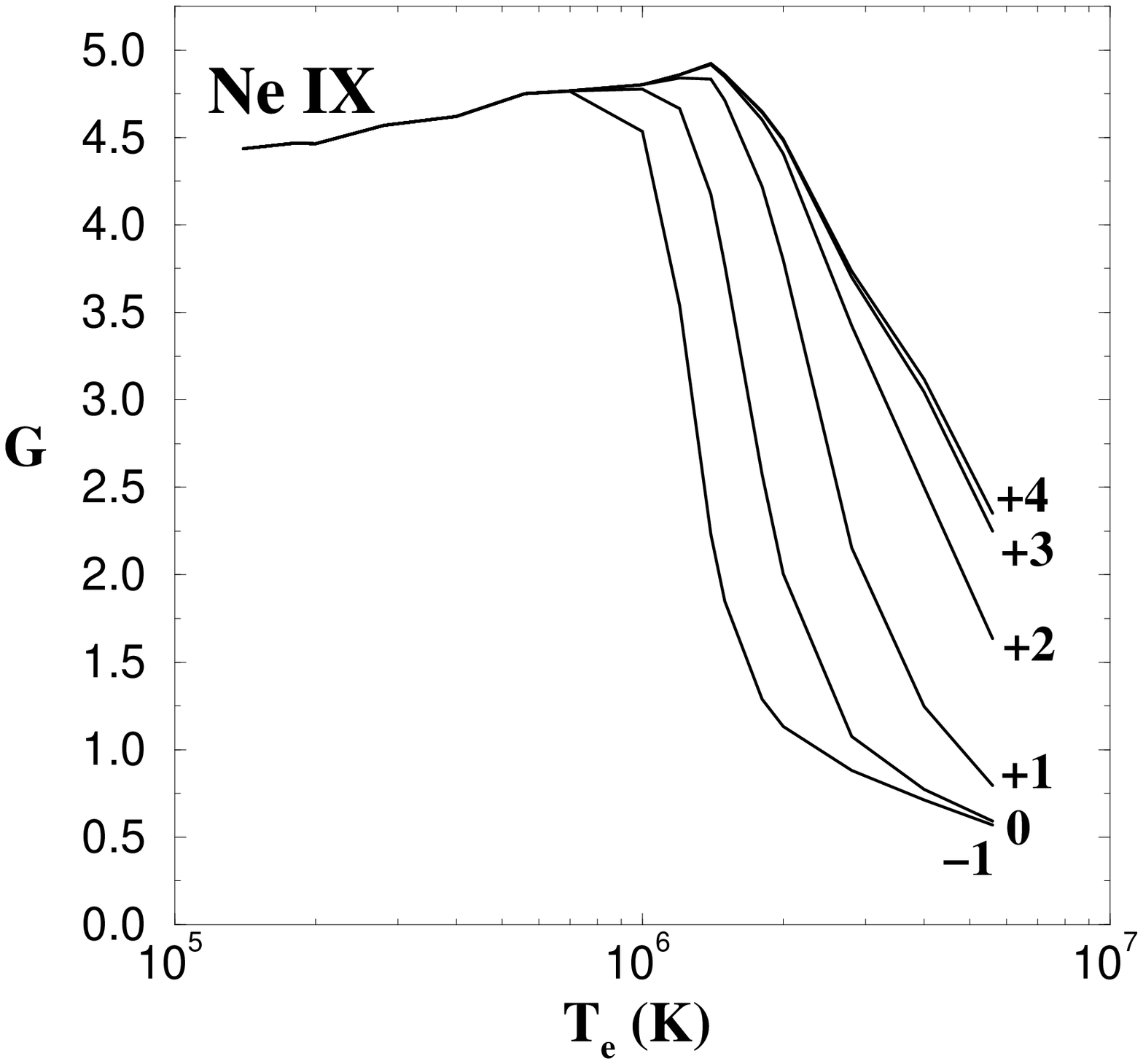}}\\
\resizebox{7.35cm}{!}{\includegraphics{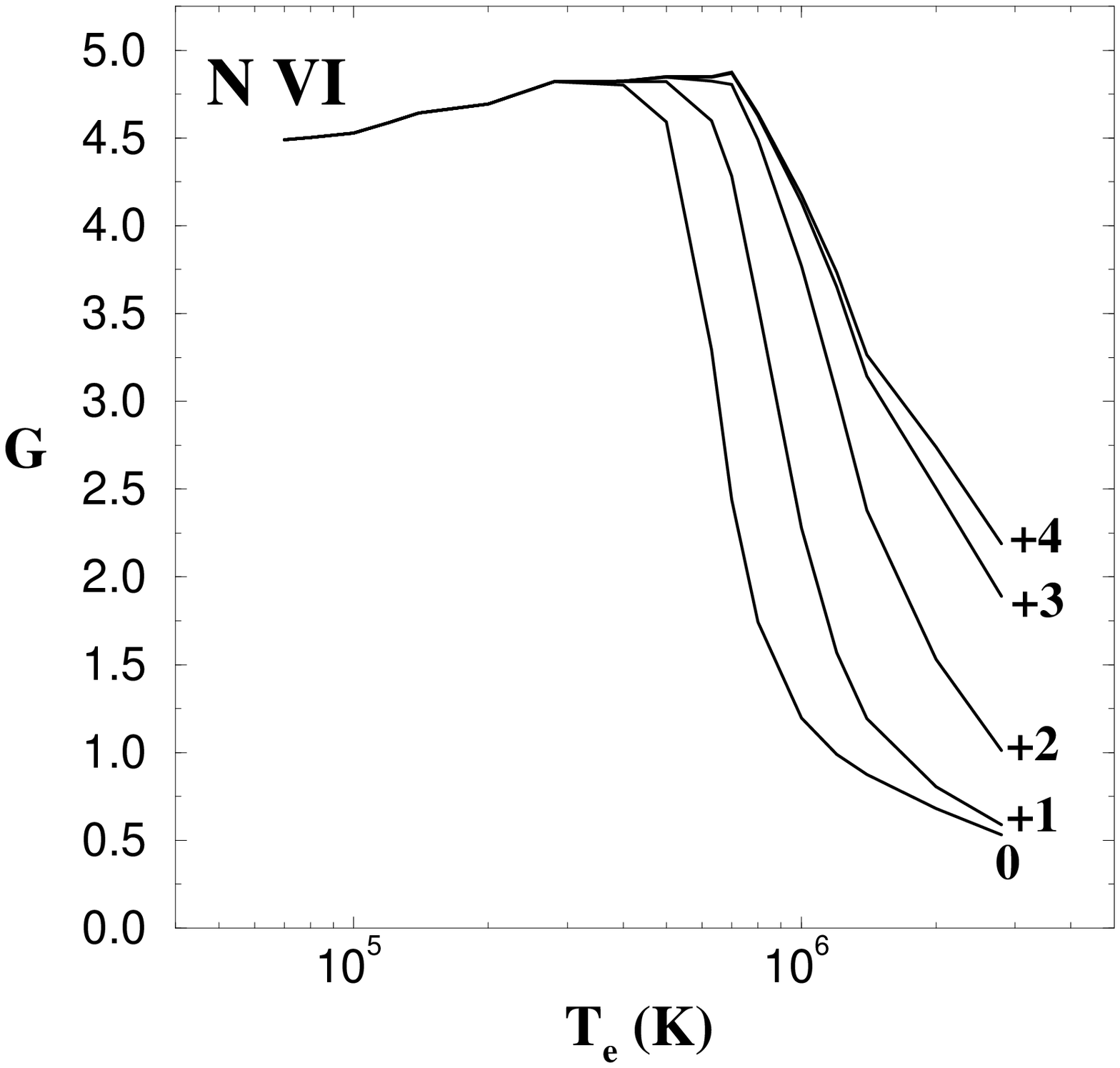}} &\resizebox{7.25cm}{!}{\includegraphics{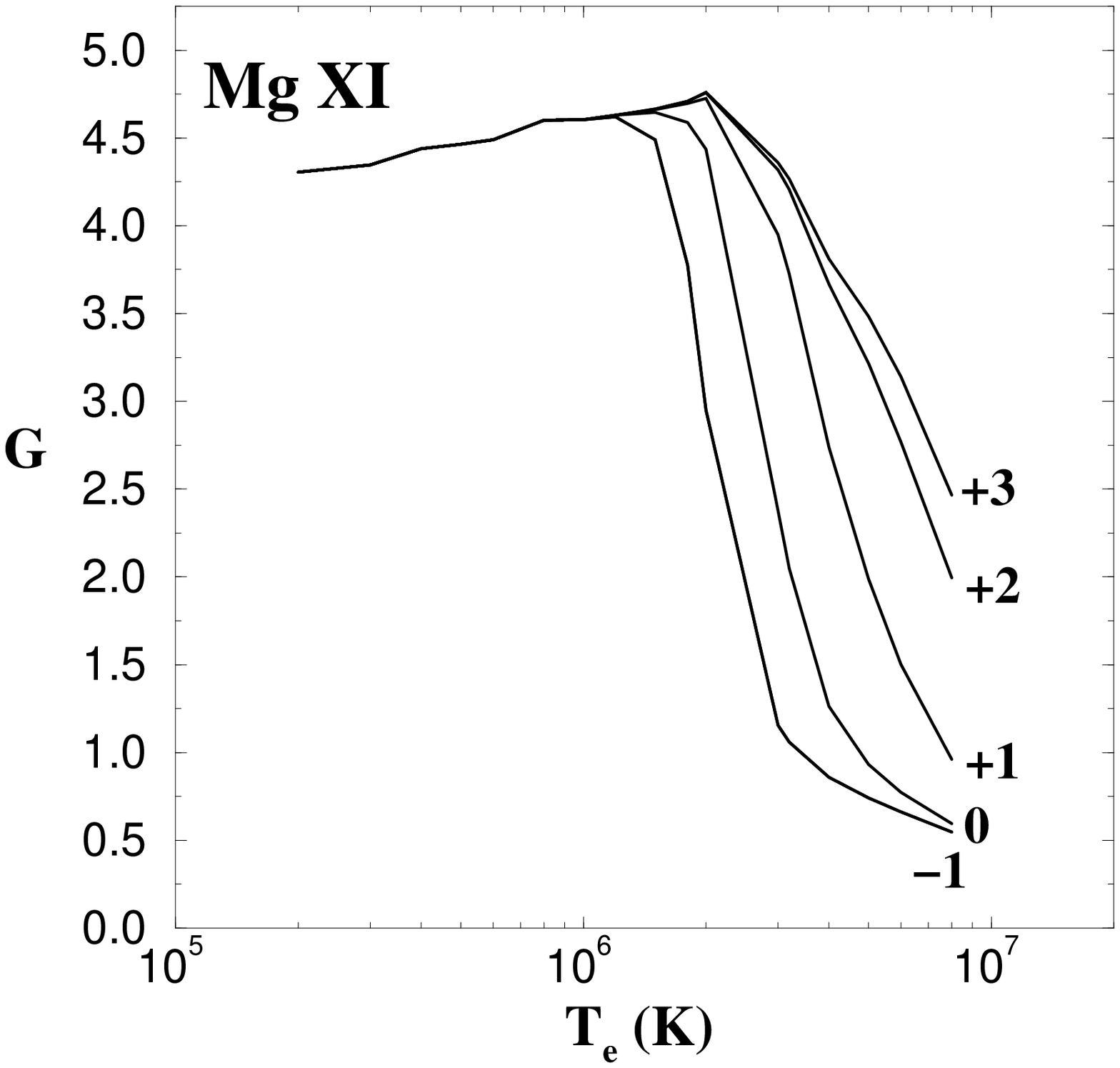}}\\
\resizebox{7.5cm}{!}{\includegraphics{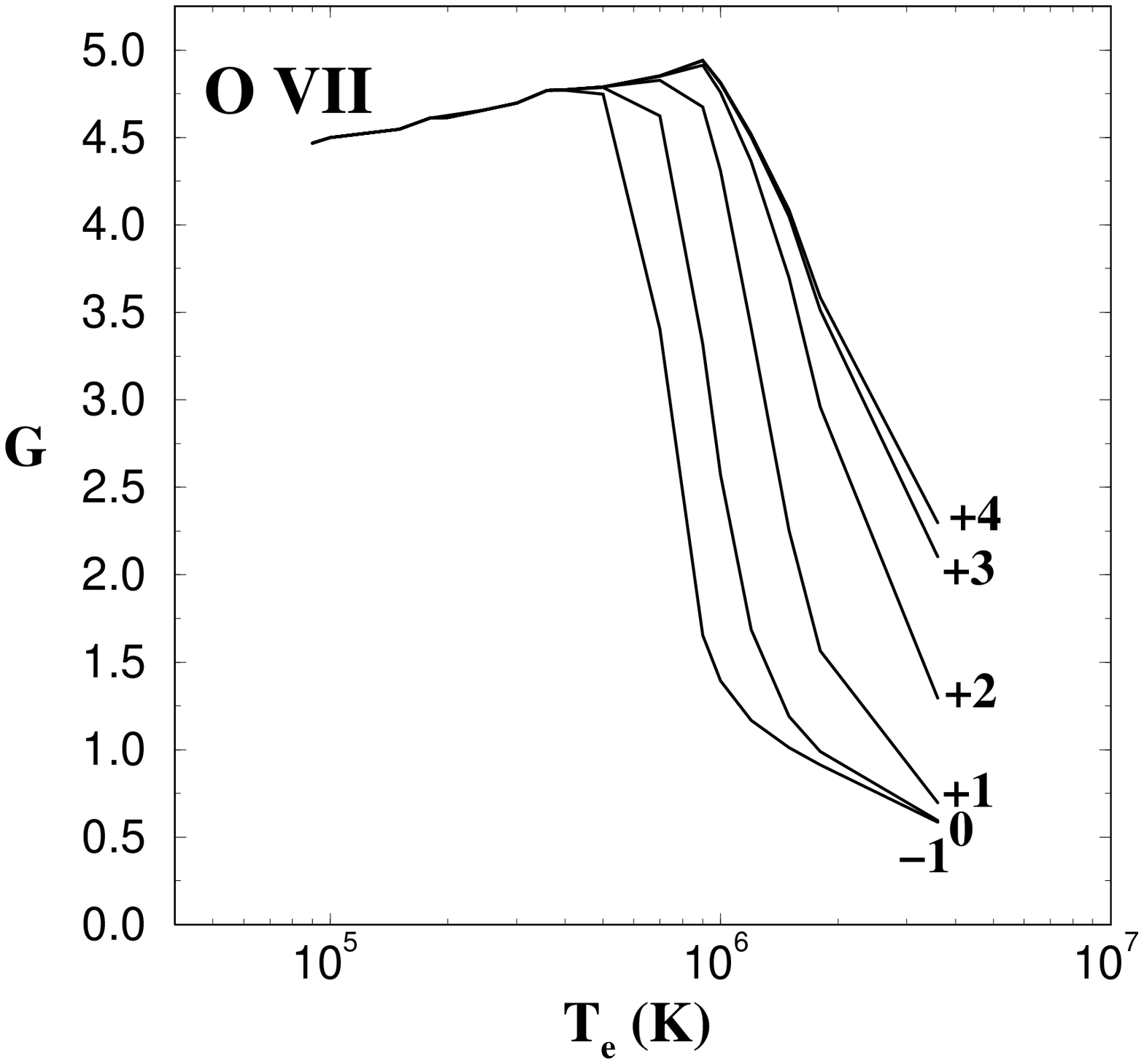}} &\resizebox{7.25cm}{!}{\includegraphics{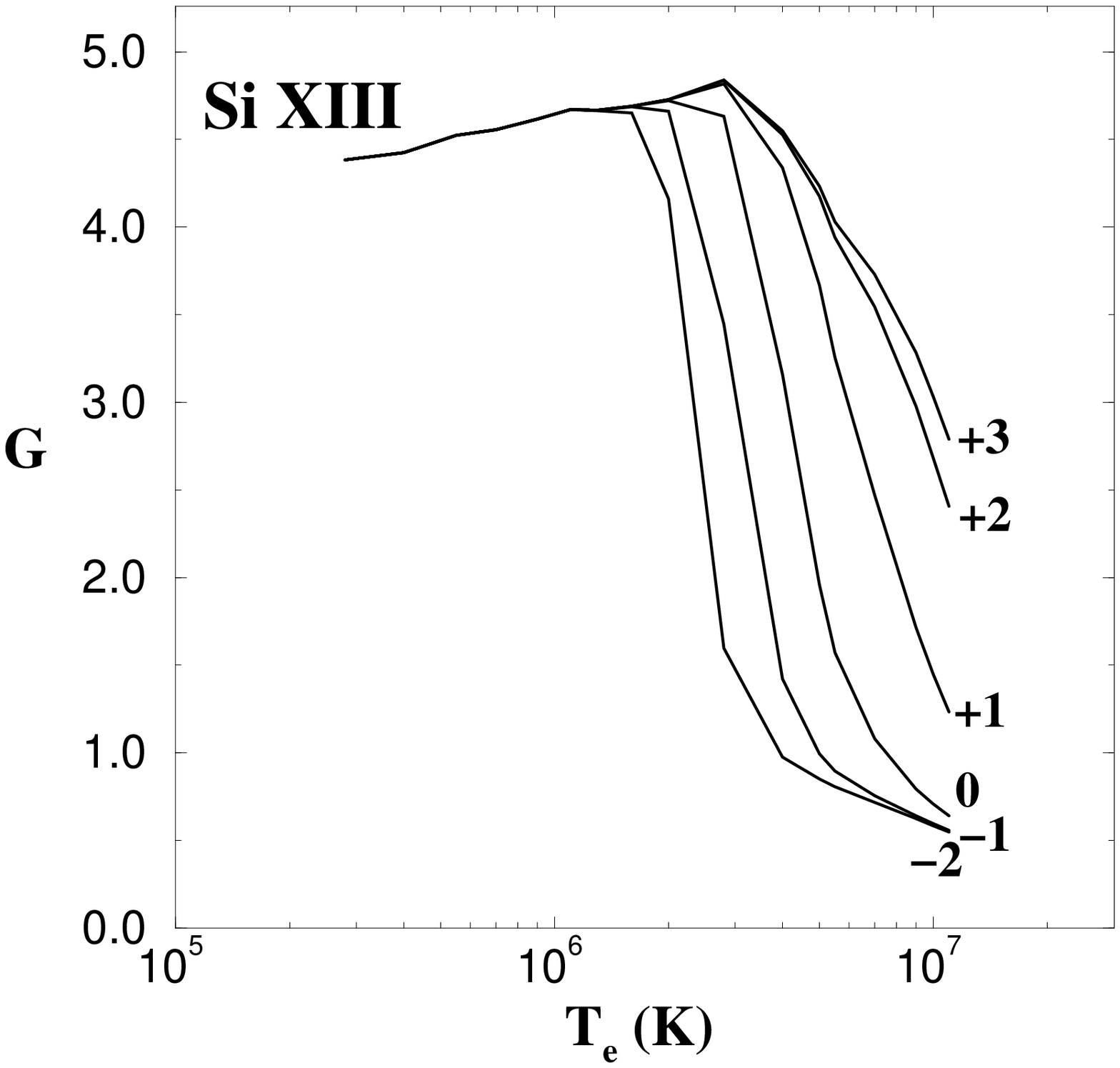}}\\
\end{tabular}
\caption{{\bf G (=(x+y+z)/w)} is reported as a function of electronic temperature (T$_{\mathrm{e}}$) for \ion{C}{v}, \ion{N}{vi}, \ion{O}{vii}, \ion{Ne}{ix}, \ion{Mg}{xi}, and \ion{Si}{xiii} in the density range where {\bf G} is not dependent on density (see $\S$\ref{sec:process}). The number ($m$) associated to each curves means {\bf X$_{\mathrm{ion}}$}=10$^{\mathrm{m}}$, where {\bf X$_{\mathrm{ion}}$} is the ratio of H-like ions over He-like ions. As an example for Oxygen ($Z$=8) it corresponds to ratio of the relative ionic abundance of \ion{O}{viii}/\ion{O}{vii} ground state population.}
\label{1surG}
\end{figure*}
\clearpage
%----------------------------------------------------------------------------------------
\begin{figure*}[h]
\begin{tabular}{cc}
\resizebox{7.15cm}{!}{\includegraphics{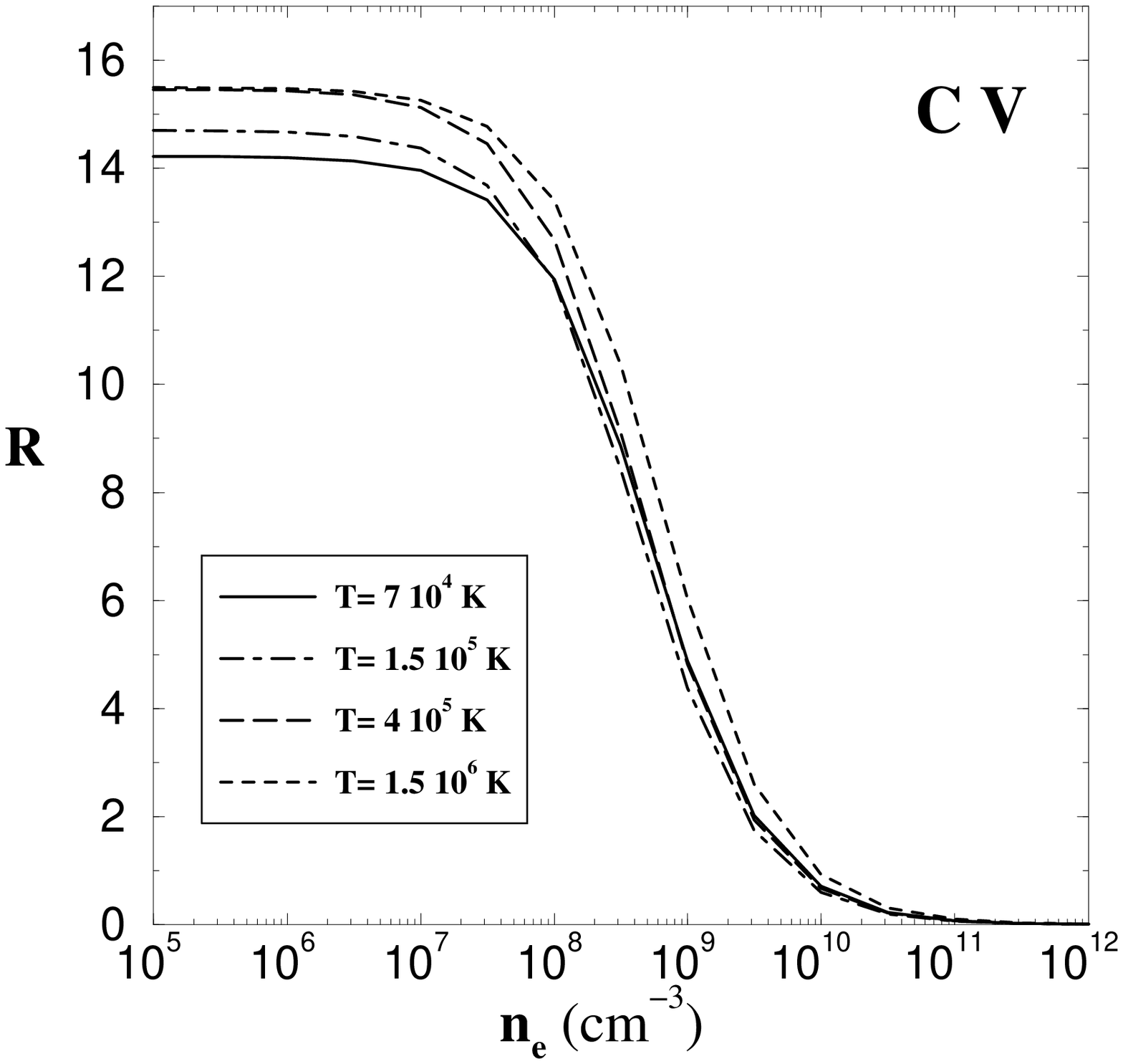}} & \resizebox{7.25cm}{!}{\includegraphics{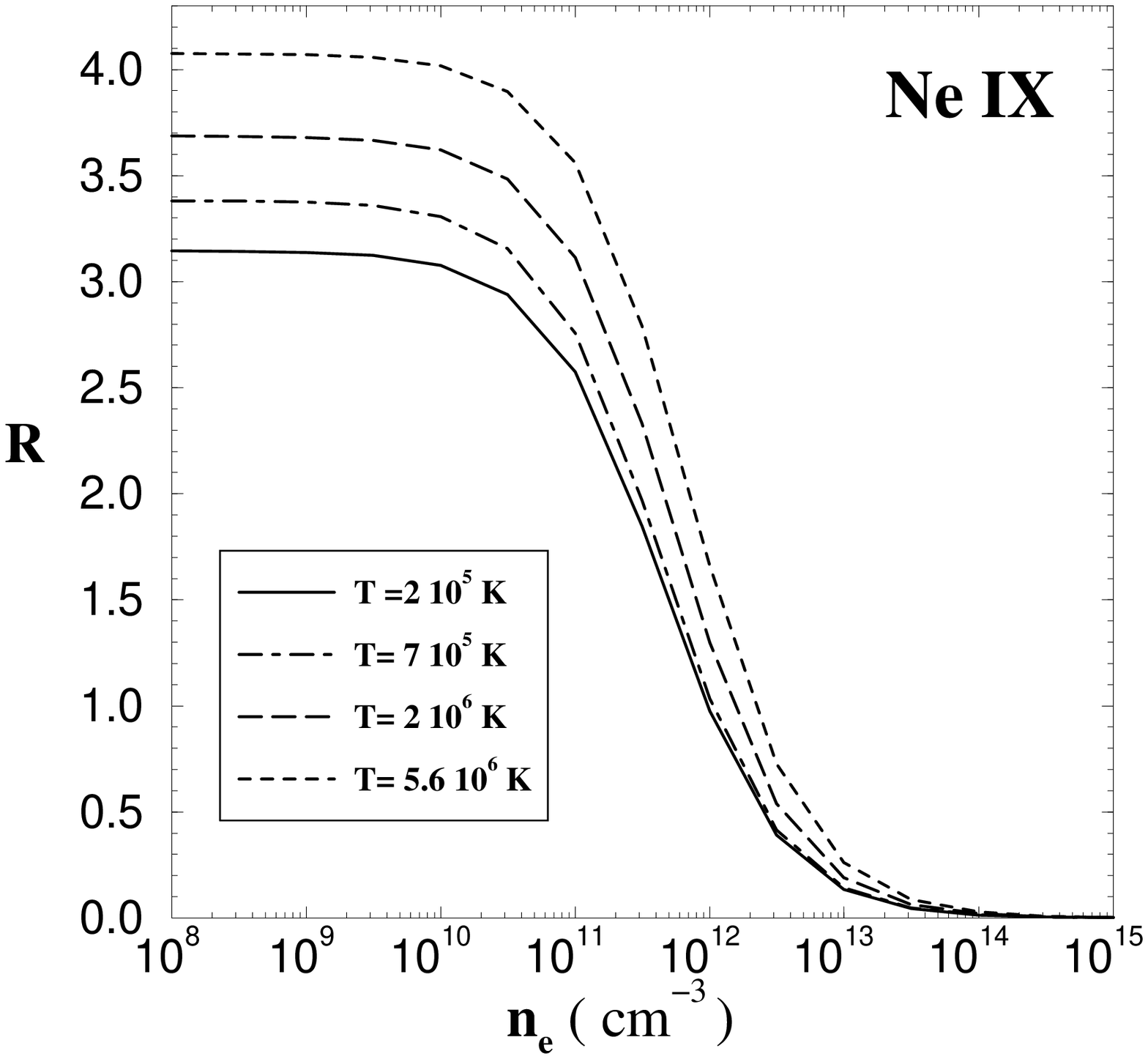}}\\ 
\resizebox{7cm}{!}{\includegraphics{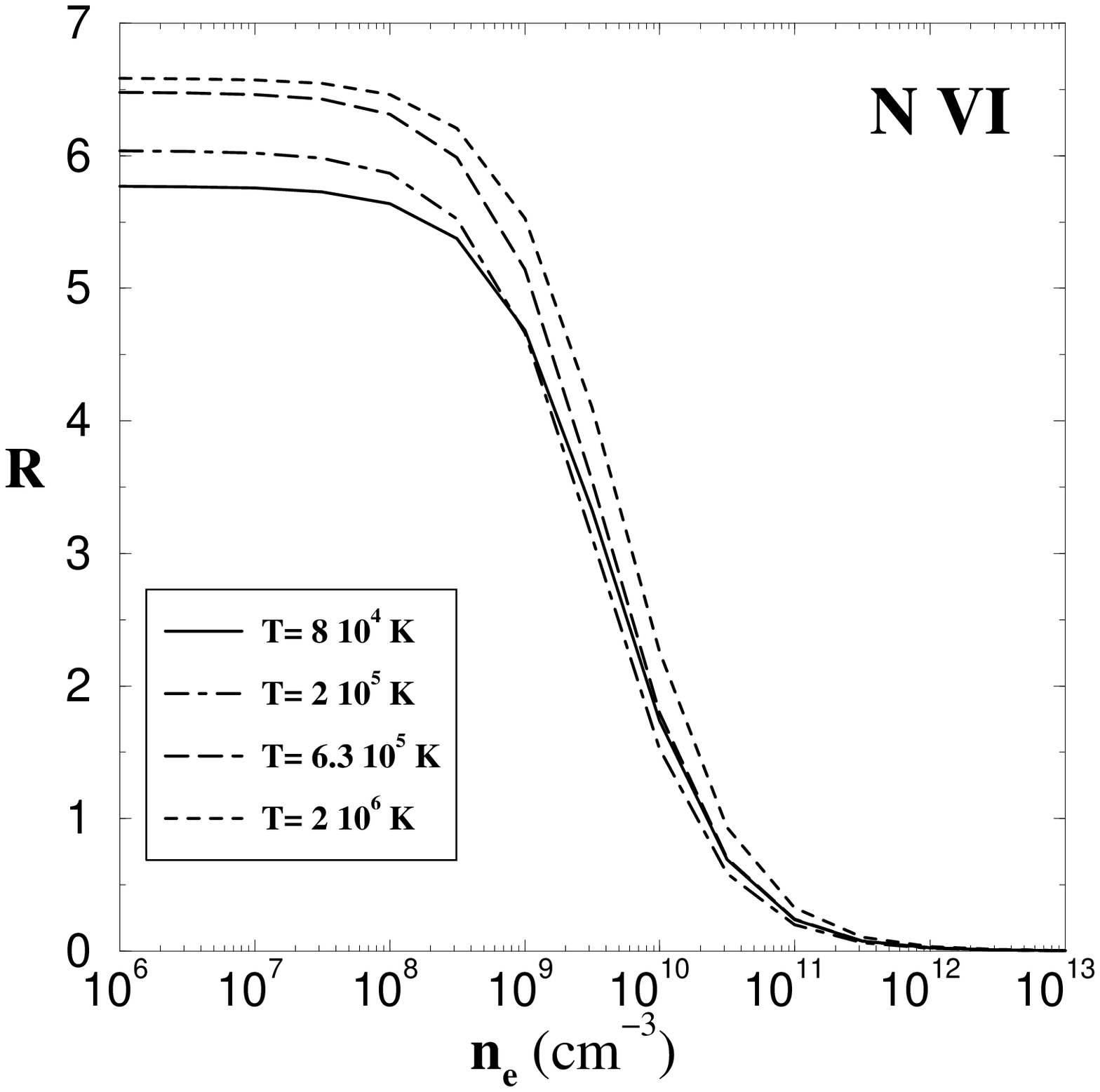}} & \resizebox{7.45cm}{!}{\includegraphics{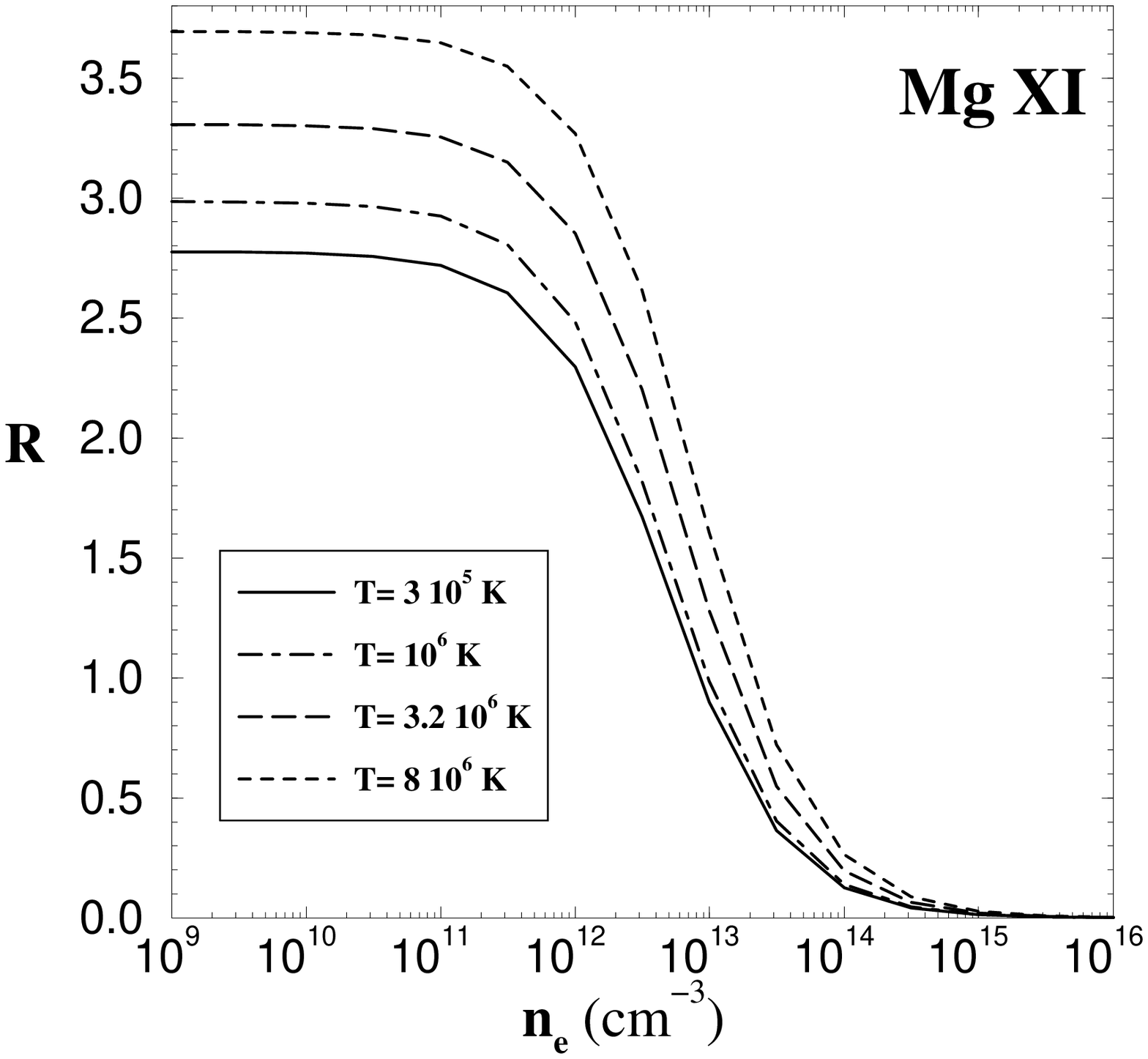}} \\
\resizebox{7.15cm}{!}{\includegraphics{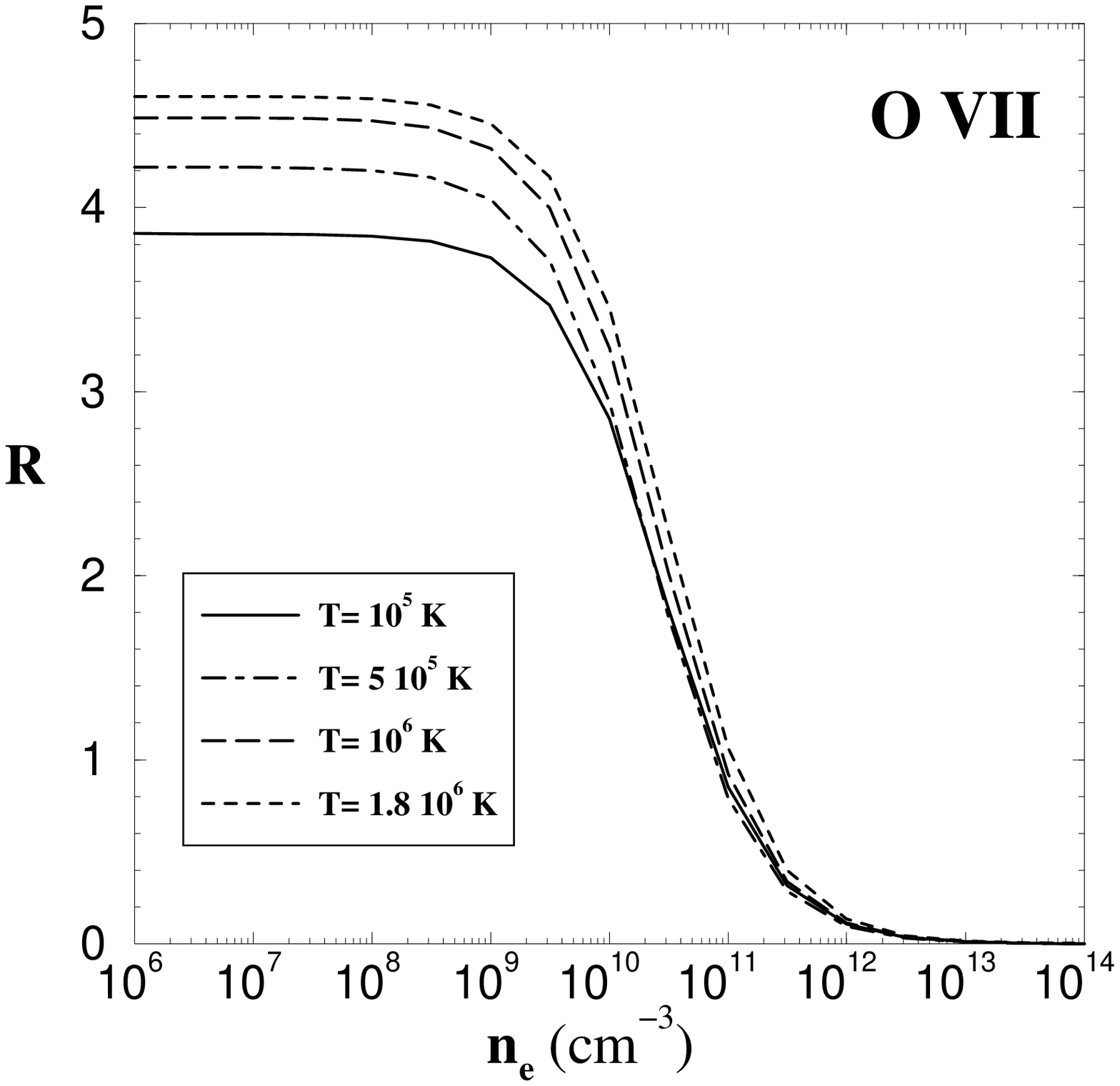}} & \resizebox{7.5cm}{!}{\includegraphics{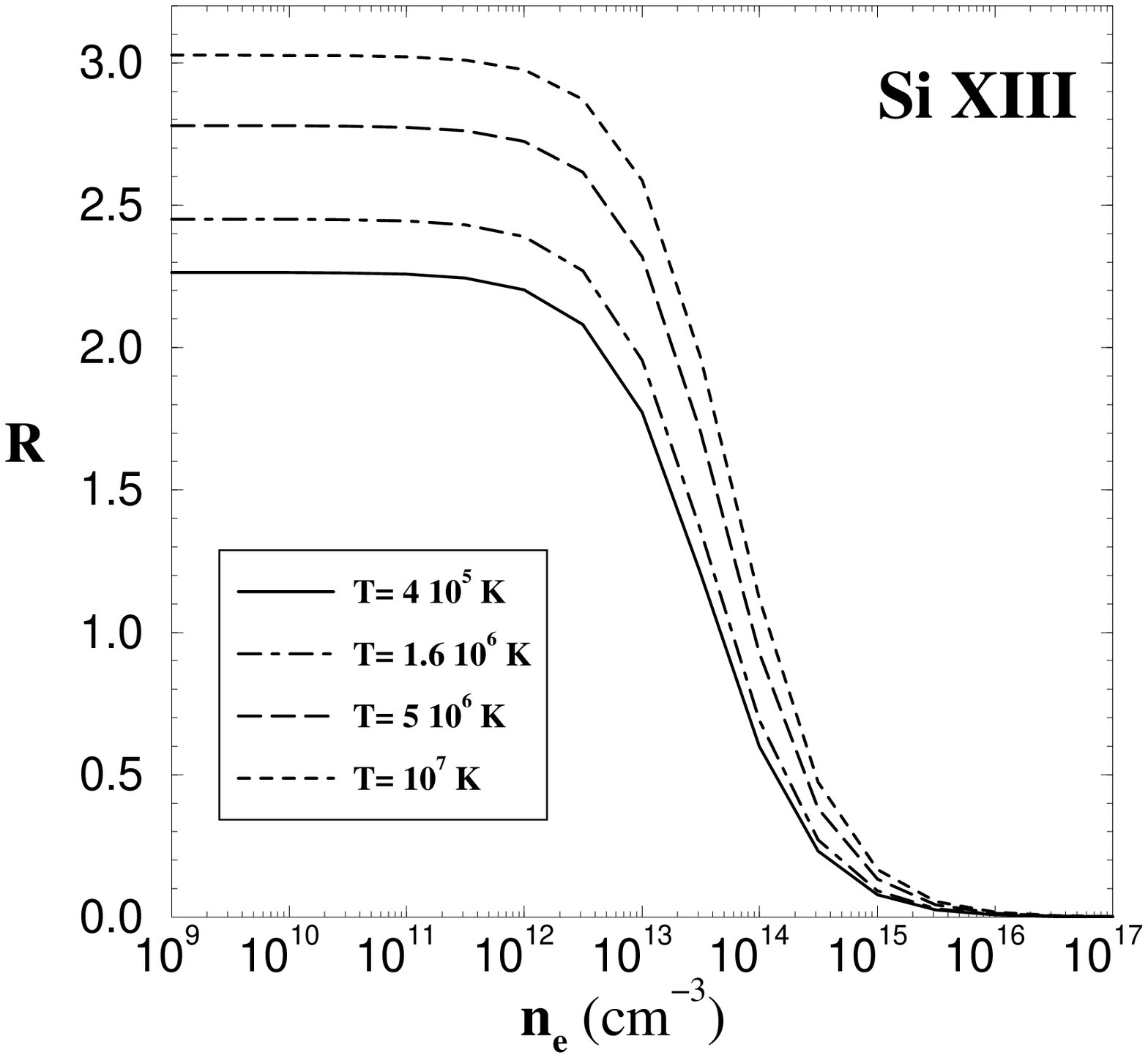}}\\ 
\end{tabular}
\caption{In case of pure photoionized plasmas (i.e. RR dominant at low temperature and DR dominant at high temperature), {\bf ratio R (=z/(x+y))} is reported as a function of n$_{\mathrm{e}}$ for \ion{C}{v}, \ion{N}{vi}, \ion{O}{vii}, \ion{Ne}{ix}, \ion{Mg}{xi}, and \ion{Si}{xiii} at different electronic temperatures (T$_{\mathrm{e}}$ in Kelvin).  For low temperatures (the two first reported here: solid curves and dot-dashed curves), the value of {\bf R} is independent of the value of {\bf X$_{\mathrm{ion}}$}. As the temperature increases, {\bf X$_{\mathrm{ion}}$} is high enough to maintain recombination dominant compared to collisional excitation from the ground level: $\sim$ 10$^{2}$ and 10$^{3-4}$ (for increasing temperature: respectively for long-dashed curves and short-dashed curves).}
\label{Rphoto}
\end{figure*}
\clearpage
%----------------------------------------------------------------------------------------
\begin{figure*}[h]
\begin{center}
\begin{tabular}{cc}
\resizebox{7cm}{!}{\includegraphics{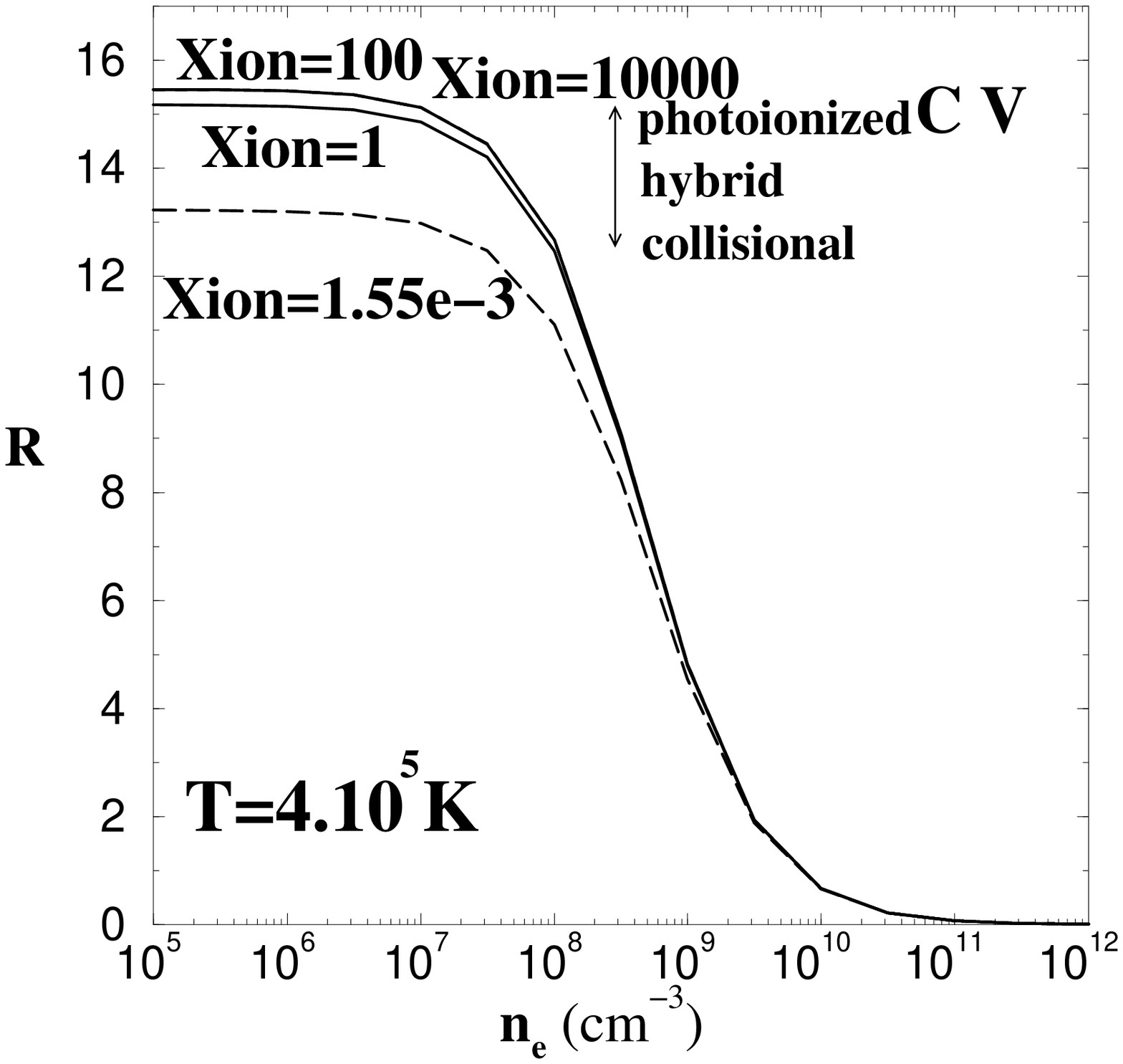}} &\resizebox{7.25cm}{!}{\includegraphics{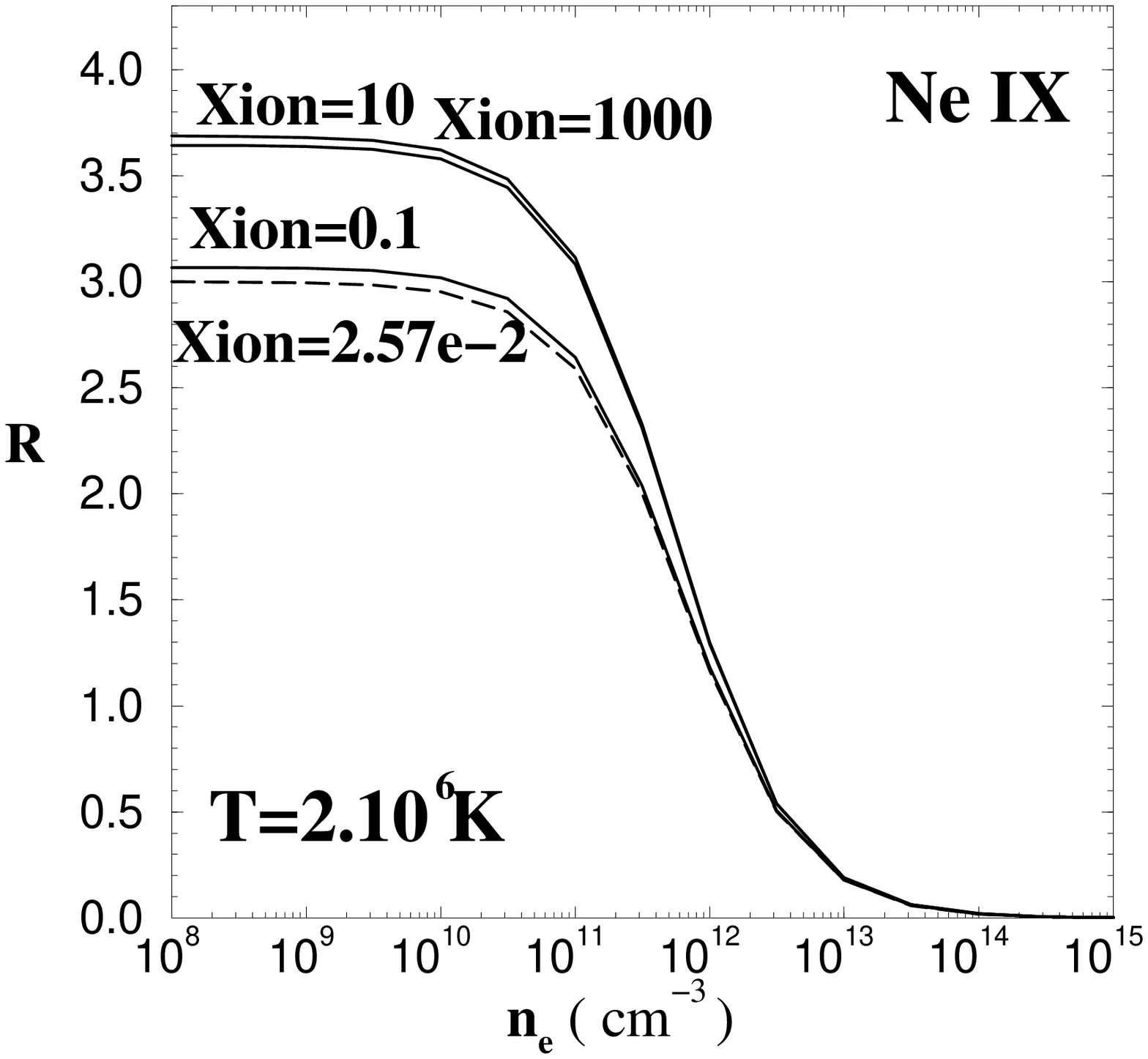}}\\
\resizebox{7cm}{!}{\includegraphics{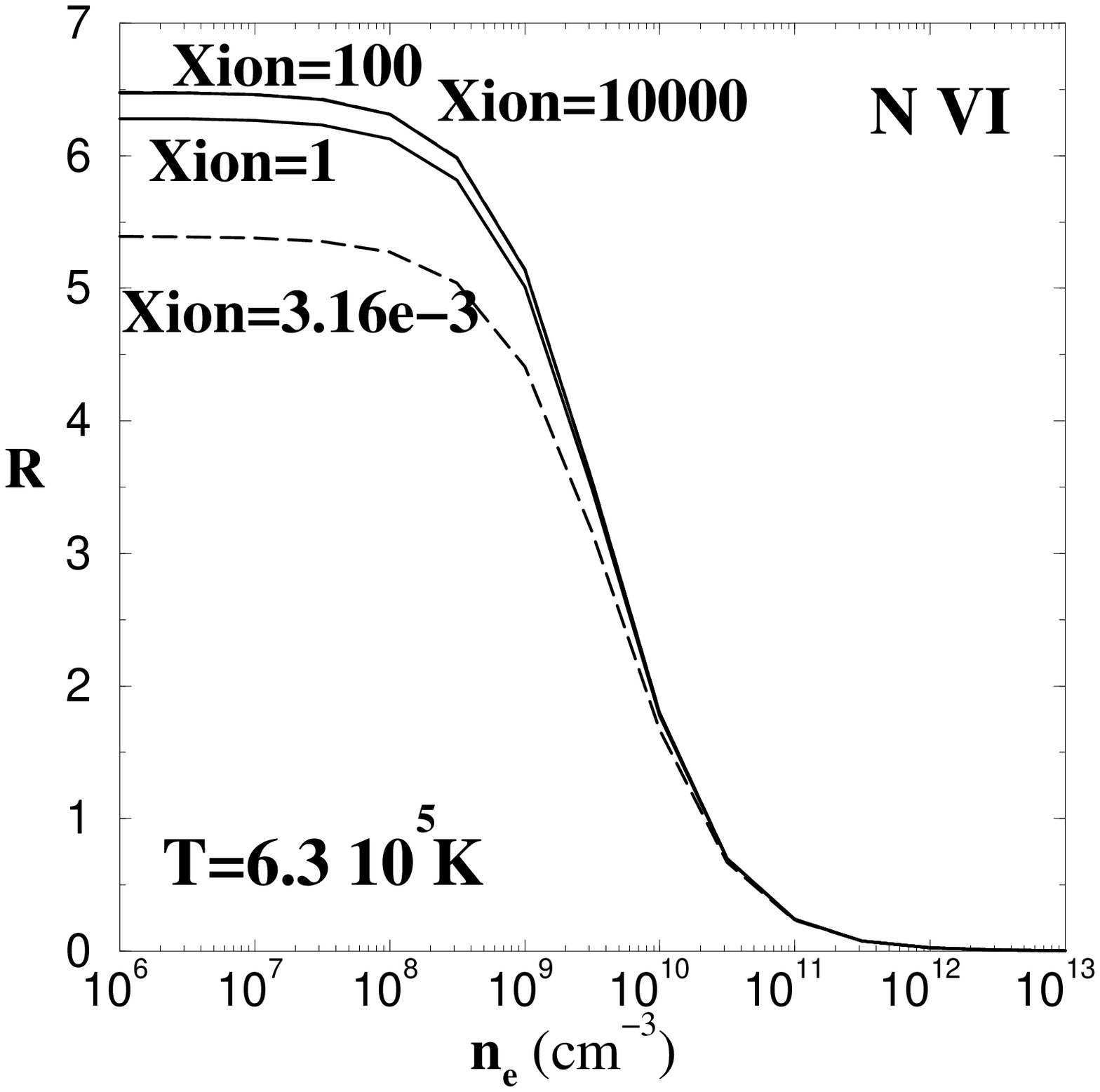}} &\resizebox{7.35cm}{!}{\includegraphics{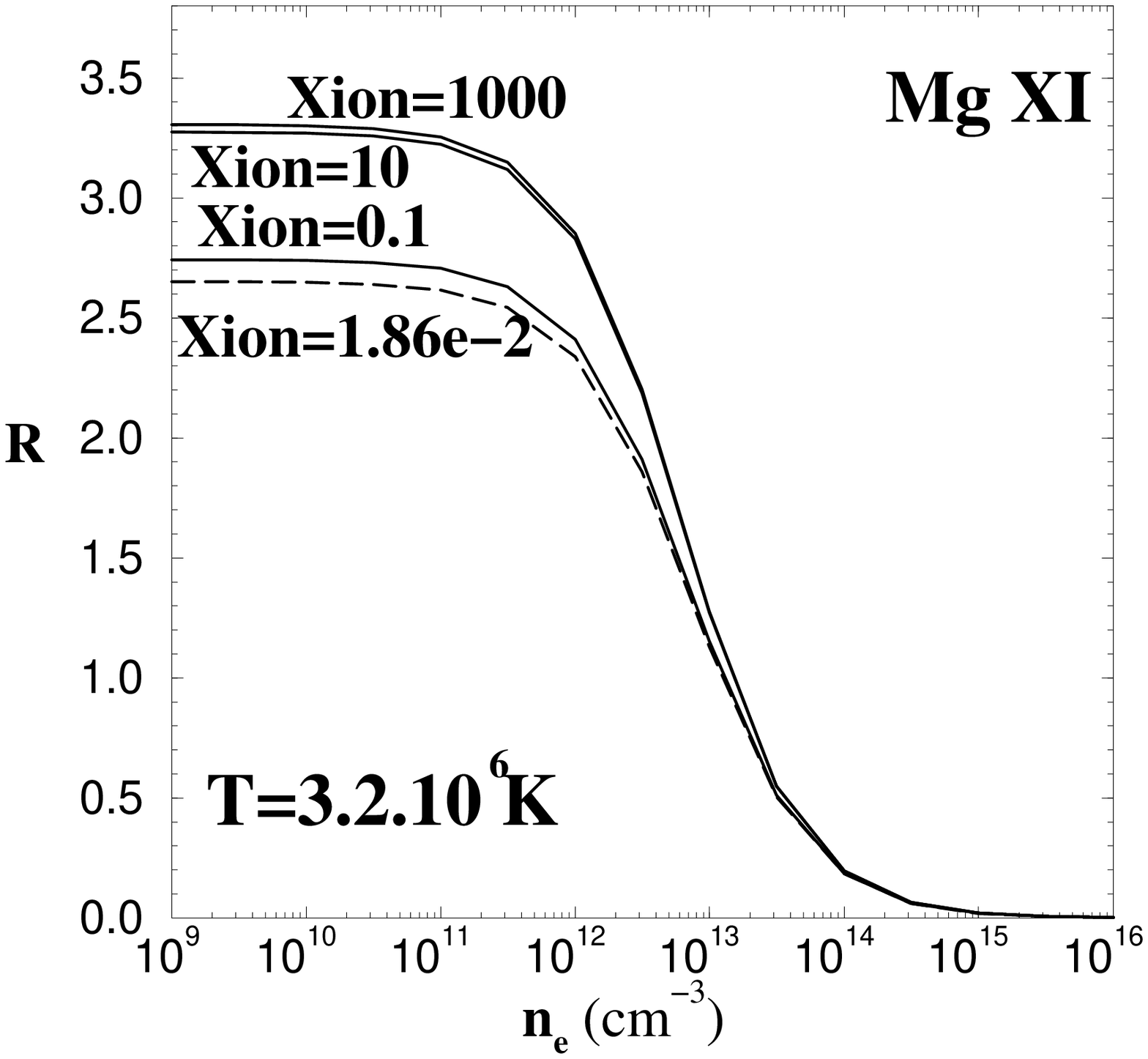}}\\ 
\resizebox{7cm}{!}{\includegraphics{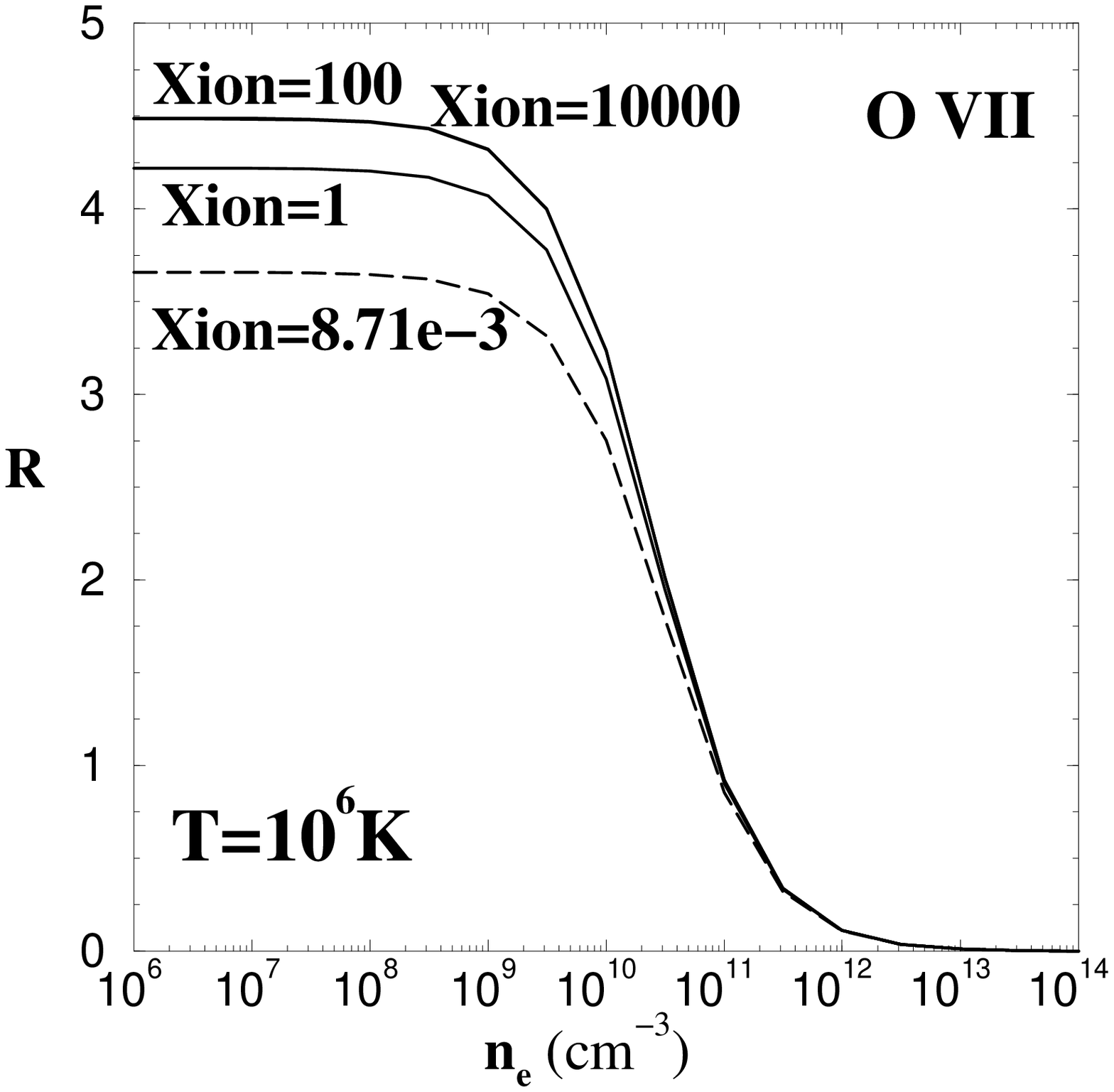}} &\resizebox{7.35cm}{!}{\includegraphics{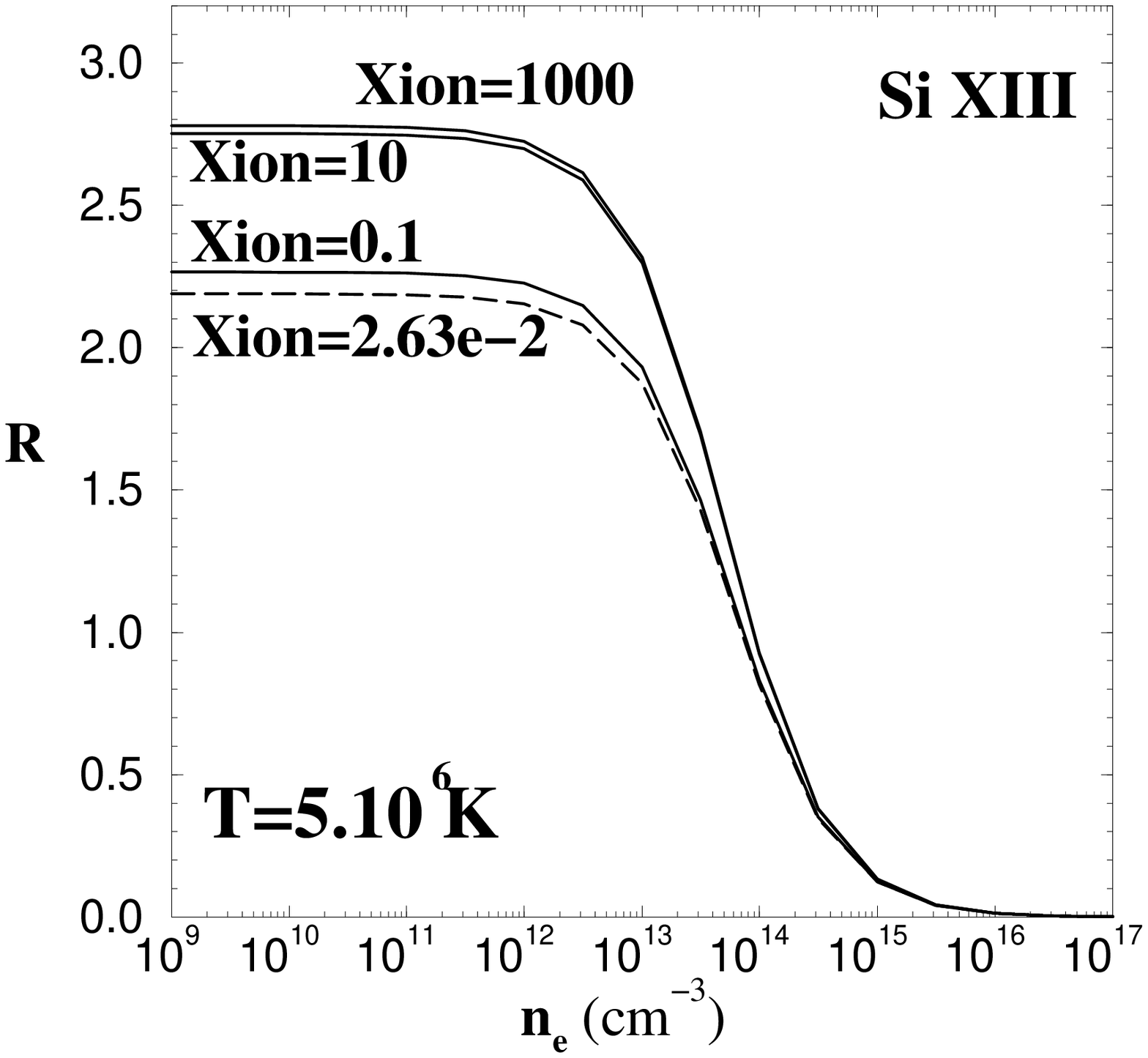}}\\ 
\end{tabular}
\end{center}
\caption{In case of hybrid plasmas (partially photoionized: recombination plus collisional excitation from the ground level), the {\bf ratio R (=z/(x+y))} is reported as a function of n$_{\mathrm{e}}$ for \ion{C}{v}, \ion{N}{vi}, \ion{O}{vii}, \ion{Ne}{ix}, \ion{Mg}{xi}, and \ion{Si}{xiii} at different values of {\bf X$_{\mathrm{ion}}$} (=H-like/He-like ionic fraction). {\bf R} is calculated at the temperature corresponding to the maximum of the He-like ion abundance for a collisional plasma (see Arnaud \& Rothenflug \cite{Arnaud85}).
Solid curves: the lowest values of {\bf X$_{\mathrm{ion}}$} corresponds to hybrid plasmas, and the highest value of {\bf X$_{\mathrm{ion}}$} to pure photoionized plasmas. Long-dashed curves: {\bf X$_{\mathrm{ion}}$} is equal to the ratio H-like/He-like in a case of collisional plasma (from Arnaud \& Rothenflug \cite{Arnaud85}). {\it Note: for \ion{C}{v}, \ion{N}{vi} and \ion{O}{vii} at these temperatures, the curves for X$_{\mathrm{ion}}$=100 and 10\,000 are indistinguishable}.}
\label{Rhybrid}
\end{figure*}
\clearpage
%----------------------------------------------------------------------------------------
\begin{figure*}[h]
\begin{tabular}{cc}
\resizebox{7cm}{!}{\includegraphics{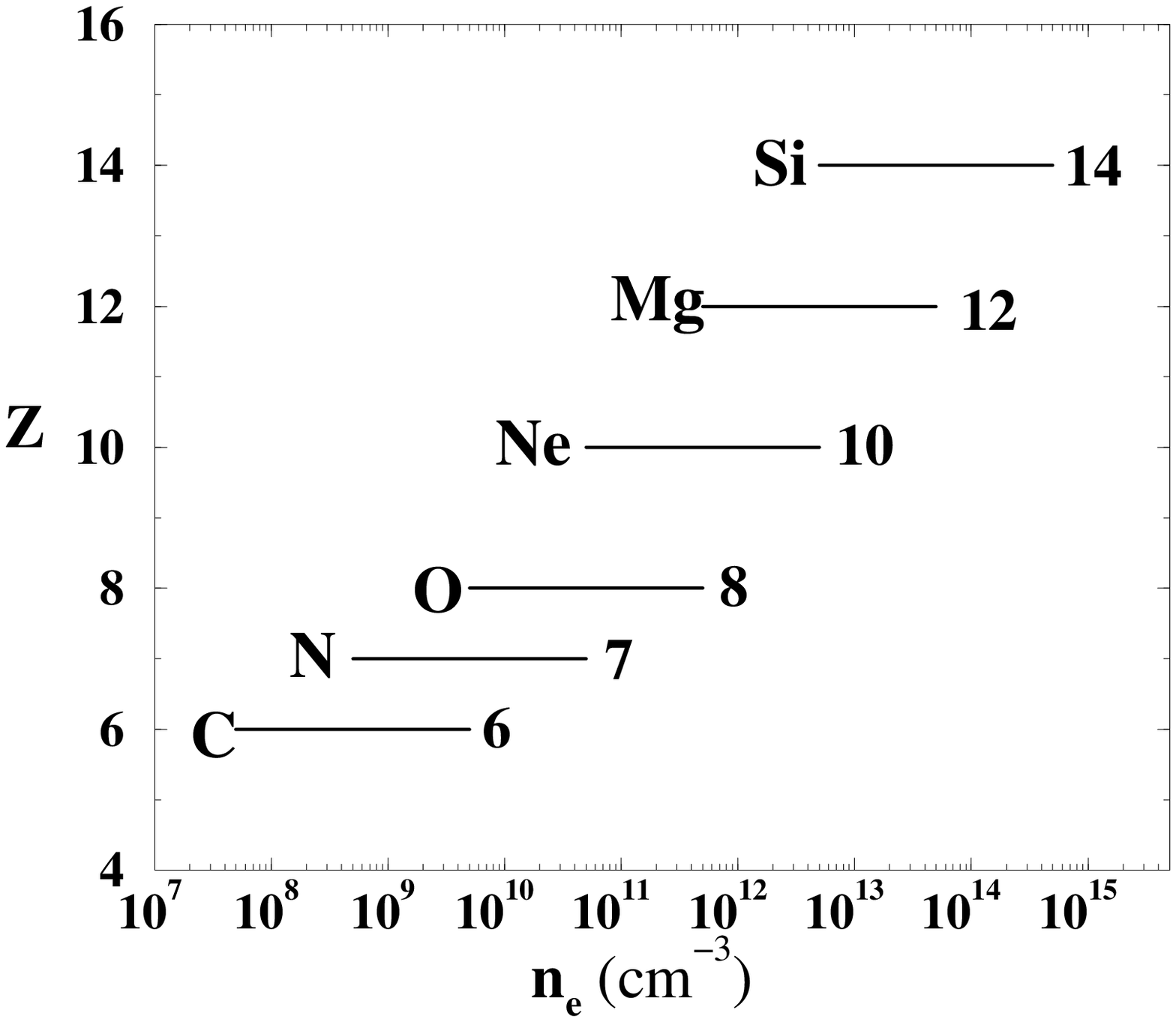}} &\resizebox{7.15cm}{!}{\includegraphics{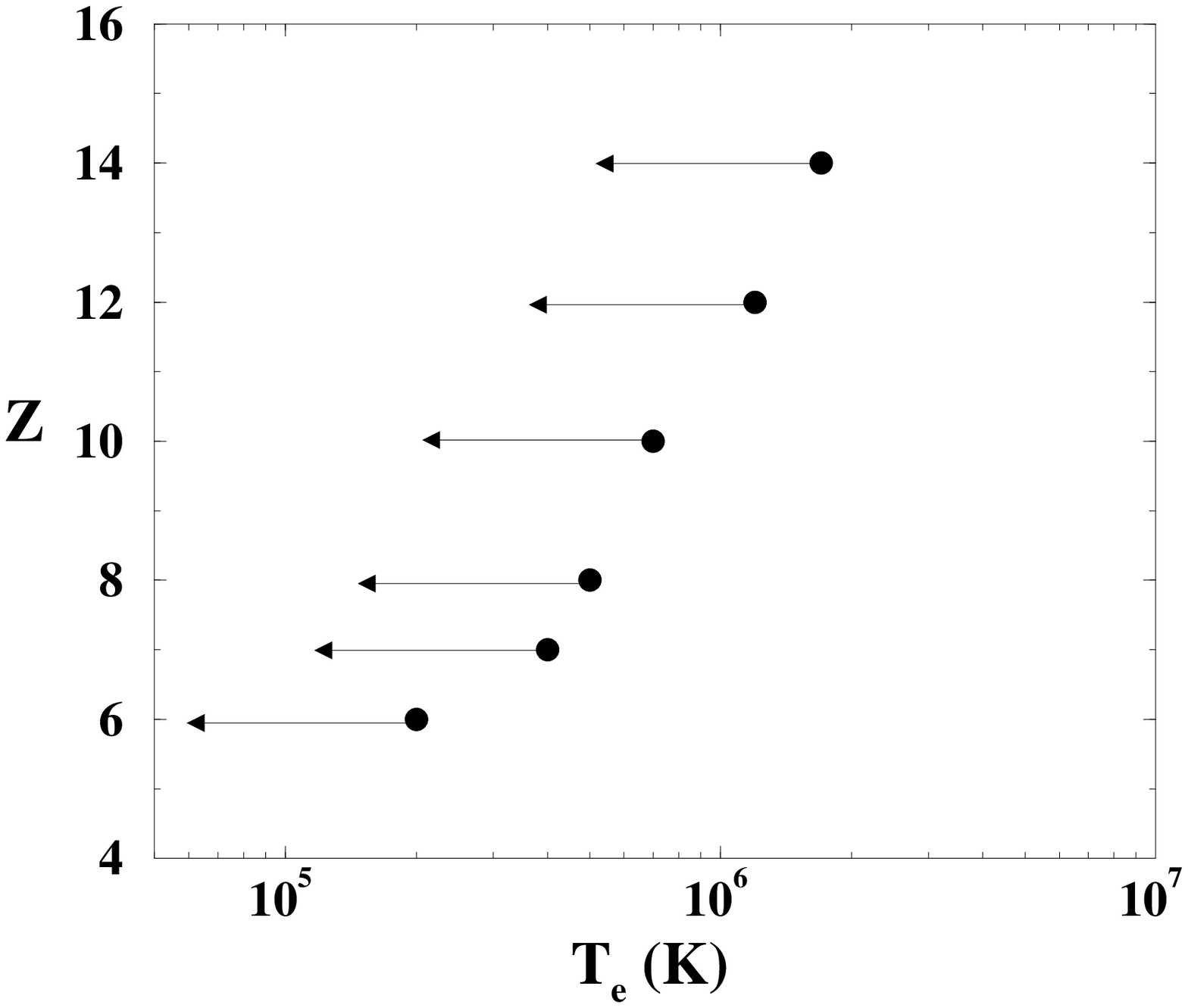}}
\end{tabular}
\caption{{\it At left}: This figure reports for each ion treated in this paper the two decades (approximatively) where the ratio {\bf R} is strongly sensitive to the density. {\it At right}: the approximative range of temperatures for each ion where the plasma can be considered purely photoionized, independent of the {\bf X$_{\mathrm{ion}}$} value.}
\label{nedomaine}
\end{figure*}
\clearpage
%----------------------------------------------------------------------------------------
\begin{figure*}[h]
\begin{tabular}{cc}
\resizebox{8cm}{!}{\includegraphics{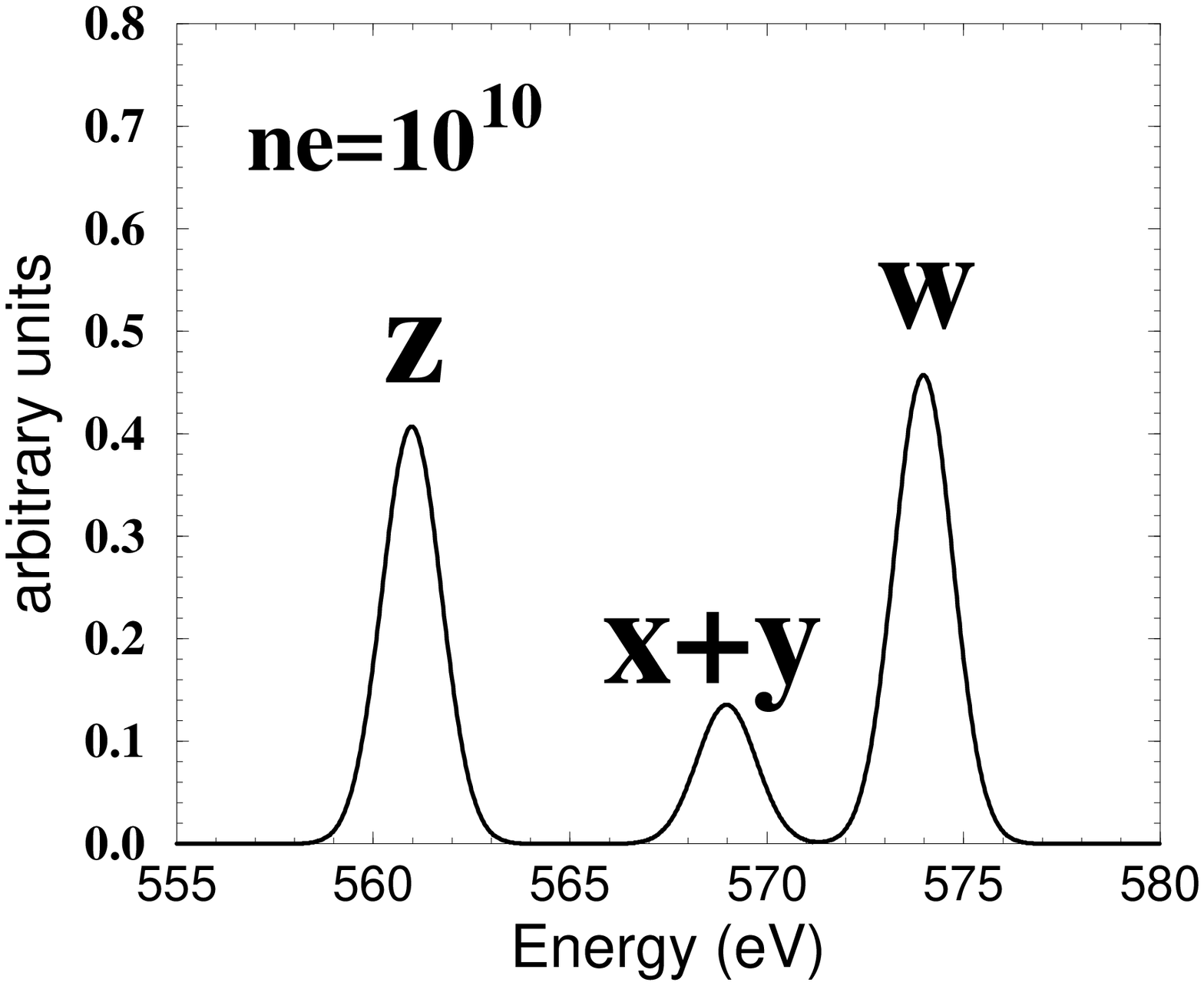}} &\resizebox{8cm}{!}{\includegraphics{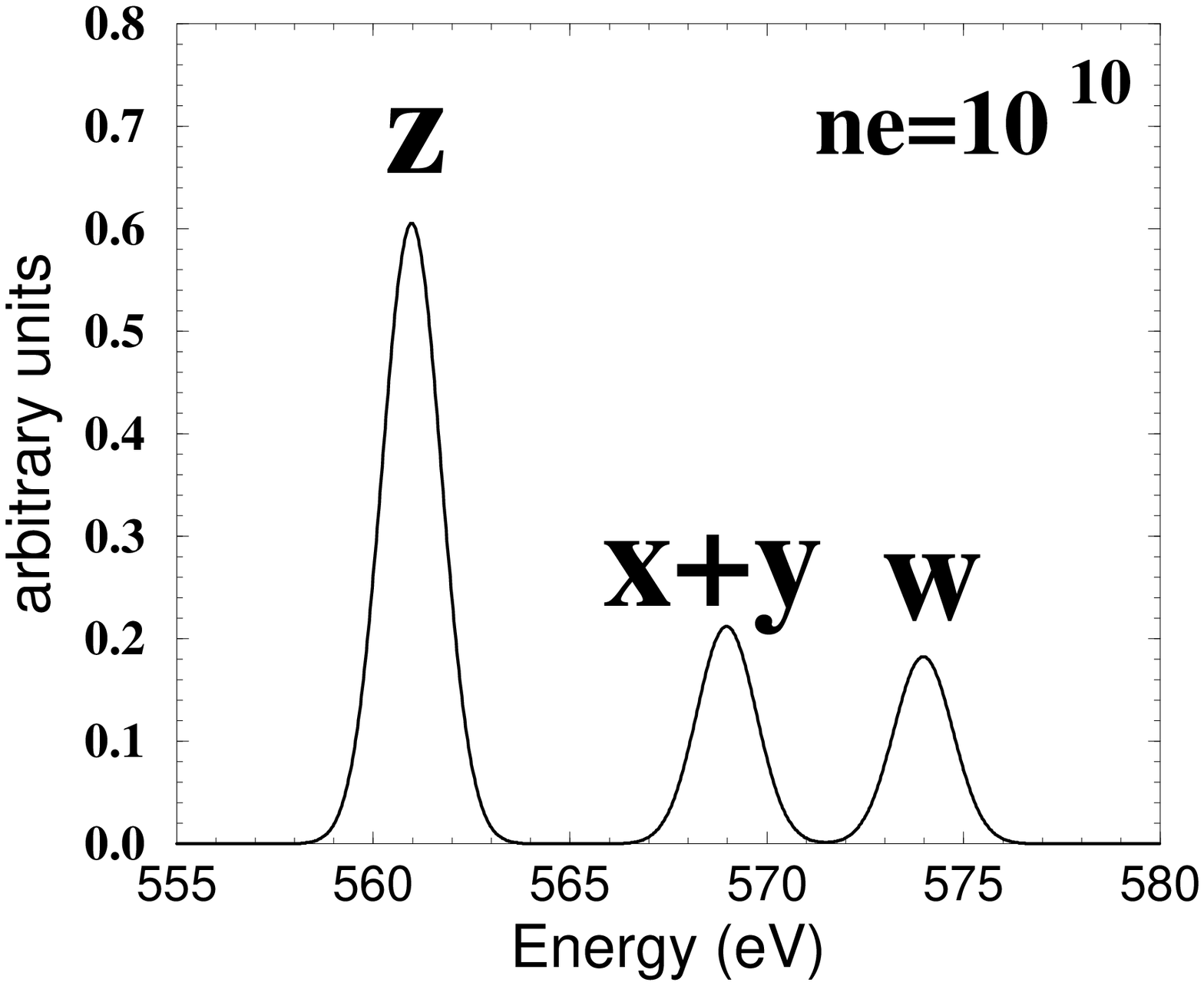}}\\
\resizebox{8cm}{!}{\includegraphics{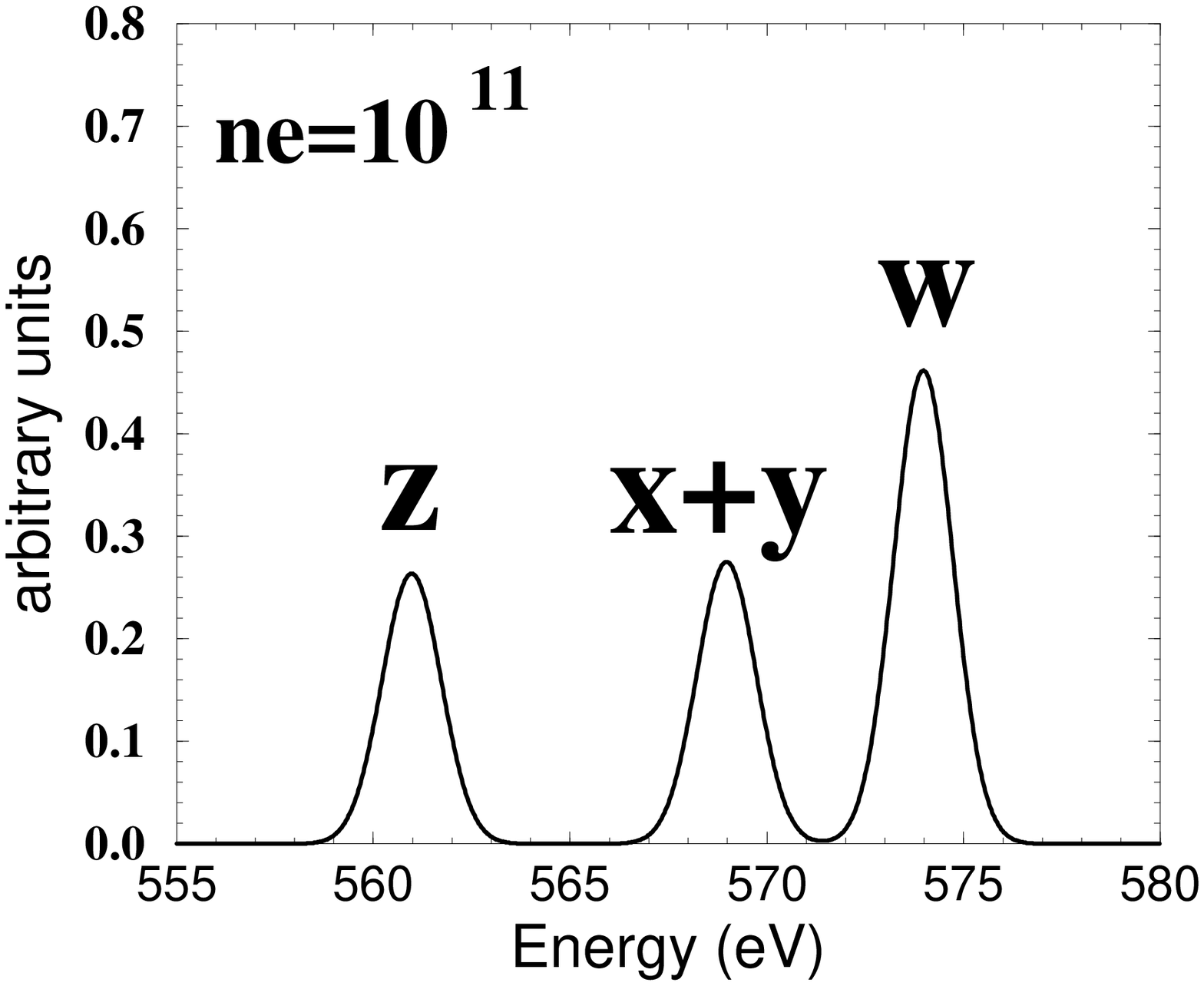}} &\resizebox{8cm}{!}{\includegraphics{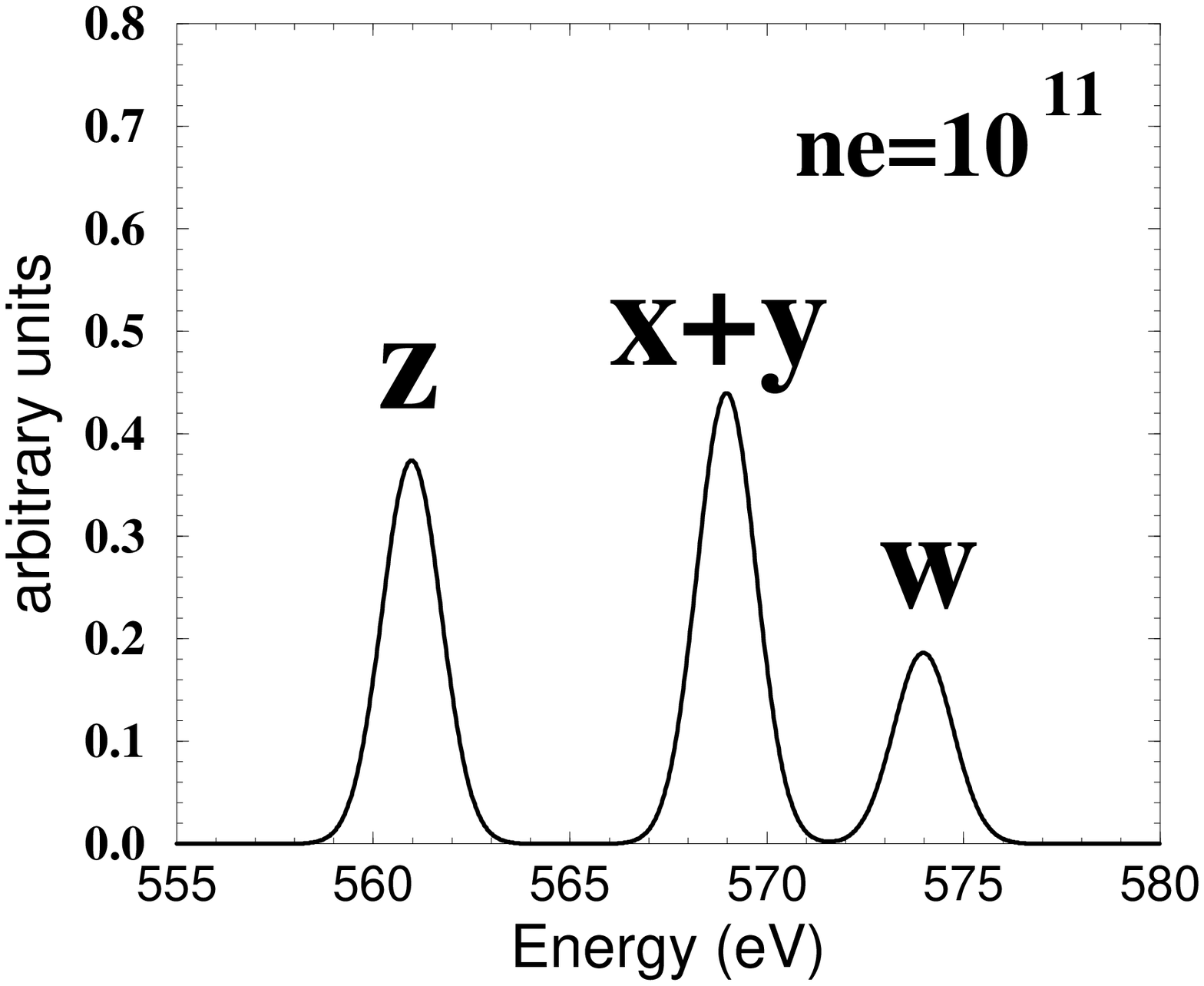}}\\
\resizebox{8cm}{!}{\includegraphics{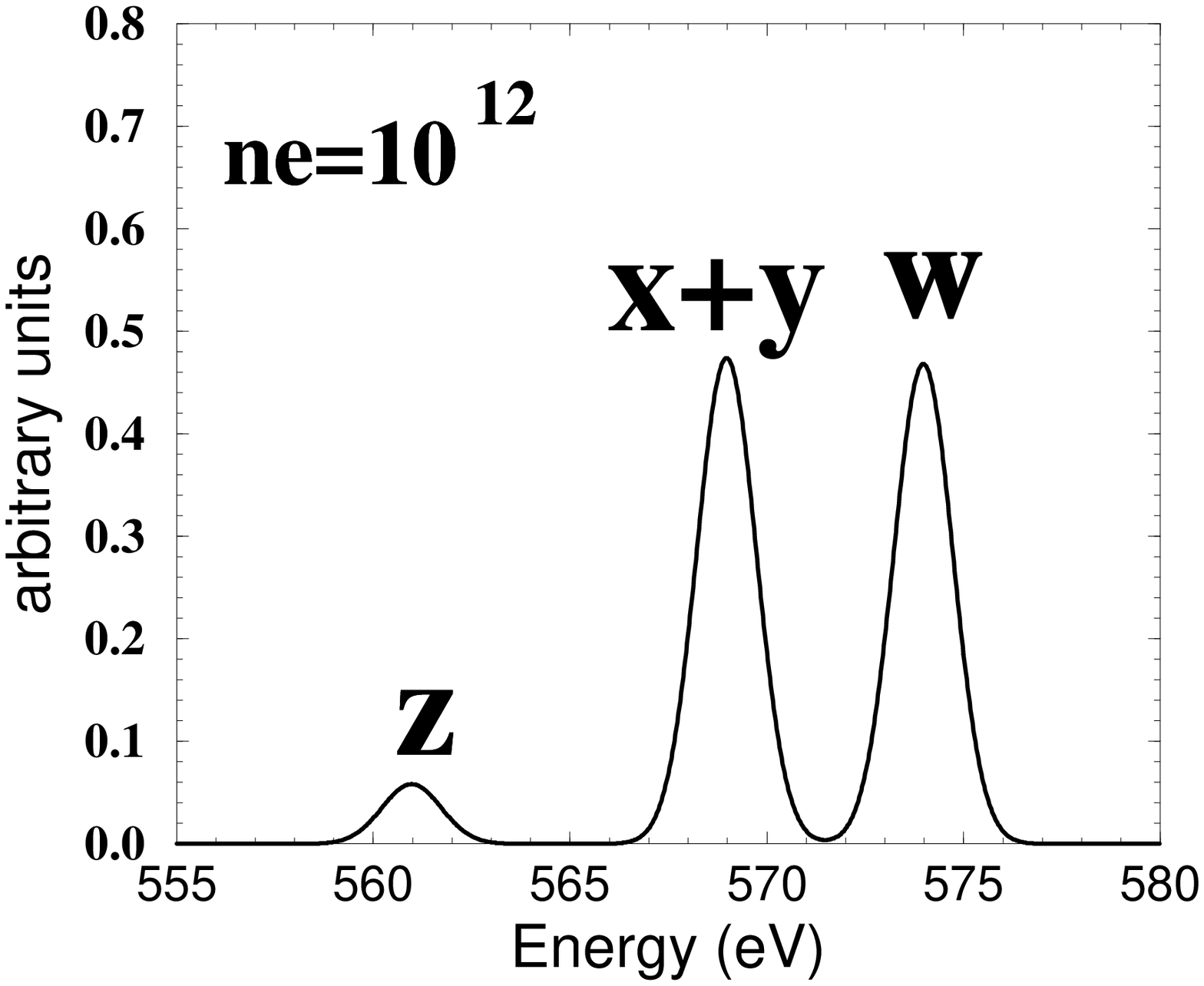}} &\resizebox{8cm}{!}{\includegraphics{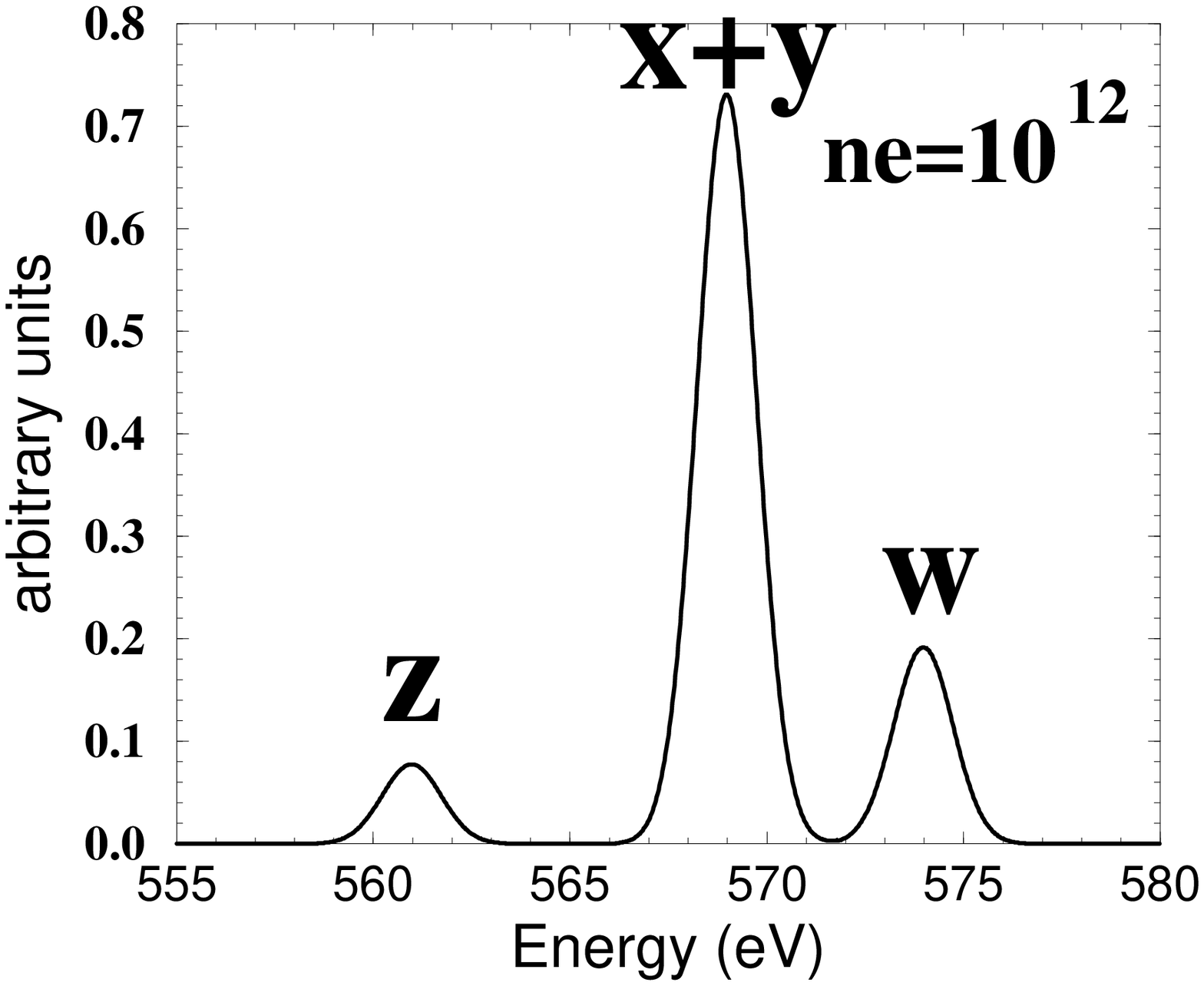}}\\
\end{tabular}
\caption{\ion{O}{vii} theoretical spectra constructed using the RGS (XMM) resolving power (E/$\Delta$E) for three values of density (in cm$^{-3}$). This corresponds (approximatively) to the range where the ratio {\bf R} is very sensitive to density. {\bf z}: forbidden lines, {\bf x+y}: intercombination lines and {\bf w}: resonance line. {\it At left}: ``hybrid plasma'' at T$_{\mathrm{e}}$=1.5\,10$^{6}$\,K and X$_{\mathrm{ion}}$=1; {\it At right}: ``pure'' photoionized plasma at T$_{\mathrm{e}}$=10$^{5}$\,K (at this temperature this part of the spectra are independent of the value of X$_{\mathrm{ion}}$, see Figure~\ref{1surG}). {\it Note: the intensities are normalized in order to have the sum of the lines equal to the unity}.}
\label{XMM}
\end{figure*}
\clearpage

%***************************Table********************************
%TABLE 1
\begin{table*}
\caption{Energy (in cm$^{-1}$) for the first 17 levels for \ion{C}{v}, \ion{N}{vi}, \ion{O}{vii}, \ion{Ne}{ix}, \ion{Mg}{xi} and \ion{Si}{xiii} calculated by the SUPERSTRUCTURE code (except for the first seven levels which are from Vainshtein \& Safronova 1985). Here X(Y) means X$\times$10$^{Y}$.}
\begin{center}
{\scriptsize
\begin{tabular}{ccccccccc}
\hline
\hline
%%          &  &             &             &              &              &              &               \\
%          &  &             &             &              &              &              &               \\
i &conf    &level   &  \ion{C}{v} & \ion{N}{vi} & \ion{O}{vii} & \ion{Ne}{ix} & \ion{Mg}{xi} & \ion{Si}{xiii}\\
%%          &  &             &             &              &              &              &               \\
\hline 
\hline
%          &  &    &             &              &              &              &               \\
1 &1s$^{2}$&$^{1}\mathrm{S}_{0}$&     0.      &   0.         &0.            &   0.         &   0.         &  0. \\
2 &1s\,2s  &$^{3}\mathrm{S}_{1}$&2.4114\,(+6)&3.3859\,(+6) &4.5253\,(+6)  &7.2996\,(+6)  &10.7358\,(+6) &14.8357\,(+6)   \\
3 &1s\,2p  &$^{3}\mathrm{P}_{0}$&2.4553\,(+6)&3.4383\,(+6) &4.5863\,(+6)  &7.3779\,(+6)  &10.8317\,(+6) &14.9495\,(+6)     \\
4 &1s\,2p  &$^{3}\mathrm{P}_{1}$&2.4552\,(+6)&3.4383\,(+6) &4.5863\,(+6)  &7.3782\,(+6)  &10.8325\,(+6) &14.9513\,(+6)     \\
5 &1s\,2p  &$^{3}\mathrm{P}_{2}$&2.4554\,(+6)&3.4386\,(+6) &4.5869\,(+6)  &7.3798\,(+6)  &10.8361\,(+6) &14.9585\,(+6)     \\
6 &1s\,2s  &$^{1}\mathrm{S}_{0}$&2.4551\,(+6)&3.4393\,(+6) &4.5884\,(+6)  &7.3824\,(+6)  &10.8385\,(+6) &14.9585\,(+6)     \\
7 &1s\,2p  &$^{1}\mathrm{P}_{1}$&2.4833\,(+6)&3.4737\,(+6) &4.6291\,(+6)  &7.4361\,(+6)  &10.9062\,(+6) &15.0417\,(+6)   \\
8 &1s\,3s  &$^{3}\mathrm{S}_{1}$&2.8239\,(+6)&3.9765\,(+6) &5.3251\,(+6)  &8.6105\,(+6)  &12.6824\,(+6) &17.5435\,(+6)     \\
9 &1s\,3p  &$^{3}\mathrm{P}_{0}$&2.8352\,(+6)&3.9902\,(+6) &5.3441\,(+6)  &8.6314\,(+6)  &12.7081\,(+6) &17.5741\,(+6)     \\
10&1s\,3p  &$^{3}\mathrm{P}_{1}$&2.8352\,(+6)&3.9903\,(+6) &5.3412\,(+6)  &8.6316\,(+6)  &12.7087\,(+6) &17.5752\,(+6)     \\
11&1s\,3p  &$^{3}\mathrm{P}_{2}$&2.8353\,(+6)&3.9904\,(+6) &5.3414\,(+6)  &8.6322\,(+6)  &12.7099\,(+6) &17.5775\,(+6)     \\
12&1s\,3s  &$^{1}\mathrm{S}_{0}$&2.8401\,(+6)&3.9953\,(+6) &5.3463\,(+6)  &8.6368\,(+6)  &12.7136\,(+6) &17.5795\,(+6)     \\
13&1s\,3d  &$^{3}\mathrm{D}_{1}$&2.8408\,(+6)&3.9973\,(+6) &5.3497\,(+6)  &8.6433\,(+6)  &12.7238\,(+6) &17.5942\,(+6)     \\
14&1s\,3d  &$^{3}\mathrm{D}_{2}$&2.8408\,(+6)&3.9973\,(+6) &5.3497\,(+6)  &8.6433\,(+6)  &12.7239\,(+6) &17.5945\,(+6)     \\
15&1s\,3d  &$^{3}\mathrm{D}_{3}$&2.8408\,(+6)&3.9973\,(+6) &5.3498\,(+6)  &8.6435\,(+6)  &12.7244\,(+6) &17.5953\,(+6)     \\
16&1s\,3d  &$^{1}\mathrm{D}_{2}$&2.8411\,(+6)&3.9977\,(+6) &5.3502\,(+6)  &8.6411\,(+6)  &12.7251\,(+6) &17.5962\,(+6)     \\
17&1s\,3p  &$^{1}\mathrm{P}_{1}$&2.8433\,(+6)&4.0004\,(+6) &5.3534\,(+6)  &8.6480\,(+6)  &12.7294\,(+6) &17.6005\,(+6)     \\
%%  &            &             &              &              &              &              \\
\hline
\hline
\end{tabular}
}
\end{center}
\label{energielevel}
\end{table*}

%TABLE 2
\begin{table*}[h]
\caption{Radiative transitions probabilities (A$_{\mathrm{ki}}$ in s$^{-1}$, i=1,7; k=2,17) for \ion{C}{v}, \ion{N}{vi}, \ion{O}{vii}, \ion{Ne}{ix}, \ion{Mg}{xi} and \ion{Si}{xiii} calculated by the SUPERSTRUCTURE code, except for marked values (a) which are from Lin et al. (1977) and (b) which are from Mewe \& Schrijver (1978a). i and k correspond respectively to the lower and the upper level of the transition.}
\begin{center}
{\scriptsize
\begin{tabular}{cccccccc}
\hline
\hline
%%            &             &            &             &              &             &               &      \\
            &             &                        \multicolumn{6}{c}{A$_{ki}$ (s$^{-1}$)}                                       \\
%%            &             &            &             &              &             &               &      \\
\cline{3-8}
%%            &             &            &             &              &             &               &      \\
     i      &     k   & \ion{C}{v} & \ion{N}{vi} & \ion{O}{vii} & \ion{Ne}{ix} & \ion{Mg}{xi} & \ion{Si}{xiii}\\
\hline
\hline
%%            &          &            &             &              &              &              &     \\
1           &  2       &4.960\,(+01)$^{\mathrm{a}}$&2.530\,(+02)$^{\mathrm{b}}$   & 1.060\,(+03)$^{\mathrm{a}}$   &1.100\,(+04)$^{\mathrm{a}}$    & 7.330\,(+04)$^{\mathrm{a}}$   &3.610\,(+05)$^{\mathrm{a}}$\\
1           &  4       &2.159\,(+07)               &1.100\,(+08)                  &4.447\,(+08)                   &4.470\,(+09)    & 2.867(+10)   & 1.345\,(+11)    \\
1           &  5       &2.650\,(+04)$^{\mathrm{a}}$&1.030\,(+05)$^{\mathrm{b}}$   &3.330\,(+05)$^{\mathrm{a}}$    &2.270\,(+06)$^{\mathrm{a}}$    & 1.060\,(+07)$^{\mathrm{a}}$   & 3.890\,(+07)$^{\mathrm{a}}$\\
1           &  6       &3.310\,(+05)$^{\mathrm{a}}$&9.430\,(+05)$^{\mathrm{b}}$    &2.310\,(+06)$^{\mathrm{a}}$    &1.000\,(+07)$^{\mathrm{a}}$    & 3.220\,(+07)$^{\mathrm{a}}$   & 8.470\,(+07)$^{\mathrm{a}}$\\
1           &  7       &9.477\,(+11)  &1.911\,(+12)   &3.467\,(+12)    &9.197\,(+12)    & 2.010\,(+13)   &3.857\,(+13)     \\
1           &  10         &6.939\,(+06)  & 3.525\,(+07)  &1.423\,(+08)    &1.429\,(+09)    & 9.141\,(+09)   &4.268\,(+10)     \\
1           &  17         &3.105\,(+11)  &6.061\,(+11)   & 1.073\,(+12)   &2.752\,(+12)    & 5.877\,(+12)   & 1.107\,(+13)    \\
2           &  3          &5.616\,(+07)  &6.717\,(+07)   &7.818\,(+07)    &1.003\,(+08)    & 1.228\,(+08)   &1.460\,(+08)     \\
2           &  4          &5.655\,(+07)  &6.794\,(+07)   &7.956\,(+07)    & 1.039\,(+08)   & 1.304\,(+08)   &1.602\,(+08)     \\
2           &  5          &5.735\,(+07)  &6.955\,(+07)   &8.249\,(+07)    &1.118\,(+08)    & 1.486\,(+08)   &1.977\,(+08)     \\
2           &  9          &1.376\,(+10)  & 2.872\,(+10)  & 5.342\,(+10)   &1.466\,(+11)    & 3.280\,(+11)   &6.406\,(+11)     \\
2           &  10         &1.375\,(+10)  &2.870\,(+10)   & 5.337\,(+10)   &1.464\,(+11)    & 3.269\,(+11)   &6.366\,(+11)     \\
2           &  11         &1.374\,(+10)  &2.867\,(+10)   &5.329\,(+10)    &1.461\,(+11)    & 3.262\,(+11)   &6.360\,(+11)     \\
2           &  17         &2.898\,(+05)  & 1.607\,(+06)  &6.902\,(+06)    &7.514\,(+07)    &  5.061\,(+08)  & 2.452\,(+09)    \\
3           &  8          &7.088\,(+08)  &1.366\,(+09)   & 2.398\,(+09)   & 6.079\,(+09)   &  1.290\,(+10)  & 2.426\,(+10)    \\
3           &  13         &2.349\,(+10)  & 4.847\,(+10)  & 8.947\,(+10)   & 2.432\,(+11)   & 5.408\,(+11)   & 1.052\,(+12)    \\
4           &  8          &2.129\,(+09) & 4.106\,(+09)& 7.211\,(+09)   &1.831\,(+10)    &3.890\,(+10)    & 7.315\,(+10)    \\
4           &  13         &1.761\,(+10)  & 3.634\,(+10)  & 6.706\,(+10)   & 1.822\,(+11)   & 4.046\,(+11)   & 7.856\,(+11)    \\
4           &  14         &3.165\,(+10)  &6.519\,(+10)   &1.199\,(+11)    & 3.213\,(+11)   & 6.967\,(+11)   &1.314\,(+12)     \\
4           &  16         &5.150\,(+07)  &2.346\,(+08)   &8.525\,(+08)    &6.862\,(+09)    & 3.311\,(+10)   & 1.073\,(+11)    \\
5           &  8          &3.557\,(+09)  & 6.870\,(+09)  & 1.208(+10)   & 3.080(+10)   & 6.583(+10)   & 1.248(+11)    \\
5           &  13         &1.174\,(+09)  & 2.421\,(+09)  & 4.468(+09)   &1.213(+10)    & 2.695(+10)   & 5.240(+10)    \\
5           &  14         &1.054\,(+10)  & 2.169\,(+10)  & 3.983(+10)   &1.061(+11)    & 2.268(+11)   & 4.173(+11)    \\
5           &  15         &4.225\,(+10)  & 8.718\,(+10)  & 1.609(+11)   &4.369(+11)    & 9.708(+11)   & 1.887(+12)    \\
5           &  16         &2.214\,(+07)  & 1.025\,(+08)  & 3.784(+08)   &3.145(+09)    &1.582(+10)     &5.438(+10)     \\
6           &  7          &5.875\,(+06)  &9.199\,(+06)   & 1.307\,(+07)   &2.266\,(+07)    & 3.541\,(+07)   & 5.286\,(+07)    \\
6           &  10         &4.013\,(+05)  &2.088\,(+06)   & 8.582\,(+06)   &8.838\,(+07)    & 5.759\,(+08)   &2.730\,(+09)     \\
6           &  17         &1.457\,(+10)  & 2.982\,(+10)  & 5.478\,(+10)   &1.482\,(+11)    & 3.286\,(+11)   & 6.371\,(+11)    \\
7           &  12         &5.646\,(+09)  &1.145\,(+10)   & 2.071\,(+10)   &5.436\,(+10)    & 1.175\,(+11)   &2.232\,(+11)     \\
7           &  13         &3.673\,(+05)  &1.940\,(+06)   & 8.054\,(+06)   &8.401\,(+07)    & 5.522\,(+08)   &2.638\,(+09)     \\
7           &  14         &6.862\,(+07)  &3.164\,(+08)   & 1.162\,(+09)   &9.519\,(+09)    & 4.675\,(+10)   &1.547\,(+11)     \\
7           &  16         &3.950\,(+10)  &8.194\,(+10)   & 1.516\,(+11)   &4.092\,(+11)    & 8.896\,(+11)   & 1.674\,(+12)    \\
\hline
\hline
\end{tabular}
}
\end{center}
\label{Aki1}
\end{table*}
%\newpage

%TABLE 3
\begin{table*}[h]
\caption{Radiative and dielectronic recombination rates (respectively RR and DR) calculated in this work (in cm$^{3}$\,s$^{-1}$) for each $n$=2 level of \ion{C}{v}.}
\label{cascadeC6}
\begin{center}
{\scriptsize
\begin{tabular}{ccccccc}
\hline
\hline
T$_{\mathrm{e}}$&$^{3}\mathrm{S}_{1}$&$^{3}\mathrm{P}_{0}$&$^{3}\mathrm{P}_{1}$&$^{3}\mathrm{P}_{2}$&$^{1}\mathrm{S}_{0}$& $^{1}\mathrm{P}_{1}$\\ 
%%             &                    &                    &                    &                    &                    &                      \\
\hline
\hline
   5.0\,(+04)  &  2.43\,(-13)$^{\mathrm{a}}$& 7.13\,(-14)& 2.14\,(-13)        &  3.57\,(-13)       & 8.09\,(-14)        &2.14\,(-13)          \\
             &  8.97\,(-13)$^{\mathrm{b}}$& 2.51\,(-13)& 7.51\,(-13)        &  1.25\,(-12)       & 1.69\,(-14)        &6.87\,(-13)           \\
             &    0$^{\mathrm{c}}$        &      0             &          0         &      0             &      0             &          0    \\
%             &                    &                    &                    &                    &                    &                      \\
\hline
%%             &                    &                    &                    &                    &                    &                      \\

   1.0\,(+05)  & 1.71\,(-13)        &4.85\,(-14)         & 1.46\,(-13)        & 2.43\,(-13)        & 5.69\,(-14)        &1.46\,(-13)           \\
             & 5.78\,(-13)        & 1.43\,(-13)        & 4.30\,(-13)        & 7.16\,(-13)        & 1.09\,(-14)        &3.89\,(-13)           \\
             &    0               &      0             &          0         &      0             &      0             &          0    \\
%             &                    &                    &                    &                    &                    &                      \\
\hline
%%             &                    &                    &                    &                    &                    &                      \\

   2.0\,(+05)  & 1.20\,(-13)        & 3.20\,(-14)        & 9.60\,(-14)        & 1.60\,(-13)        & 3.99\,(-14)        & 9.60\,(-14)          \\
             & 3.58\,(-13)        & 7.80\,(-14)        & 2.34\,(-13)        & 3.89\,(-13)        & 6.70\,(-15)        & 2.09\,(-13)          \\
             & 4.32\,(-19)        &1.34\,(-20)       & 3.81\,(-20)           &  5.13\,(-20)          & 1.33\,(-19)     &  5.61(-19) \\
%             &                    &                    &                    &                    &                    &                      \\
\hline
%%             &                    &                    &                    &                    &                    &                      \\

   5.0\,(+05)  & 7.35\,(-14)        & 1.71\,(-14)        & 5.12\,(-14)        & 8.54\,(-14)        & 2.45\,(-14)        & 5.12\,(-14)          \\
             & 1.75\,(-13)        & 3.19\,(-14)        & 9.58\,(-14)        & 1.60\,(-13)        & 3.30\,(-15)        & 8.38\,(-14)          \\
             & 2.75\,(-15)       & 4.72\,(-16)         & 1.28\,(-15)        & 1.55\,(-15)         & 6.77\,(-16)       &  4.56\,(-15)                    \\
%             &                    &                    &                    &                    &                    &                      \\
\hline
%%             &                    &                    &                    &                    &                    &                      \\

   1.0\,(+06)  & 4.96\,(-14)        & 9.77\,(-15)        & 2.93\,(-14)        & 4.89\,(-14)        & 1.65\,(-14)        &  2.93\,(-14)         \\
             & 9.64\,(-14)        & 1.53\,(-14)        & 4.60\,(-14)        & 7.61\,(-14)        & 1.80\,(-15)        &  3.95\,(-14)         \\
             & 3.72\,(-14)        & 8.94\,(-15)         & 2.42\,(-14)        & 2.85\,(-14)           & 8.17\,(-15)    &   6.92\,(-14)                   \\
%             &                    &                    &                    &                    &                    &                      \\
\hline
%%             &                    &                    &                    &                    &                    &                      \\

   2.0\,(+06)  & 3.24\,(-14)        & 5.10\,(-15)        & 1.53\,(-14)        & 2.55\,(-14)        & 1.08\,(-14)        &  1.53\,(-14)         \\
             & 4.98\,(-14)        & 7.10\,(-15)        & 2.12\,(-14)        & 3.53\,(-14)        & 9.00\,(-16)        &  1.79\,(-14)         \\
             & 8.57\,(-14)       &  2.32\,(-14)          &  6.26\,(-14)     &  7.31\,(-14)      & 1.80\,(-14)        &  1.68\,(-13)                    \\
\hline
\hline
\end{tabular}
}
\end{center}
\begin{list}{}{}
\item[$^{\mathrm{a}}$] RR direct contribution. 
\item[$^{\mathrm{b}}$] RR upper level radiative cascade contribution from the n$>$2 levels. 
\item[$^{\mathrm{c}}$] DR direct plus upper level radiative cascade from the n$>$2 levels contributions (when the value is equal to zero this means that the DR rate is negligible compared to the RR rates).
\end{list}
Note: a+b+c represent the total recombination rates.
\end{table*}
%\clearpage

%TABLE 4
\begin{table*}[h]
\caption{Same as Table~\ref{cascadeC6} for \ion{N}{vi}.}
\label{cascadeC6}
\begin{center}
{\scriptsize
\begin{tabular}{ccccccc}
\hline
\hline
T$_{\mathrm{e}}$&$^{3}\mathrm{S}_{1}$&$^{3}\mathrm{P}_{0}$&$^{3}\mathrm{P}_{1}$&$^{3}\mathrm{P}_{2}$&$^{1}\mathrm{S}_{0}$& $^{1}\mathrm{P}_{1}$\\ 
%%             &                    &                    &                    &                    &                    &                      \\
\hline
\hline
   7.0\,(+04)&  2.86\,(-13)         & 8.42\,(-14)      & 2.53\,(-13)        &  4.21\,(-13)       & 9.55\,(-14)        &2.53\,(-13)          \\
             &  1.07\,(-12)         & 2.94\,(-13)      & 8.77\,(-13)        &  1.47\,(-12)       & 2.35\,(-14)        &8.07\,(-13)           \\
             &    0                 &      0             &          0         &      0             &      0             &          0    \\
%             &                    &                    &                    &                    &                    &                      \\
\hline
%%             &                    &                    &                    &                    &                    &                      \\

   1.4\,(+05)& 2.02\,(-13)        &5.73\,(-14)         & 1.72\,(-13)        & 2.86\,(-13)        & 6.72\,(-14)        &1.72\,(-13)           \\
             & 6.91\,(-13)        & 1.68\,(-13)        & 5.04\,(-13)        & 8.44\,(-13)        & 1.52\,(-14)        &4.58\,(-13)           \\
             &    0                 &      0             &          0         &      0             &      0             &          0    \\
%             &                    &                    &                    &                    &                    &                      \\
\hline
%%             &                    &                    &                    &                    &                    &                      \\

   2.8\,(+05)& 1.41\,(-13)        & 3.78\,(-14)        & 1.13\,(-13)        & 1.89\,(-13)        & 4.71\,(-14)        & 1.13\,(-13)          \\
             & 4.28\,(-13)        & 9.12\,(-14)        & 2.74\,(-13)        & 4.56\,(-13)        & 9.30\,(-15)        & 2.46\,(-13)          \\
             & 8.56\,(-18)        &3.67\,(-20)         & 1.34\,(-19)        & 2.38\,(-19)       & 2.90\,(-18)        &  1.17\,(-17)      \\
%             &                    &                    &                    &                    &                    &                      \\
\hline
%%             &                    &                    &                    &                    &                    &                      \\

   7.0\,(+05)  & 8.68\,(-14)        & 2.02\,(-14)        & 6.05\,(-14)        & 1.01\,(-13)        & 2.89\,(-14)        & 6.05\,(-14)          \\
               & 2.10\,(-13)        & 3.72\,(-14)        & 1.12\,(-13)        & 1.86\,(-13)        & 4.60\,(-15)        & 9.85\,(-14)          \\
               & 6.58\,(-15)        & 5.98\,(-16)        & 1.50\,(-15)        & 1.75\,(-15)        & 2.00\,(-15)        & 9.76\,(-15)    \\
%             &                    &                    &                    &                    &                    &                      \\
\hline
%%             &                    &                    &                    &                    &                    &                      \\

   1.4\,(+06)  & 5.86\,(-14)        & 1.15\,(-14)        & 3.46\,(-14)        & 5.76\,(-14)        & 1.95\,(-14)        &  3.46\,(-14)         \\
               & 1.15\,(-13)        & 1.79\,(-14)        & 5.36\,(-14)        & 8.94\,(-14)        & 2.50\,(-15)        &  4.64\,(-14)         \\
               & 4.88\,(-14)        & 1.03\,(-14)        & 2.54\,(-14)        & 2.82\,(-14)        & 1.26\,(-14)        &  8.57\,(-14)     \\
%             &                    &                    &                    &                    &                    &                      \\
\hline
%%             &                    &                    &                    &                    &                    &                      \\

   2.8\,(+06)  & 3.82\,(-14)        & 6.02\,(-15)        & 1.81\,(-14)        & 3.01\,(-14)        & 1.27\,(-14)        &  1.81\,(-14)         \\
               & 5.97\,(-14)        & 8.18\,(-15)        & 2.46\,(-14)        & 4.11\,(-14)        & 1.30\,(-15)        &  2.09\,(-14)         \\
               & 9.17\,(-14)        & 2.56\,(-14)        & 6.27\,(-14)        & 6.85\,(-14)        & 2.16\,(-14)        &  1.76\,(-13)   \\
%%             &                    &                    &                    &                    &                    &                      \\
%             &                    &                    &                    &                    &                    &                      \\
\hline
\hline
\end{tabular}
}
\end{center}
\end{table*}
%\clearpage

%TABLE 5
\begin{table*}[h]
\caption{Same as Table~\ref{cascadeC6} for \ion{O}{vii}.}
\label{cascadeO8}
\begin{center}
{\scriptsize
\begin{tabular}{ccccccc}
\hline
\hline
T$_{\mathrm{e}}$&$^{3}\mathrm{S}_{1}$&$^{3}\mathrm{P}_{0}$&$^{3}\mathrm{P}_{1}$&$^{3}\mathrm{P}_{2}$&$^{1}\mathrm{S}_{0}$& $^{1}\mathrm{P}_{1}$\\ 
%%             &                    &                    &                    &                    &                    &                      \\
\hline
\hline
  9.0\,(+04)   &  3.36\,(-13)       & 9.90\,(-14)        & 2.97\,(-13)        & 4.95\,(-13)        & 1.12\,(-13)        & 2.97\,(-13)          \\
             &  1.24\,(-12)       & 3.51\,(-13)        & 1.05\,(-12)        & 1.75\,(-12)        & 2.50\,(-14)        & 9.63\,(-13)          \\
             &    0                 &      0             &          0         &      0             &      0             &          0    \\
\hline
%%             &                    &                    &                    &                    &                    &                      \\
 
 1.8\,(+05)    & 2.37\,(-13)        & 6.74\,(-14)        & 2.02\,(-13)        & 3.37\,(-13)        & 7.89\,(-14)        &2.02\,(-13)           \\
             & 8.03\,(-13)        & 2.02\,(-13)        & 6.05\,(-13)        & 1.01\,(-12)        &  1.63\,(-14)       &5.50\,(-13)           \\
             &    0                 &      0             &          0         &      0             &      0             &          0    \\
\hline
%%             &                    &                    &                    &                    &                    &                      \\
 
3.6\,(+05)     & 1.66\,(-13)        & 4.45\,(-14)        & 1.34\,(-13)        & 2.23\,(-13)        & 5.53\,(-14)        & 1.34\,(-13)          \\
             & 4.95\,(-13)        & 1.10\,(-13)        & 3.29\,(-13)        & 5.50\,(-13)        & 1.01\,(-14)        & 2.96\,(-13)          \\
             & 9.67\,(-19)           & 1.86\,(-20)           & 5.19\,(-20)           & 6.73\,(-20)           & 3.42\,(-19)           & 1.36\,(-18)                     \\
\hline
%%             &                    &                    &                    &                    &                    &                      \\
 
 9.0\,(+05)    & 1.02\,(-13)        & 2.39\,(-14)        & 7.16\,(-14)        & 1.19\,(-13)        & 3.40\,(-14)        & 7.16\,(-14)          \\
             & 2.44\,(-13)        & 4.52\,(-14)        & 1.35\,(-13)        & 2.27\,(-13)        & 4.90\,(-15)        & 1.19\,(-13)          \\
             & 3.55\,(-15)           & 4.90(-16)           & 1.32\,(-15)           & 1.54\,(-15)           & 1.05\,(-15)           & 6.02\,(-15)                     \\
\hline
%%             &                    &                    &                    &                    &                    &                      \\
 
 1.8\,(+06)    & 6.90\,(-14)        &1.37\,(-14)         & 4.11\,(-14)        & 6.85\,(-14)        & 2.30\,(-14)        & 4.11\,(-14)          \\
             &  1.34\,(-13)       & 2.18\,(-14)        & 6.49\,(-14)        &  1.09\,(-13)       & 2.70\,(-15)        & 5.65\,(-14)          \\
             & 3.90\,(-14)           & 8.49\,(-15)           & 2.26\,(-14)           &2.57\,(-14)            & 1.01\,(-14)           & 7.43\,(-14)                     \\
\hline
%%             &                    &                    &                    &                    &                    &                      \\
 
  3.6\,(+06)   &4.51\,(-14)         & 7.19\,(-15)        & 2.16\,(-14)        & 3.60\,(-14)        & 1.50\,(-14)        & 2.16\,(-14)          \\
             &6.99\,(-14)         & 1.00\,(-14)        & 3.01\,(-14)        & 5.02\,(-14)        & 1.40\,(-15)        & 2.56\,(-14)          \\
             &8.19\,(-14)            & 2.11\,(-14)           & 5.60\,(-14)           & 6.28\,(-14)           &  1.98\,(-14)          & 1.65\,(-13)                     \\
%%             &                    &                    &                    &                    &                    &                      \\
\hline
\hline
\end{tabular}
}
\end{center}
\end{table*}

%TABLE 6
\begin{table*}[h]
\caption{Same as Table~\ref{cascadeC6} for \ion{Ne}{ix}.}
\label{cascadeNe9}
\begin{center}
{\scriptsize
\begin{tabular}{ccccccc}
\hline
\hline
             &                    &                    &                    &                    &                    &                      \\
T$_{\mathrm{e}}$&$^{3}\mathrm{S}_{1}$&$^{3}\mathrm{P}_{0}$&$^{3}\mathrm{P}_{1}$&$^{3}\mathrm{P}_{2}$&$^{1}\mathrm{S}_{0}$& $^{1}\mathrm{P}_{1}$\\ 
%%             &                    &                    &                    &                    &                    &                      \\
\hline
\hline
  1.4\,(+05)   &4.33\,(-13)         & 1.27\,(-13)        &  3.82\,(-13)       & 6.37\,(-13)        & 1.44\,(-13)        &  3.82\,(-13)         \\
             & 1.59\,(-12)        & 4.57\,(-13)        &  1.37\,(-12)       & 2.28\,(-12)        & 3.40\,(-14)        &  1.26\,(-12)         \\
             &    0                 &      0             &          0         &      0             &      0             &          0    \\
\hline
 
  2.8\,(+05)   &3.05\,(-13)         & 8.69\,(-14)        & 2.61\,(-13)        & 4.35\,(-13)        & 1.02\,(-13)        &  2.61\,(-13)         \\
             & 1.02\,(-12)        & 2.62\,(-13)        & 7.89\,(-13)        & 1.31\,(-12)        & 2.20\,(-14)        &  7.19\,(-13)         \\
             &    0                 &      0             &          0         &      0             &      0             &          0    \\
\hline
%%             &                    &                    &                    &                    &                    &                      \\
 
  5.6\,(+05)   & 2.14\,(-13)        & 5.75\,(-14)        & 1.73\,(-13)        &2.88\,(-13)         & 7.12\,(-14)        & 1.73\,(-13)          \\
             & 6.35\,(-13)        & 1.43\,(-13)        & 4.31\,(-13)        & 7.22\,(-13)        & 1.36\,(-14)        & 3.88\,(-13)          \\
             & 1.22\,(-18)           & 1.19\,(-20)           & 3.74\,(-20)           & 5.73\,(-20)           & 5.29\,(-19)           &  2.01\,(-18)                    \\
\hline
%%             &                    &                    &                    &                    &                    &                      \\
 
  1.4\,(+06)   & 1.31\,(-13)        & 3.09\,(-14)        & 9.28\,(-14)        & 1.55\,(-13)        &  4.38\,(-14)       & 9.28\,(-14)          \\
             & 3.15\,(-13)        & 5.93\,(-14)        & 1.78\,(-13)        & 2.97\,(-13)        &  6.70\,(-15)       & 1.57\,(-13)          \\
             & 3.65\,(-15)        &  2.68\,(-16)       &  8.31\,(-16)       &  1.19\,(-15)       &  1.38\,(-15)       &  6.78\,(-15)                    \\
\hline
%%             &                    &                    &                    &                    &                    &                      \\
 
  2.8\,(+06)   & 8.90\,(-14)        & 1.78\,(-14)        &  5.35\,(-14)       & 8.91\,(-14)        &  2.97\,(-14)       & 5.35\,(-14)          \\
             & 1.73\,(-13)        & 2.86\,(-14)        &  8.55\,(-14)       & 1.43\,(-13)        &  3.60\,(-15)       & 7.45\,(-14)          \\
             & 3.66\,(-14)        & 4.47\,(-15)        & 1.38\,(-14)          & 1.96\,(-14)         &  1.23\,(-14)        & 7.45\,(-14)                     \\
\hline
%%              &                    &                    &                    &                    &                   &                      \\
 
  5.6\,(+06)   & 5.83\,(-14)        & 9.39\,(-15)        &  2.82\,(-14)       &  4.70\,(-14)       &  1.94\,(-14)       & 2.82\,(-14)          \\
             & 8.97\,(-14)        &  1.32\,(-14)       &  3.96\,(-14)       &  6.60\,(-14)       &  1.90\,(-15)       & 3.40\,(-14)          \\
             & 7.37\,(-14)        & 1.09\,(-14)          &  3.37\,(-14)       &  4.75\,(-14)       &  2.32\,(-14)          & 1.57\,(-13)                     \\
%%             &                    &                    &                    &                    &                    &                      \\
\hline
\hline
\end{tabular}
}
\end{center}
\end{table*}

%TABLE 7
\begin{table*}[h]
\caption{Same as Table~\ref{cascadeC6} for \ion{Mg}{xi}.}
\label{cascadeMg11}
\begin{center}
{\scriptsize
\begin{tabular}{ccccccc}
\hline
\hline
T$_{\mathrm{e}}$&$^{3}\mathrm{S}_{1}$&$^{3}\mathrm{P}_{0}$&$^{3}\mathrm{P}_{1}$&$^{3}\mathrm{P}_{2}$&$^{1}\mathrm{S}_{0}$& $^{1}\mathrm{P}_{1}$\\ 
%%             &                    &                    &                    &                    &                    &                      \\
\hline
\hline
  2.0\,(+05)   &5.30\,(-13)         & 1.56\,(-13)        & 4.69\,(-13)        & 7.82\,(-13)        & 1.77\,(-13)        & 4.69\,(-13)          \\
             &1.94\,(-12)         &  5.65\,(-13)       & 1.70\,(-12)        & 2.83\,(-12)        & 4.30\,(-14)        & 1.56\,(-12)          \\
             &    0                 &      0             &          0         &      0             &      0             &          0    \\
\hline
%%             &                    &                    &                    &                    &                    &                      \\
 
    4.0\,(+05) & 3.74\,(-13)        & 1.07\,(-13)        & 3.20\,(-13)        & 5.34\,(-13)        &1.25\,(-13)         &3.20\,(-13)           \\
             & 1.25\,(-12)        & 3.25\,(-13)        & 9.80\,(-13)        & 1.63\,(-12)        & 2.80\,(-14)        &8.90\,(-13)           \\
             &    0                 &      0             &          0         &      0             &      0             &          0    \\
\hline
%%             &                    &                    &                    &                    &                    &                      \\
 
   8.0\,(+05)  &  2.62\,(-13)       &   7.07\,(-14)      & 2.12\,(-13)        & 3.54\,(-13)        & 8.74\,(-14)        & 2.12\,(-13)          \\
             &  7.78\,(-13)       &   1.78\,(-13)      & 5.36\,(-13)        & 8.96\,(-13)        & 1.76\,(-14)        & 4.85\,(-13)          \\
             & 1.19\,(-18)          & 7.95\,(-21)        & 2.92\,(-20)         & 4.91\,(-20)        & 6.68\,(-19)          &   2.45\,(-18)                   \\
\hline
%%             &                    &                    &                    &                    &                    &                      \\
 
   2.0\,(+06)  &  1.61\,(-13)       & 3.81\,(-14)        & 1.14\,(-13)        &  1.91\,(-13)       &  5.38\,(-14)       & 1.14\,(-13)          \\
             &  3.85\,(-13)       &7.39\,(-14)         & 2.22\,(-13)        &  3.70\,(-13)       &  8.60\,(-15)       & 1.98\,(-13)          \\
             &  3.36\,(-15)          &  1.71\,(-16)          & 6.40\,(-16)          & 9.60\,(-16)           & 1.61\,(-15)           & 7.10\,(-15)                     \\
\hline
%%             &                    &                    &                    &                    &                    &                      \\
 
   4.0\,(+06)  & 1.09\,(-13)        & 2.21\,(-14)        &  6.62\,(-14)       &  1.10\,(-13)       &  3.64\,(-14)       & 6.62\,(-14)          \\
             & 2.13\,(-13)        & 3.56\,(-14)        &  1.07\,(-13)       &  1.79\,(-13)       &  4.80\,(-15)       & 9.38\,(-14)          \\
             & 3.31\,(-14)           & 2.83\,(-15)           &  1.07\,(-14)          & 1.59\,(-14)           & 1.36\,(-14)           & 7.11\,(-14)                     \\
\hline
%%             &                    &                    &                    &                    &                    &                      \\
 
   8.0\,(+06)  & 7.17\,(-14)        &  1.17\,(-14)       & 3.50\,(-14)        &  5.83\,(-14)       & 2.39\,(-14)        & 3.50\,(-14)          \\
             & 1.11\,(-13)        & 1.64\,(-14)        & 4.95\,(-14)        &   8.27\,(-14)      & 2.50\,(-15)        & 4.27\,(-14)          \\
             &  6.62\,(-14)          & 6.90(\,-15)           &  2.62\,(-14)          & 3.88\,(-14)           & 2.47\,(-14)       &    1.44\,(-13)                   \\
%\hline
%%             &                    &                    &                    &                    &                    &                      \\
\hline
\hline
\end{tabular}
}
\end{center}
\end{table*}

%TABLE 8
\begin{table*}[h]
\caption{Same as Table~\ref{cascadeC6} for \ion{Si}{xiii}.}
\label{cascadeSi13}
\begin{center}
{\scriptsize
\begin{tabular}{ccccccc}
\hline
\hline
T$_{\mathrm{e}}$&$^{3}\mathrm{S}_{1}$&$^{3}\mathrm{P}_{0}$&$^{3}\mathrm{P}_{1}$&$^{3}\mathrm{P}_{2}$&$^{1}\mathrm{S}_{0}$& $^{1}\mathrm{P}_{1}$\\ 
%%             &                    &                    &                    &                    &                    &                      \\
\hline
\hline
 2.8\,(+05)    & 6.23\,(-13)        & 1.84\,(-13)        &  5.52\,(-13)       & 9.19\,(-13)        &  2.08\,(-13)       &  5.52\,(-13)         \\
             & 2.25\,(-12)        &  6.65\,(-13)       &  2.00\,(-12)       & 3.33\,(-12)        &  5.30\,(-14)       &  1.85\,(-12)         \\
             &    0                 &      0             &          0         &      0             &      0             &          0    \\
\hline
%%             &                    &                    &                    &                    &                    &                      \\
 
  5.5\,(+05)   & 4.39\,(-13)        & 1.25\,(-13)        & 3.76\,(-13)        & 6.27\,(-13)        & 1.46\,(-13)        & 3.76\,(-13)          \\
             & 1.44\,(-12)        &  3.84\,(-13)       & 1.15\,(-12)        &  1.92\,(-12)       & 3.50\,(-14)        & 1.05\,(-12)          \\
             &    0                 &      0             &          0         &      0             &      0             &          0    \\
\hline
%%             &                    &                    &                    &                    &                    &                      \\
 
 1.1\,(+06)    & 3.08\,(-13)        & 8.32\,(-14)        & 2.49\,(-13)        &  4.16\,(-13)       & 1.03\,(-13)        &  2.49\,(-13)         \\
             & 8.92\,(-13)        & 2.10\,(-13)        & 6.32\,(-13)        &  1.05\,(-12)       & 2.10\,(-14)        &  5.74\,(-13)         \\
             & 1.29\,(-18)        & 8.47\,(-21)        &3.32\,(-20)          & 6.78\,(-20)       &9.70\,(-19)         &  3.46\,(-18)                    \\
\hline
%%             &                    &                    &                    &                    &                    &                      \\
 
  2.8\,(+06)   & 1.90\,(-13)        & 4.49\,(-14)        &  1.35\,(-13)       & 2.24\,(-13)        & 6.32\,(-14)        & 1.35\,(-13)          \\
             & 4.45\,(-13)        & 8.71\,(-14)        &  2.61\,(-13)       & 4.37\,(-13)        & 1.06\,(-14)        &2.34\,(-13)           \\
             & 3.43\,(-15)       &  1.77\,(-16)       &6.55\,(-16)          &9.75\,(-16)         & 2.09\,(-15 )       & 8.56\,(-15)                     \\
\hline
%%             &                    &                    &                    &                    &                    &                      \\
 
  5.5\,(+06)   & 1.29\,(-13)        & 2.59\,(-14)        & 7.78\,(-14)        & 1.30\,(-13)        &  4.28\,(-14)       &  7.78\,(-14)         \\
             & 2.46\,(-13)        & 4.20\,(-14)        & 1.26\,(-13)        & 2.11\,(-13)        & 5.90\,(-15)        &  1.11\,(-13)         \\
             & 2.97\,(-14)        & 2.50\,(-15)        & 9.28\,(-15)        &  1.35\,(-14)       &   1.48\,(-14)       & 6.93\,(-14) \\                  
\hline
%%             &                    &                    &                    &                    &                    &                      \\
 
  1.1(+07)   &  8.43\,(-14)       & 1.37\,(-14)        & 4.12\,(-14)        &6.86\,(-14)         &  2.81\,(-14)       &4.12\,(-14)           \\
             &  1.29\,(-13)       & 1.94\,(-14)        & 5.83\,(-14)        & 9.74\,(-14)        &  3.10\,(-15)       &5.07\,(-14)           \\
             &  5.84\,(-14)       &   5.93\,(-15)      & 2.21\,(-14)        & 3.20\,(-14)        & 2.55\,(-14)        &  1.30\,(-13)                    \\
%\hline
%%             &                    &                    &                    &                    &                    &                      \\
\hline
\hline
\end{tabular}
}
\end{center}
\end{table*}

%TABLE 9
\begin{table*}
\caption{Effective collisions strengths ($\Upsilon$) for each  1s$^{2}$--1s2l transition of \ion{C}{v}.}
\begin{center}
{\scriptsize
\begin{tabular}{ccccccc}
\hline
\hline
T$_{\mathrm{e}}$/$Z^3$&$^{3}\mathrm{S}_{1}$&$^{3}\mathrm{P}_{0}$&$^{3}\mathrm{P}_{1}$&$^{3}\mathrm{P}_{2}$&$^{1}\mathrm{S}_{0}$& $^{1}\mathrm{P}_{1}$\\ 
\hline
\hline
  400     &8.48\,(-03)$^{\mathrm{a}}$& 4.95\,(-03)        & 1.48\,(-02)        &2.47\,(-02)         & 1.42\,(-02)        &4.05\,(-02)           \\
          &7.53\,(-06)$^{\mathrm{b}}$& 7.67\,(-07)        & 2.31\,(-06)        &3.88\,(-06)         & 4.32\,(-07)        &9.86\,(-06)           \\
\hline
%%             &                    &                    &                    &                    &                    &                      \\

 600         &9.09\,(-03)         & 5.04\,(-03)        & 1.51\,(-02)        &2.51\,(-02)         & 1.46\,(-02)        &4.26\,(-02)           \\
             &8.96\,(-05)         & 6.94\,(-06)        & 2.09\,(-05)        &3.50\,(-05)         & 4.40\,(-06)        &7.09\,(-05)           \\
\hline
%%             &                    &                    &                    &                    &                    &                      \\

 900         &9.46\,(-03)         & 5.02\,(-03)        & 1.50\,(-02)        &2.51\,(-02)         & 1.50\,(-02)        &4.48\,(-02)            \\
             &4.76\,(-04)         & 3.12\,(-05)        & 9.37\,(-05)        &1.57\,(-04)         & 2.28\,(-05)        &2.73\,(-04)           \\
\hline
%%             &                    &                    &                    &                    &                    &                      \\

 1\,350        &9.39\,(-03)         & 4.85\,(-03)        & 1.46\,(-02)        &2.42\,(-02)         & 1.52\,(-02)        &4.76\,(-02)           \\
             &1.45\,(-03)         & 8.59\,(-05)        & 2.58\,(-04)        &4.32\,(-04)         & 7.53\,(-05)        &6.95\,(-04)           \\
\hline
%%             &                    &                    &                    &                    &                    &                      \\

 2\,000        &8.91\,(-03)         & 4.56\,(-03)        & 1.37\,(-02)        & 2.27\,(-02)        & 1.55\,(-02)        &5.13\,(-02)           \\
             &2.97\,(-03)         & 1.64\,(-04)        & 4.93\,(-04)        & 8.26\,(-04)        & 1.80\,(-04)        &1.31\,(-03)           \\
\hline
%%             &                    &                    &                    &                    &                    &                      \\

 3\,000        &8.14\,(-03)         & 4.15\,(-03)        & 1.25\,(-02)        & 2.08\,(-02)        & 1.58\,(-02)        & 5.65\,(-02)          \\
             &4.75\,(-03)         & 2.51\,(-04)        & 7.54\,(-04)        & 1.26\,(-03)        & 3.62\,(-04)        & 2.10\,(-03)          \\
\hline
%%             &                    &                    &                    &                    &                    &                      \\

 4\,500        &7.23\,(-03)         & 3.68\,(-03)        & 1.11\,(-02)        & 1.84\,(-02)        & 1.62\,(-02)        & 6.40\,(-02)          \\
             &6.23\,(-03)         & 3.20\,(-04)        & 9.60\,(-04)        & 1.61\,(-03)        & 6.36\,(-04)        & 2.97\,(-03)          \\
\hline
%%             &                    &                    &                    &                    &                    &                      \\

 6\,700        &6.30\,(-03)         & 3.20\,(-03)        & 9.60\,(-03)        & 1.60\,(-02)        & 1.68\,(-02)        & 7.38\,(-02)          \\
             &7.08\,(-03)         & 3.56\,(-04)        & 1.07\,(-03)        & 1.79\,(-03)        & 1.01\,(-03)        & 3.88\,(-03)          \\
\hline
%%             &                    &                    &                    &                    &                    &                      \\

 10\,000     &5.37\,(-03)         & 2.70\,(-03)        & 8.09\,(-03)        & 1.34\,(-02)        & 1.76\,(-02)        & 8.70\,(-02)          \\
             &7.23\,(-03)         & 3.59\,(-04)        & 1.08\,(-03)        &  1.80\,(-03)       & 1.50\,(-03)        & 4.82\,(-03)          \\
%             &                    &                    &                    &                    &                    &                      \\
\hline
\hline
\end{tabular}
}
\end{center}
\begin{list}{}{}
\item[$^{\mathrm{a}}$] direct + resonance contribution inferred from the data for \ion{O}{vii} (from Zhang \& Sampson 1987, see Table~\ref{collcascadeO8}) with the scaling reported in Figure~\ref{figure1s21s2l} . 
\item[$^{\mathrm{b}}$] cascade contribution calculated in this paper.
\item Note: here a+b corresponds to the total collision strength which populates the level considered.  
\end{list}
\label{collcascadeC5}
\end{table*}

%TABLE 10
\begin{table*}
\caption{Effective collisions strengths ($\Upsilon$) for each  1s$^{2}$--1s2l transition of \ion{O}{vii}.}
\begin{center}
{\scriptsize
\begin{tabular}{ccccccc}
\hline
\hline
T$_{\mathrm{e}}$/$Z^3$&$^{3}\mathrm{S}_{1}$&$^{3}\mathrm{P}_{0}$&$^{3}\mathrm{P}_{1}$&$^{3}\mathrm{P}_{2}$&$^{1}\mathrm{S}_{0}$& $^{1}\mathrm{P}_{1}$\\ 
\hline
\hline
          &4.06\,(-03)$^{\mathrm{a}}$& 2.51\,(-03)        & 7.50\,(-03)        &1.25\,(-02)         & 7.35\,(-03)        &2.10\,(-02)           \\
 400      &5.01\,(-04)$^{\mathrm{b}}$& 1.53\,(-04)        & 4.60\,(-04)        &7.71\,(-04)         & 2.92\,(-04)        &7.53\,(-04)           \\
          &1.82\,(-05)$^{\mathrm{c}}$& 1.32\,(-06)        & 3.95\,(-06)        &6.67\,(-06)         & 9.42\,(-07)        &1.26\,(-05)           \\
\hline
%%             &                    &                    &                    &                    &                    &                      \\

             &4.06\,(-03)         & 2.48\,(-03)        & 7.41\,(-03)        &1.23\,(-02)         & 7.45\,(-03)        &2.19\,(-02)           \\
 600         &8.27\,(-04)         & 2.31\,(-04)        & 6.97\,(-04)        &1.16\,(-03)         & 4.06\,(-04)        &9.84\,(-04)           \\
             &1.33\,(-04)         & 8.38\,(-06)        & 2.51\,(-05)        &4.23\,(-05)         & 6.68\,(-06)        &6.98\,(-05)           \\
\hline
%%             &                    &                    &                    &                    &                    &                      \\

             &4.04\,(-03)         & 2.42\,(-03)        & 7.26\,(-03)        &1.21\,(-02)         & 7.59\,(-03)        &2.30\,(-02)            \\
 900         &1.05\,(-03)         & 2.75\,(-04)        & 8.31\,(-04)        &1.38\,(-03)         & 4.56\,(-04)        &1.06\,(-03)           \\
             &5.05\,(-04)         & 2.94\,(-05)        & 8.82\,(-05)        &1.48\,(-04)         & 2.71\,(-05)        &2.26\,(-04)           \\
\hline
%%             &                    &                    &                    &                    &                    &                      \\

             &3.96\,(-03)         & 2.33\,(-03)        & 7.00\,(-03)        &1.16\,(-02)         & 7.75\,(-03)        &2.46\,(-02)           \\
 1350        &1.09\,(-03)         & 2.74\,(-04)        & 8.29\,(-04)        &1.37\,(-03)         & 4.38\,(-04)        &9.83\,(-04)           \\
             &1.22\,(-03)         & 6.80\,(-05)        & 2.04\,(-04)        &3.42\,(-04)         & 7.61\,(-05)        &5.12\,(-04)           \\
\hline
%%             &                    &                    &                    &                    &                    &                      \\

             &3.80\,(-03)         & 2.21\,(-03)        & 6.63\,(-03)        & 1.10\,(-02)        & 7.94\,(-03)        &2.68\,(-02)           \\
 2000        &9.86\,(-04)         & 2.42\,(-04)        & 7.31\,(-04)        & 1.20\,(-03)        & 3.78\,(-04)        & 8.31\,(-04)          \\
             &2.14\,(-03)         & 1.15\,(-04)        & 3.45\,(-04)        & 5.79\,(-04)        & 1.63\,(-04)        &8.96\,(-04)           \\
\hline
%%             &                    &                    &                    &                    &                    &                      \\

             &3.58\,(-03)         & 2.04\,(-03)        & 6.12\,(-03)        & 1.02\,(-02)        & 6.26\,(-03)        & 2.97\,(-02)          \\
 3000        &8.04\,(-04)         & 1.93\,(-04)        & 5.86\,(-04)        & 9.73\,(-04)        & 2.28\,(-04)        & 6.47\,(-04)          \\
             &3.04\,(-03)         & 1.60\,(-04)        & 4.81\,(-04)        & 8.06\,(-04)        & 3.04\,(-04)        & 1.37\,(-03)          \\
\hline
%%             &                    &                    &                    &                    &                    &                      \\

             &3.28\,(-03)         & 1.83\,(-03)        & 5.51\,(-03)        & 9.18\,(-03)        & 8.51\,(-03)        & 3.39\,(-02)          \\
 4500        &6.09\,(-04)         & 1.45\,(-04)        & 4.42\,(-04)        & 7.32\,(-04)        & 2.23\,(-04)        & 4.81\,(-04)          \\
             &3.67\,(-03)         & 1.90\,(-04)        & 5.72\,(-04)        & 9.57\,(-04)        & 5.05\,(-04)        & 1.88\,(-03)          \\
\hline
%%             &                    &                    &                    &                    &                    &                      \\

             &2.94\,(-03)         & 1.61\,(-03)        & 4.84\,(-03)        & 8.06\,(-03)        & 8.89\,(-03)        & 3.93\,(-02)          \\
 6\,700      &4.51\,(-04)         & 1.06\,(-04)        & 3.22\,(-04)        & 5.33\,(-04)        & 1.61\,(-04)        & 3.45\,(-04)          \\
             &3.91\,(-03)         & 2.01\,(-04)        & 6.04\,(-04)        & 1.01\,(-03)        & 7.70\,(-04)        & 2.41\,(-03)          \\
\hline
%%             &                    &                    &                    &                    &                    &                      \\

             &2.57\,(-03)         & 1.37\,(-03)        & 4.12\,(-03)        & 6.85\,(-03)        & 9.34\,(-03)        & 4.65\,(-02)          \\
 10\,000     &3.16\,(-04)         & 7.53\,(-05)        & 2.28\,(-04)        & 3.76\,(-04)        & 1.14\,(-04)        & 2.42\,(-04)          \\
             &3.81\,(-03)         & 1.94\,(-04)        & 5.85\,(-04)        &  9.77\,(-04)       & 1.11\,(-03)        & 2.96\,(-03)          \\
%             &                    &                    &                    &                    &                    &                      \\
\hline
\hline
\end{tabular}
}
\end{center}
\begin{list}{}{}
\item[$^{\mathrm{a}}$] direct contribution (from Zhang \& Sampson 1987). 
\item[$^{\mathrm{b}}$] resonance contribution (from Zhang \& Sampson 1987).
\item[$^{\mathrm{c}}$] cascade contribution calculated in this paper.
\item Note: here a+b+c corresponds to the total collision strength which populates the level considered.  
\end{list}
\label{collcascadeO8}
\end{table*}

%TABLE 11
\begin{table*}
\caption{Same as Table~\ref{collcascadeO8} but for the \ion{Ne}{ix}.}
\begin{center}
{\scriptsize
\begin{tabular}{ccccccc}
\hline
\hline
%%             &                    &                    &                    &                    &                    &                      \\
T$_{\mathrm{e}}$/$Z^3$&$^{3}\mathrm{S}_{1}$&$^{3}\mathrm{P}_{0}$&$^{3}\mathrm{P}_{1}$&$^{3}\mathrm{P}_{2}$&$^{1}\mathrm{S}_{0}$& $^{1}\mathrm{P}_{1}$\\ 
\hline
\hline

             &2.53\,(-03)         & 1.56\,(-03)        & 4.67\,(-03)        & 7.77\,(-03)        &4.79\,(-03)         & 1.45\,(-02)          \\
 400         &5.05\,(-04)         & 1.41\,(-04)        & 4.23\,(-04)        & 7.04\,(-04)        &2.44\,(-04)         & 5.92\,(-04)          \\
             &3.14\,(-05)         & 1.97\,(-06)        & 5.91\,(-06)        & 9.99\,(-06)        &1.63\,(-06)         & 1.64\,(-05)          \\
\hline
%%             &                    &                    &                    &                    &                    &                      \\

             &2.53\,(-03)         & 1.53\,(-03)        & 4.59\,(-03)        &  7.65\,(-03)       & 4.88\,(-03)        & 1.51\,(-02)          \\
 600         &6.75\,(-04)         &1.77\,(-04)         & 5.35\,(-04)        &  8.89\,(-04)       & 2.93\,(-04)        & 6.76\,(-04)          \\
             &1.64\,(-04)         &9.55\,(-06)         & 2.87\,(-05)        &  4.84\,(-05)       & 8.87\,(-06)        & 7.27\,(-05)          \\
\hline
%%             &                    &                    &                    &                    &                    &                      \\

             & 2.50\,(-03)        &1.49\,(-03)         & 4.47\,(-03)        &  7.43\,(-03)       & 4.96\,(-03)        &1.59\,(-02)           \\
 900         & 7.34\,(-04)        &1.85\,(-04)         & 5.62\,(-04)        &  9.29\,(-04)       & 2.95\,(-04)        &6.63\,(-04)           \\
             & 4.96\,(-04)        &2.77\,(-05)         & 8.32\,(-05)        &  1.40\,(-04)       & 3.02\,(-05)        &2.03\,(-04)           \\
\hline
%%             &                    &                    &                    &                    &                    &                      \\

             &2.43\,(-03)         &1.42\,(-03)         & 4.27\,(-03)        &  7.10\,(-03)       & 5.08\,(-03)        &1.71\,(-02)           \\
 1350        &6.89\,(-04)         &1.69\,(-04)         & 5.11\,(-04)        &  8.44\,(-04)       & 2.62\,(-04)        &5.78\,(-04)           \\
             &1.03\,(-03)         &5.59\,(-05)         & 1.68\,(-04)        &  2.83\,(-04)       & 7.49\,(-05)        &4.15\,(-04)           \\
\hline
%%             &                    &                    &                    &                    &                    &                      \\

             &2.31\,(-03)         &1.33\,(-03)         & 4.00\,(-03)        &  6.64\,(-03)       & 5.21\,(-03)        & 1.87\,(-02)          \\
2000         &5.78\,(-04)         &1.40\,(-04)         & 4.25\,(-04)        &  6.96\,(-04)       & 2.14\,(-04)        & 4.67\,(-04)          \\
             &1.60\,(-03)         &8.61\,(-05)         & 2.59\,(-04)        &  4.35\,(-04)       & 1.48\,(-04)        & 6.83\,(-04)         \\
\hline
%%             &                    &                    &                    &                    &                    &                      \\

             &2.15\,(-03)         &1.21\,(-03)         & 3.65\,(-03)        & 6.06\,(-03)        &5.39\,(-03)         & 2.09\,(-02)          \\
3000         &4.49\,(-04)         &1.08\,(-04)         & 3.26\,(-04)        & 5.37\,(-04)        & 1.63\,(-04)        & 3.52\,(-04)          \\
             &2.10\,(-03)         &1.11\,(-04)         & 3.35\,(-04)        & 5.63\,(-04)        & 2.61\,(-04)        & 9.97\,(-04)          \\
\hline
%%             &                    &                    &                    &                    &                    &                      \\

             &1.95\,(-03)         &1.08\,(-03)         & 3.24\,(-03)        & 5.39\,(-03)        & 5.61\,(-03)        & 2.40\,(-02)          \\
4500         &3.30\,(-04)         &7.91\,(-05)         & 2.39\,(-04)        & 3.94\,(-04)        & 1.19\,(-04)        & 2.56\,(-04)          \\
             &2.39\,(-03)         &1.25\,(-04)         & 3.78\,(-04)        & 6.33\,(-04)        & 4.17\,(-04)        & 1.34\,(-03)          \\
\hline
%%             &                    &                    &                    &                    &                    &                      \\

             &1.72\,(-03)         &9.32\,(-04)         & 2.81\,(-03)        & 4.66\,(-03)        & 5.86\,(-03)        &2.80\,(-02)           \\
6700         &2.36\,(-04)         &5.65\,(-05)         & 1.70\,(-04)        & 2.80\,(-04)        &  8.50\,(-05)       &1.81\,(-04)           \\
             &2.43\,(-03)         &1.27\,(-04)         & 3.85\,(-04)        & 6.42\,(-04)        & 6.18\,(-04)        &1.68\,(-03)           \\
\hline
%%             &                    &                    &                    &                    &                    &                      \\

             &1.49\,(-03)         &7.83\,(-04)         & 2.36\,(-03)        & 3.91\,(-03)        & 6.17\,(-03)        &  3.32\,(-02)         \\
10000        &1.64\,(-04)         &3.93\,(-05)         & 1.19\,(-04)        & 1.97\,(-04)        & 5.92\,(-05)        &  1.26\,(-04)         \\
             &2.29\,(-03)         &1.19\,(-04)         & 3.62\,(-04)        & 6.01\,(-04)        & 8.73\,(-04)        & 2.05\,(-03)          \\
%%\hline
%%             &                    &                    &                    &                    &                    &                      \\
\hline
\hline
\end{tabular}
}
\end{center}
\label{collcascadeNe9}
\end{table*}

%TABLE 12
\begin{table*}
\caption{Same as Table~\ref{collcascadeO8} but for the \ion{Mg}{xi}.}
\begin{center}
{\scriptsize
\begin{tabular}{ccccccc}
\hline
\hline
%%             &                    &                    &                    &                    &                    &                      \\
T$_{\mathrm{e}}$/$Z^3$&$^{3}\mathrm{S}_{1}$&$^{3}\mathrm{P}_{0}$&$^{3}\mathrm{P}_{1}$&$^{3}\mathrm{P}_{2}$&$^{1}\mathrm{S}_{0}$& $^{1}\mathrm{P}_{1}$\\ 
\hline
\hline

             & 1.74\,(-03)        & 1.05\,(-03)        & 3.18\,(-03)        & 5.28\,(-03)        & 3.39\,(-03)        &1.06\,(-02)           \\
 400         & 4.41\,(-04)        & 1.16\,-04)         & 3.51\,(-04)        & 5.86\,(-04)        &  1.93\,(-04)       &4.49\,(-04)           \\
             & 4.40\,(-05)        & 2.59\,(-06)        & 7.75\,(-06)        & 1.32\,(-05)        & 2.40\,(-06)        &2.00\,(-05)           \\
\hline
%%             &                    &                    &                    &                    &                    &                      \\

             & 1.72\,(-03)        & 1.04\,(-03)        &3.11\,(-03)         & 5.18\,(-03)        & 3.44\,(-03)        &1.10\,(-02)           \\
 600         & 5.17\,(-04)        & 1.32\,(-04)        &3.97\,(-04)         & 6.61\,(-04)        & 2.10\,(-04)        &4.74\,(-04)           \\
             & 1.81\,(-04)        & 1.02\,(-05)        & 3.06\,(-05)        & 5.21\,(-05)        & 1.07\,(-05)        &7.46\,(-05)           \\
\hline
%%             &                    &                    &                    &                    &                    &                      \\

             & 1.69\,(-03)        & 1.00\,(-03)        &  3.01\,(-03)       &  5.00\,(-03)       & 3.50\,(-03)        &  1.18\,(-02)         \\
 900         & 5.11\,(-04)        & 1.28\,(-04)        &  3.84\,(-04)       &  6.35\,(-04)       & 1.97\,(-04)        &  4.39\,(-04)         \\
             & 4.65\,(-04)        & 2.56\,(-05)        & 7.68\,(-05)        & 1.30\,(-04)        & 3.19\,(-05)        &1.85\,(-04)           \\
\hline
%%             &                    &                    &                    &                    &                    &                      \\

             & 1.63\,(-03)        & 9.50\,(-04)        &  2.85\,(-03)       &  4.73\,(-03)       & 3.58\,(-03)        & 1.26\,(-02)          \\
 1350        & 4.45\,(-04)        & 1.10\,(-04)        &  3.30\,(-04)       &  5.45\,(-04)       & 1.67\,(-04)        & 3.65\,(-04)          \\
             & 8.56\,(-04)        & 4.66\,(-05)        & 1.40\,(-04)        & 2.37\,(-04)        & 7.25\,(-05)        &3.50\,(-04)           \\
\hline
%%             &                    &                    &                    &                    &                    &                      \\

             & 1.53\,(-03)        &8.77\,(-04)         & 2.65\,(-03)        &  4.38\,(-03)       & 3.69\,(-03)        & 1.39\,(-02)          \\
2000         & 3.60\,(-04)        &8.75\,(-05)         & 2.63\,(-04)        &  4.35\,(-04)       & 1.31\,(-04)        & 2.87\,(-04)          \\
             & 1.24\,(-03)        &6.66\,(-05)         &2.01\,(-04)         & 3.38\,(-04)        & 1.35\,(-04)        & 5.46\,(-04)          \\
\hline
%%             &                    &                    &                    &                    &                    &                      \\

             & 1.41\,(-03)        &7.92\,(-04)         & 2.40\,(-03)        &   3.96\,(-03)      & 3.82\,(-02)        & 1.57\,(-02)          \\
3000         & 2.70\,(-04)        &6.57\,(-05)         & 1.85\,(-04)        &  3.25\,(-04)       & 9.77\,(-05)        & 2.12\,(-04)          \\
             & 1.52\,(-03)        &8.15\,(-05)         & 2.47\,(-04)        &  4.14\,(-04)       & 2.28\,(-04)        & 7.72\,(-04)          \\
\hline
%%             &                    &                    &                    &                    &                    &                      \\

             & 1.27\,(-03)        & 6.96\,(-04)        & 2.11\,(-03)        & 3.48\,(-03)        & 3.99\,(-03)        & 1.80\,(-02)          \\
4500         & 1.94\,(-04)        & 4.72\,(-05)        & 1.42\,(-04)        & 2.34\,(-4)         & 7.04\,(-05)        & 1.52\,(-04)          \\
             & 1.65\,(-03)        & 8.78\,(-05)        & 2.68\,(-04)        & 4.47\,(-04)        & 3.54\,(-04)        & 1.01\,(-03)          \\
\hline
%%             &                    &                    &                    &                    &                    &                      \\

             & 1.11\,(-03)        & 5.95\,(-04)        & 1.81\,(-03)        & 2.98\,(-03)        & 4.18\,(-03)        & 2.12\,(-02)          \\
6700         & 1.37\,(-04)        & 3.33\,(-05)        & 9.97\,(-05)        & 1.65\,(-04)        & 4.97\,(-05)        & 1.07\,(-04)          \\
             & 1.63\,(-03)        & 8.60\,(-05)        & 2.66\,(-04)        & 4.40\,(-04)        & 5.13\,(-04)        & 1.26\,(-03)          \\
\hline
%%             &                    &                    &                    &                    &                    &                      \\

             & 9.45\,(-04)        & 4.93\,(-04)        & 1.51\,(-03)        & 2.47\,(-03)        & 4.39\,(-03)        & 2.51\,(-02)          \\
10000        & 9.47\,(-05)        & 2.31\,(-05)        & 6.91\,(-05)        & 1.14\,(-04)        & 3.42\,(-05)        & 7.37\,(-05)          \\
             & 1.49\,(-03)        & 7.83\,(-05)        & 2.46\,(-04)        & 4.02\,(-04)        & 7.12\,(-04)        & 1.52\,(-03)          \\
%%\hline
%%             &                    &                    &                    &                    &                    &                      \\
\hline
\hline
\end{tabular}
}
\end{center}
\label{collcascadeMg11}
\end{table*}

%TABLE 13
\begin{table*}
\caption{Same as Table~\ref{collcascadeO8} but for the \ion{Si}{xiii}.}
\begin{center}
{\scriptsize
\begin{tabular}{ccccccc}
%\hline
\hline
\hline
%%             &                    &                    &                    &                    &                    &                      \\
T$_{\mathrm{e}}$/$Z^3$&$^{3}\mathrm{S}_{1}$&$^{3}\mathrm{P}_{0}$&$^{3}\mathrm{P}_{1}$&$^{3}\mathrm{P}_{2}$&$^{1}\mathrm{S}_{0}$& $^{1}\mathrm{P}_{1}$\\ 
\hline
\hline

             &1.26\,(-03)         & 7.63\,(-04)        &2.31\,(-03)         &3.81\,(-03)         & 2.51\,(-03)        & 8.07\,(-03)          \\
 400         &3.62\,(-04)         & 9.39\,(-05)        &2.82\,(-04)         &4.69\,(-04)         & 1.50\,(-04)        & 3.41\,(-04)          \\
             &5.44\,(-05)         & 3.11\,(-06)        &9.32\,(-06)         &1.60\,(-05)         & 3.18\,(-06)        & 2.31\,(-05)          \\
\hline
%%             &                    &                    &                    &                    &                    &                      \\

             &1.24\,(-03)         & 7.45\,(-04)        &2.25\,(-03)         &3.71\,(-03)         & 2.56\,(-03)        & 8.46\,(-03)          \\
 600         &3.88\,(-04)         & 9.84\,(-05)        &2.95\,(-04)         &4.88\,(-04)         & 1.52\,(-04)        & 3.39\,(-04)          \\
             &1.87\,(-04)         & 1.04\,(-05)        &3.13\,(-05)         &5.34\,(-05)         & 1.22\,(-05)        & 7.49\,(-05)          \\
\hline
%%             &                    &                    &                    &                    &                    &                      \\

             &1.21\,(-03)         & 7.13\,(-04)        &2.16\,(-03)         &3.57\,(-03)         & 2.60\,(-03)        & 9.00\,(-03)          \\
 900         &3.59\,(-04)         & 9.00\,(-05)        &2.68\,(-04)         & 4.45\,(-04)        & 1.36\,(-04)        & 3.00\,(-04)          \\
             &4.23\,(-04)         & 2.34\,(-05)        &7.01\,(-05)         &1.19\,(-04)         & 3.28\,(-05)        & 1.69\,(-04)          \\
\hline
%%             &                    &                    &                    &                    &                    &                      \\

             &1.16\,(-03)         & 6.70\,(-04)        &2.04\,(-03)         &3.34\,(-03)         & 2.67\,(-03)        & 9.79\,(-03)          \\
 1350        &3.01\,(-04)         & 7.44\,(-05)        &2.22\,(-04)         &3.66\,(-04)         & 1.11\,(-04)        &  2.44\,(-04)         \\
             &7.14\,(-04)         & 3.91\,(-05)        &1.18\,(-04)         &2.00\,(-04)         & 6.96\,(-05)        & 3.00\,(-04)          \\
\hline
%%             &                    &                    &                    &                    &                    &                      \\

             &1.09\,(-03)         & 6.16\,(-04)        & 1.88\,(-03)        &3.08\,(-03)         & 2.75\,(-03)        &1.08\,(-02)           \\
2000         &2.34\,(-04)         & 5.78\,(-05)        & 1.72\,(-04)        &2.84\,(-04)         & 8.54\,(-05)        &1.87\,(-04)           \\
             &9.68\,(-04)         & 5.27\,(-05)        &1.60\,(-04)         & 2.69\,(-04)        & 1.24\,(-04)        &4.51\,(-04)           \\
\hline
%%             &                    &                    &                    &                    &                    &                      \\

             &9.88\,(-04)         & 5.52\,(-04)        &1.69\,(-03)         &2.75\,(-03)         & 2.86\,(-03)        &1.22\,(-02)           \\
3000         &1.72\,(-04)         & 4.25\,(-05)        &1.27\,(-04)         &2.09\,(-04)         & 6.26\,(-05)        & 1.36\,(-04)          \\
             &1.14\,(-03)         & 6.17\,(-05)        &1.89\,(-04)         &3.16\,(-04)         & 2.02\,(-04)        & 6.20\,(-04)          \\
\hline
%%             &                    &                    &                    &                    &                    &                      \\

             &8.78\,(-04)         & 4.80\,(-04)        &1.48\,(-03)         &2.39\,(-03)         & 2.99\,(-03)        &1.41\,(-02)           \\
4500         &1.22\,(-04)         & 3.03\,(-05)        &9.01\,(-05)         &1.48\,(-04)         & 4.35\,(-05)        &9.61\,(-05)           \\
             &1.19\,(-03)         & 6.42\,(-05)        &2.00\,(-04)         & 3.30\,(-04)        & 3.06\,(-04)        & 7.98\,(-04)          \\
\hline
%%             &                    &                    &                    &                    &                    &                      \\

             &7.66\,(-04)         & 4.06\,(-04)        & 1.28\,(-03)        &2.03\,(-03)         & 3.13\,(-03)        &1.66\,(-02)           \\
6700         &8.54\,(-05)         & 2.12\,(-05)        & 6.33\,(-05)        &1.03\,(-04)         & 3.10\,(-05)        &6.75\,(-05)           \\
             &1.14\,(-03)         & 6.13\,(-05)        & 1.95\,(-04)        &3.17\,(-04)         & 4.36\,(-04)        &9.81\,(-04)           \\
\hline
%%             &                    &                    &                    &                    &                    &                      \\

             &6.47\,(-04)         &3.32\,(-04)         & 1.07\,(-03)        & 1.66\,(-03)        & 3.29\,(-03)        &1.99\,(-02)           \\
10000        &5.88\,(-05)         &1.46\,(-05)         & 4.35\,(-05)        & 7.15\,(-05)        & 2.14\,(-05)        &4.67\,(-05)           \\
             &1.02\,(-03)         &5.46\,(-05)         & 1.80\,(-04)        & 2.86\,(-04)        & 5.97\,(-04)        &1.17\,(-03)           \\
%%\hline
%%             &                    &                    &                    &                    &                    &                      \\
\hline
\hline
\end{tabular}
}
\end{center}
\label{collcascadeSi13}
\end{table*}

%TABLE 14
\begin{table*}[h]
\caption{Energy of the three main X-ray lines of \ion{C}{v}, \ion{N}{vi}, \ion{O}{vii}, \ion{Ne}{ix}, \ion{Mg}{xi} and \ion{Si}{xiii}, as well as the corresponding wavelength in \AA, in parentheses. 
{\bf w} corresponds to the resonance line, {\bf x+y} corresponds to the intercombination lines (here too close to be separated) and {\bf z} corresponds to the forbidden line.}
\label{lambda}
\begin{center}
{\scriptsize
\begin{tabular}{ccccccc}
\hline
\hline
%          &           &           &              &            &               &            \\  
Multiplet & \ion{C}{v}&\ion{N}{vi}&\ion{O}{vii}  &\ion{Ne}{ix} &\ion{Mg}{xi}  &\ion{Si}{xiii}\\
%          &           &           &              &              &             &            \\  
\hline
w        & 307.88    & 430.65    & 574.00       &  921.82      & 1357.07     & 1864.44    \\  
         &(40.27)    &(28.79)    &(21.60)       & (13.45)      &(9.17)       & (6.65)        \\
x+y      & 304.41    & 426.36    & 568.74       & 915.02       &1343.28      & 1853.29    \\  
         &(40.73)    &(29.08)    &(21.80)       &(13.55)       &(9.23)       &(6.69)               \\
z        & 298.97    & 419.86    & 561.02       & 905.00       & 1331.74     & 1839.54    \\  
         & (41.47)   &(29.53)    &(22.10)       &(13.70)       &(9.31)      &(6.74)      \\
\hline
\hline
\end{tabular}
}
\end{center}
\end{table*}

%***************************Table********************************

%------------------------------------------------------------------------------------


\begin{thebibliography}{}
\bibitem[1985]{Arnaud85} 
Arnaud, M. \& Rothenflug, R. 1985, A\&AS, 60, 425 
\bibitem[1982a]{Bely-Dubau82a} 
Bely-Dubau, F., Faucher, P., Dubau, J., Gabriel, A. H. 1982a, MNRAS, 198, 239 
\bibitem[1982b]{Bely-Dubau82b} 
Bely-Dubau, F., Faucher, P., Steenman-Clark, L., Dubau, J., Loulergue, M., Gabriel, A. H., Antonucci, E., Volonte, S., Rapley, C. G. 1982b, MNRAS, 201, 1155 
\bibitem[1972]{Blumenthal72}
Blumenthal, G. R., Drake, G. W. F., Tucker, W. H. 1972, ApJ, 172, 205 
\bibitem[1958]{Burgess58}
Burgess, A. 1958, MNRAS, 118, 477
\bibitem[1960]{BurgessSeaton60}
Burgess, A., Seaton, M. J. 1960, MNRAS, 121, 471 
\bibitem[1993]{Cunto93} 
Cunto, W., Mendoza, C., Ochsenbein, F., Zeippen, C. J. 1993, A\&A, 275, L5 
\bibitem[1980]{Doyle80} 
Doyle, J. G. 1980, A\&A, 87, 184 
\bibitem[1982]{Doyle82}
Doyle, J. G., Schwob, J. L. 1982, J. Phys B, 15, 813
\bibitem[1994]{Dubau94}
Dubau, J. 1994, ADNDT, 57, 21
\bibitem[1980]{DubauVolonte80} 
Dubau, J. \& Volonte, S. 1980, Reports of Progress in Physics, 43, 199
\bibitem[1974]{Eissner74}
Eissner, W., Jones M., Nussbaumer H., 1974, Comput. Phys. Commun., 8, 270
\bibitem[1987]{Fernley87}   
Fernley, J.A., Taylor, K. T., Seaton, M. J. 1987, J. Phys. B: At. Mol. Phys., 20, 6457
\bibitem[1969]{Gabriel69}
Gabriel, A. H., Jordan, C. 1969, MNRAS, 145, 241
\bibitem[1972]{GabrielJordan72}
Gabriel, A. H., Jordan, C. 1972, in ``Case studies in atomic collision physics'', vol.2, p209, McDaniel, McDowell (eds.).
\bibitem[1973]{Gabriel73}
Gabriel, A. H., Jordan, C. 1973, ApJ, 186, 327 
\bibitem[1995]{George95}
George, I. M., Turner, T. J., Netzer, H. 1995, ApJ 438, L67
\bibitem[1998]{George98} 
George, I. M., Turner, T. J., Netzer, H. , Nandra, K., Mushotzky, R. F., Yaqoob, T. 1998, ApJS, 114, 73 
\bibitem[1984]{Halpern84}
Halpern, J. C. 1984, ApJ, 281, 90
\bibitem[1997]{Iwasawa97} 
Iwasawa, K., Fabian, A. C., Matt, G. 1997, MNRAS, 289, 443
\bibitem[1977]{Jacobs77} 
Jacobs, V. L., Davis, J., Kepple, P. C. \& Blaha, M. 1977, ApJ, 211, 605 
\bibitem[1997]{Leighly97} 
Leighly, K. M., Matsuoka, M. , Cappi, M., Mihara, T.  1997, IAU Symposia 188, 249 
\bibitem[1999]{Liedahl99}
Liedahl, D.A. 1999, in X-rays Spectroscopy in Astrophysics, EADN School proceeding, 1997, J.A. van Paradijs, J.A.M Bleeker (eds.), p.189
\bibitem[1977]{Lin77}
Lin, C. D., Johnson, W. R., Dalgarno, A. 1977, Physical Review A, vol. 15, 154
\bibitem[1998]{Mazzotta98} Mazzotta, P., Mazzitelli, G., Colafrancesco, S., Vittorio, N. 1998, A\&AS, 133, 403 
\bibitem[1935]{MenzelPekeris35}
Menzel, D.H, Pekeris, C.L. 1935, MNRAS, 96, 77
\bibitem[1978a]{Mewe78a} 
Mewe, R. \& Schrijver, J. 1978a, A\&A, 65, 99
\bibitem[1978b]{Mewe78b} 
Mewe, R. \& Schrijver, J. 1978b, A\&AS, 33, 311 
\bibitem[1999]{Mewe99}
Mewe, R. 1999, in X-rays Spectroscopy in Astrophysics, EADN School proceeding, 1997, J.A. van Paradijs, J.A.M Bleeker (eds.), p.109
 \bibitem[1999]{Nahar99} 
Nahar, S. N. 1999, ApJS, 120, 31 
\bibitem[1994]{Nandra94}
Nandra, K., Pounds, K. A. 1994, MNRAS, 268, 405
\bibitem[1993]{Netzer93}
Netzer, H., 1993, ApJ 411, 594 
\bibitem[1997]{Netzer97}
Netzer, H., Turner, T. J. 1997, ApJ, 488, 694 
\bibitem[1999]{Nicastro99} 
Nicastro, F. , Fiore, F. , Perola, G. C., Elvis, M.  1999, ApJ, 512, 184 
\bibitem[1996]{Otani96}
Otani, C. , Kii, T. , Reynolds, C. S., Fabian, A. C., Iwasawa, K. , Hayashida, K. , Inoue, H. , Kunieda, H., Makino, F. , Matsuoka, M., Tanaka, Y.  1996, PASJ, 48, 211 
\bibitem[1998]{Paerelsetal98}
Paerels, F. et al. 1998, in Proceedings of the First XMM Workshop on "Science with XMM", held at Noordwijk, The Netherlands, M.Dahlem (ed.), URL http://astro.estec.esa.nl/XMM/news/ws1/ws1$_{}$papers.html
\bibitem[1999]{Paerels99}
Paerels, F. 1999, in X-rays Spectroscopy in Astrophysics, EADN School proceeding, 1997, p.347, J.A. van Paradijs, J.A.M Bleeker (eds.)
\bibitem[1991]{Pequignot91} 
Pequignot, D., Petitjean, P. \& Boisson, C. 1991, A\&A, 251, 680
\bibitem[1998]{PorquetDumont98}
Porquet, D., Dumont, A.M. 1998, in "Structure and Kinematics of Quasar Broad Line Regions", Eds C. M. Gaskell, W. N. Brandt, M. Dietrich, D. Dultzin-Hacyan, and M. Eracleous, ASP Conf. Ser., in press
\bibitem[1998]{Porquet98}
Porquet, D., Dumont, A.M., Mouchet, M. 1998, Author, I. 1998, in Proceedings of the First XMM Workshop on "Science with XMM", held at Noordwijk, The Netherlands, M. Dahlem (ed.), URL http://astro.estec.esa.nl/XMM/news/ws1/ws1$_{}$papers.html 
\bibitem[1999]{Porquet99}
Porquet, D., Dumont, A.M., Collin, S., Mouchet, M. 1999, A\&A, 341, 58
\bibitem[1985]{Pradhan85} 
Pradhan, A. K. 1985, ApJ, 288, 824 
\bibitem[1981]{Pradhan81} 
Pradhan, A. K., Shull, J. M. 1981, ApJ, 249, 821 
\bibitem[1997]{Piro97}
Piro, L., Balucinska-Church, M., Fink, H., Fiore, F., Matsuoka, M., Perola, G. C., Soffitta, P. 1997, A\&A, 319, 74 
\bibitem[1997]{Reynolds97}
Reynolds, C. S., 1997, MNRAS, 286, 513 
\bibitem[1983]{Sampson83}
Sampson, D. H., Goett, S. J., Clarck, R. E. H. 1983, ADNDT, 29, 467
\bibitem[1959]{Seaton59}
Seaton, M. J. 1959, MNRAS, 119, 81
\bibitem[1997]{Turner97} 
Turner, M. J. L., Bleeker, J. A. M., Hasinger, G., Peacock, A., Truemper, J. 1997, IAU Symposia, 188, E20
\bibitem[1994]{Ueno94}
Ueno, S., Mushotzky, R. F., Koyama, K., Iwasawa, K., Awaki, H., Hayashi, I. 1994, PASJ, 46, L71 
\bibitem[1985]{Vainshtein85}
Vainshtein, L. A. \& Safronova U. I. 1985, Physica Scripta, vol. 31, 519
\bibitem[1981]{Winkler81} 
Winkler, P. F., Clark, G. W., Markert, T. H., Petre, R., Canizares, C. R. 1981, ApJ, 245, 574 
\bibitem[1987]{Zhang87}
Zhang, H. \& Sampson, D. H. 1987, ApJS, 63, 487
\end{thebibliography}
\end{document}